\documentclass[aps,prd,nofootinbib,print,floatfix,superscriptaddress]{revtex4-1}
\usepackage{amsfonts}
\usepackage{amsmath}
\usepackage{hyperref}
\usepackage{amssymb}
\usepackage[english]{babel}
\usepackage{graphicx}
\usepackage{epsfig}
\usepackage{bm}
\usepackage{verbatim}
\usepackage[utf8]{inputenc}
\usepackage{color}
\usepackage[dvipsnames]{xcolor}
\usepackage{subfig}
\usepackage[normalem]{ulem}
\usepackage{hyperref}

\usepackage{feynmp-auto}

\usepackage{slashed}

\usepackage{multirow}  
\usepackage{makecell} 
\usepackage{booktabs}

\usepackage{tikz}
\usetikzlibrary{shapes.geometric,arrows}

\hypersetup{colorlinks=true, linkcolor=red, urlcolor=red, citecolor=red}
\newcommand{\mathsym}[1]{{}}

\newcommand{\bmx}{\left(\begin{array}}
\newcommand{\emx}{\end{array}\right)}

\DeclareUnicodeCharacter{2212}{-}
\begin{document}
	
	\title{Collider signatures of vector-like fermions from a flavor symmetric model}

	\author{Cesar Bonilla}
	\email{cesar.bonilla@ucn.cl}
	\affiliation{Departamento de Física, Universidad Católica del Norte, 
		Avenida Angamos 0610, Casilla 1280, Antofagasta, Chile}

	\author{A. E. C\'{a}rcamo Hern\'{a}ndez}
	\email{antonio.carcamo@usm.cl}
	\affiliation{Departamento de F\'{\i}sica, Universidad T\'ecnica Federico Santa Mar\'{\i}a\\
		Casilla 110-V, Valpara\'{\i}so, Chile\\
		}
		\affiliation{Centro Cient\'{\i}fico-Tecnol\'{o}gico de Valpara\'{\i}so, Casilla 110-V, Valpara\'{\i}so, Chile}
		\affiliation{Millennium Institute for Subatomic Physics at the High-Energy Frontier, SAPHIR, Chile}

	\author{João Gonçalves}
	\email{jpedropino@ua.pt}
	\affiliation{ Departamento de F\'{i}sica da Universidade de Aveiro, 3810-183 Aveiro, Portugal \\
	}
	\affiliation{Centre for Research and Development in Mathematics and Applications (CIDMA), 3810-183 Aveiro, Portugal  
	}
	
	\author{Felipe F. Freitas}
	\email{felipefreitas@ua.pt}
	\affiliation{ Departamento de F\'{i}sica da Universidade de Aveiro, 3810-183 Aveiro, Portugal \\
	}
	\affiliation{Centre for Research and Development in Mathematics and Applications (CIDMA), 3810-183 Aveiro, Portugal  
	}
	
	\author{António P.~Morais}
	\email{aapmorais@ua.pt}
	\affiliation{ Departamento de F\'{i}sica da Universidade de Aveiro, 3810-183 Aveiro, Portugal \\
	}
	\affiliation{Centre for Research and Development in Mathematics and Applications (CIDMA), 3810-183 Aveiro, Portugal  
	}

	\author{R. Pasechnik}
	\email{roman.pasechnik@thep.lu.se}
	    \affiliation{
		\sl Department of Astronomy and Theoretical Physics, Lund University, S\"{o}lvegatan 14A, SE 223-62 Lund, Sweden 
		}

	\date{\today }
	
	\begin{abstract}

	 We propose a model with two Higgs doublets and several $SU(2)$ scalar singlets with a global non-Abelian flavor symmetry $\mathcal{Q}_6\times\mathcal{Z}_2$. This discrete group accounts for the observed pattern of fermion masses and mixing angles after spontaneous symmetry breaking.  In this scenario only the third generation of fermions get their masses as in the Standard Model (SM). The masses of the remaining fermions are generated through a seesaw-like mechanism. To that end, the matter content of the model is enlarged by introducing electrically charged vector-like fermions (VLFs), right handed Majorana neutrinos and several SM scalar singlets. Here we study the processes involving VLFs  that are within the reach of the Large Hadron Collider (LHC). We perform collider studies for vector-like leptons (VLLs) and vector-like quarks (VLQs), focusing on double production channels for both cases, while for VLLs single production topologies are also included. Utilizing genetic algorithms for neural network optimization, we determine the statistical significance for a hypothetical discovery at future LHC runs. In particular, we show that we can not safely exclude VLLs for masses greater than $200~\mathrm{GeV}$. For VLQ's in our model, we show that we can probe their masses up to 3.8 TeV, if we take only into account the high-luminosity phase of the LHC. Considering Run-III luminosities, we can also exclude VLQs for masses up to $3.4~\mathrm{TeV}$. We also show how the model with predicted VLL masses accommodates the muon anomalous magnetic moment.
	
	\end{abstract}
	
	\maketitle
	
	\section{Introduction}
	
	\label{sec:intro}

The Standard Model (SM) provides a successful framework to describe three out of four known fundamental forces of nature, i.e. the electromagnetic, nuclear strong and weak interactions. However, it does not account for the number of fermion generations and lacks a natural explanation for the tremendous hierarchy in the fermion sector, which is extended over a range of 13 orders of magnitude from the light active neutrino mass scale up to the top quark mass. Moreover, there is no assertion for the smallness of the quark mixing angles, which is in contrast with the sizable values of two of the three leptonic mixing angles. This set of issues is the so called flavor problem which, among others, motivates the construction of models where the SM particle content and symmetries are enlarged. One way to tackle the flavor problem is offered in SM extensions that assume the existence of discrete flavor symmetries\footnote{For more details about flavor symmetry groups see, for example  \cite{King:2013eh,Altarelli:2010gt,Ishimori:2010au,King:2015aea}}. These scenarios feature relations in the Yukawa sector and typically predict correlations between the observed fermion mixing patterns as well as the fermion mass relations.

With the use of flavor symmetries it can be suggested that the existing number of fermion families is because they transform as components of a three-dimensional irreducible representation (irrep) of a non-Abelian discrete group, such as $A_4$. Another option is to have the heaviest fermion transforming as one-dimensional irrep and the lighter ones being the components of a doublet under the symmetry group. The smaller non-Abelian discrete groups, containing one- and two-dimentional irreps, are \cite{Ishimori:2010au}: $S_{3}$, $Q_{4}$, $D_4$ and $Q_6$\footnote{Pioneer works using these symmetries to tackle the flavor problem can be found in \cite{Gerard:1982mm,Frampton:1994rk,Grimus:2003kq,Kubo:2003iw,Kubo:2003pd,Babu:2004tn,Kajiyama:2005rk,Lovrekovic:2012bz}}. Both assumptions point to the idea on how to account for the three generations of quarks and leptons but not for their (very strong) mass hierarchies. 

For this reason, apart of the flavor symmetry, the distribution of the fermion mass spectrum could suggest the existence of new particles, resulting in phenomenologically richer setups. One can add more scalars to the SM with non-zero vacuum expectation values (vevs) whose contributions to the fermion masses or to different mass matrix elements are restricted by the additional symmetry. Similar to the role of a $\mathcal{Z}_2$ symmetry in a 2-Higgs doublet model (2HDM) \cite{Branco:2011iw}. One fashionable approach to explain the fermion mass and mixing hierarchies is by using the  Froggatt–Nielsen (FN) mechanism~\cite{Froggatt:1978nt}, where vector-like fermions (VLFs) are introduced to the SM and transform under a new $\mathrm{U(1)_F}$ symmetry which is spontaneously broken by the vevs of $\mathrm{SU(2)}$ scalar singlets (flavons). The new energy scale is much bigger than the electroweak one as well as the VLFs are heavier than the SM ones, then all the new fields are effectively integrated out.
 
Here, we present a framework that somehow gathers the previous ideas. We consider a multiscalar model with the flavor symmetry $\mathcal{Q}_{6}\times\mathcal{Z}_{2}$, where the $\mathcal{Z}_{2}$ symmetry assigns one Higgs to the up-type fermion sector and the other to the down-type, both scalars are $\mathcal{Q}_{6}$ singlets. In contrast, fermions transform as ($doublet$+$singlet$) under $\mathcal{Q}_6$, preventing the Yukawa interaction between light fermions and the $\mathrm{SU(2)}$ scalar doublets. Therefore, one Higgs doublet furnishes the top quark with a no-null mass, whereas the other one generates the bottom quark and tau mass. In order to generate the mass of light fermions through a seesaw (or FN) mechanism, we introduce VLFs and three different flavons with a non-trivial transformation under the flavor symmetry. We also introduce right-handed (RH) Majorana neutrinos to generate the small neutrino masses via a type-I seesaw mechanism.

In contrast to the FN mechanism, in our model, the VLFs are not decoupled. For this reason, we study the processes involving VLFs that are within the reach of the Large Hadron Collider (LHC). We perform collider studies for vector-like leptons (VLLs) and vector-like quarks (VLQs), focusing on double production channels for both cases, while for VLLs the single production topologies are also included. Furthermore, it has been known that the experimentally measured muon anomalous magnetic moment deviates from the SM prediction. The longstanding non-compliance of the muon $(g-2)$ with the SM was first observed by the Brookhaven E821 experiment at BNL~\cite{Bennett:2006fi} and it has been recently confirmed by the Muon $(g-2)$ experiment at FERMILAB \cite{Abi:2021gix}. Hence, we also show how the muon $(g-2)$ anomaly is accommodated within our theory with the predicted VLLs.

The layout of the remainder of the paper is as follows. In Sec.~\ref{sec:model} we describe the model, i.e. we provide the invariant Yukawa Lagrangian, the scalar potential and the particle mass spectrum. Afterwards, in Sec.~\ref{section:Muon_g2}, the consequences of our model in the muon anomalous magnetic moment are analyzed. In Sec.~\ref{section:VLLs_collider} we detail the methodology for the collider analysis of the VLFs, for both the quark and lepton counterparts, with the numerical results being showcased in the Sec.~\ref{section:Results}. We finalize this paper with Sec.~\ref{sec:conclusions}, where we take our conclusions. 

\section{Model Description}
\label{sec:model}

We propose a model where the SM gauge group is extended with a global flavor symmetry 
group, i.e. the complete description is given by the symmetry, $\mathrm{SU(3)_C \times SU(2)_L \times U(1)_Y} \times \mathcal{Q}_{6}\times \mathcal{Z}_{2}$. This theory adds to the SM particle content, a second $\mathrm{SU(2)}$ scalar doublet, three flavon fields, two RH neutrinos, a flavor doublet of VLQs and flavor doublet of VLLs. The charge assignments of the particle content under the flavor group are shown in Tables~\ref{tab:model} and \ref{tab:model2}.
	
	\begin{table}[t!]
		\begin{tabular}{|c|cc|cc|cc|cc|cc|cc|}
			\hline
			& $H_{1}$ & $H_{2}$ & $Q_{L_D}$ & $Q_{L_3}$ & $u_{R_D}$ & ${u}_{R_3}$ & $d_{R_D}$ & ${d}_{R_3}$ & $L_{L_D}$ & $ L_{L_3} $ & ${\ell }_{R_D}$ & $\ell_{R_3}$ \\ \hline\hline
			$SU(2)_L$ & \bf{2} & \bf{2} & \bf{2} & \bf{2} & \bf{1} & \bf{1} & \bf{1} & \bf{1}& \bf{2} & \bf{2} & \bf{1}& \bf{1}\\
			$U(1)_Y$ & 1/2 & 1/2 & 1/6 & 1/6 & 2/3 & 2/3 & -1/3 & -1/3& -1/2 & -1/2 & -1 & -1\\
			$\mathcal{Q}_{6}$ & $\mathbf{1}_{+-}$ & $\mathbf{1}_{+-}$ & $\mathbf{2}_{2}$
			& $\mathbf{1}_{++}$ & $\mathbf{2}_{2}$ & $\mathbf{1_{+-}}$ & $\mathbf{2}_{2}$ & $\mathbf{1}_{+-}$ & $\mathbf{2}_{2}$ & $\mathbf{1}_{++}$ & $\mathbf{2}_{2}$ & 
			$\mathbf{1}_{+-}$ \\ 
			$\mathcal{Z}_{2}$ & $-1$ & $+1$ & $+1$ & $+1$ & $-1$ & $-1$ & $+1$ & $+1$ & $%
			+1$ & $+1$ & $+1$ & $+1$ \\ \hline
		\end{tabular}%
		\caption{Charge assignments of the SU(2) scalar doublets and SM fermions
			under the symmetry, $\mathcal{Q}_{6}\times \mathcal{Z}_{2}$.
			 We have arranged the $\mathcal{Q}_{6}$ doublets as, $Q_{L_D}\equiv (Q_{L_1},Q_{L_2})^T$, $u_{R_D}\equiv (u_{R_1},{u}_{R_2})^T$, $d_{R_D}\equiv ({d}_{R_1},{d}_{R_2})^T$, 
			 $L_{L_D}\equiv (L_{L_1},L_{L_2})^T$ and ${\ell}_{R_D}\equiv ({\ell }_{R_1},{\ell}_{R_2})^T$. 
			}
		\label{tab:model}
	\end{table}
	\begin{table}[t!]
		{\small 
			\begin{tabular}{|c|ccc|cc|cc|cc|cc|}
				\hline
				& $\sigma _{1}$ & $\sigma _{2}$ & $\xi $ & $N_{R_1}$ & $N_{R_2}$ & $T_{L}$ 
				& $T_{R}$ & $ B_{L}$ & $B_{R}$ & $E_{L}$ & $E_{R} $ \\ \hline\hline
				$SU(2)_L$ & \bf{1} & \bf{1} & \bf{1} & \bf{1} & \bf{1} & \bf{1} & \bf{1} & \bf{1}& \bf{1} & \bf{1} & \bf{1} \\
				$U(1)_Y$ & 0 & 0 & 0 & 0 & 0 & 2/3 & 2/3 & -1/3& -1/3 & -1 & -1 \\
				$\mathcal{Q}_{6}$ & $\mathbf{1}_{++}$ & $\mathbf{1}_{+-}$ & $\mathbf{2}_{2}$
				& $\mathbf{1}_{+-}$ & $\mathbf{1}_{+-}$ & $\mathbf{2}_{1}$ & $\mathbf{2}_{1}$ & $\mathbf{2}%
				_{1} $ & $\mathbf{2}_{1}$ & $\mathbf{2}_{1}$ & $\mathbf{2}_{1}$ \\ 
				$\mathcal{Z}_{2}$ & $-1$ & $-1$ & $-1$ & $+1$ & $+1$ & $+1$ & $-1$ & $-1$ & $%
				+1$ & $-1$ & $+1$ \\ \hline
			\end{tabular}%
		} 
		\caption{Assignments of the singlet scalars and exotic fermions under the $%
			\mathcal{Q}_{6}$ flavor symmetry irreps. For convenience we have not 
			included a subindex $D$ for $\mathcal{Q}_{6}$ doublets.}
		\label{tab:model2}
	\end{table}
	
Given the matter content in our theory, the invariant Yukawa Lagrangian is formed by the contribution from each sector as,
	\begin{equation}
	   \mathcal{L}_{Y}= \mathcal{L}_{u} + \mathcal{L}_{d} + \mathcal{L}_{\ell} + \mathcal{L}_{\nu} 
	\end{equation}
	where  
	\begin{equation}
	\mathcal{L}_{u}=
	  y_{u3}\overline{Q}_{L_3}u_{R_3}\widetilde{H}_{1}
	 +y_{T}\overline{Q}_{D} T_{R} \widetilde{H}_{1}
	 +y_{T1}\overline{T}_{L}T_{R}\sigma _{1}
	 +y_{T2}\overline{T}_{L} u_{R_D} \sigma_{2}
	 +y_{T3}\overline{T}_{L} u_{R_3} \xi	 
	 +y_{T4}\overline{T}_{L}T_{R}\xi + \mathrm{H.c.},
	\label{Lyu}
	\end{equation}
	\begin{equation}
	\mathcal{L}_d = 
	  y_{d3}\overline{Q}_{L_3}d_{R_3}H_{2}
	 +y_{B}\overline{Q}_{D}B_{R} H_{2}
	 +y_{B1}\overline{B}_{L}B_{R}\sigma _{1}
	 +y_{B2}\overline{B}_{L}d_{R_D}\sigma_{2}
	 +y_{B3}\overline{B}_{L}d_{R_3} \xi 
	 +y_{B4}\overline{B}_{L}B_{R}\xi +\mathrm{H.c.},
	\label{Lyd}
	\end{equation}
	\begin{equation}
	\mathcal{L}_{\ell}=
	 y_{\ell3}\overline{L}_{L_3}\ell _{R_3}H_{2}
    +y_{E}\overline{L}_{L_D}E_{R}H_{2}
    +y_{E1}\overline{E}_{L}E_{R}\sigma _{1}
    +y_{E2}\overline{E}_{L}\ell _{R_D}\sigma _{2}
    +y_{E3}\overline{E}_{L}\ell_{R_3}\xi
    +y_{E4}\overline{E}_{L}E_{R}\xi + \mathrm{H.c.},  
    \label{Lyl}
	\end{equation}
	and
	\begin{equation}
	\mathcal{L}_{\nu}=\sum_{i=1}^{2}\frac{1}{\Lambda }
	\left( 
	 y_{\nu _{i}}\overline{L}_{L_3}N_{R_i}\widetilde{H}_{1}\sigma_{1}
    +y'_{\nu _{i}}\overline{L}_{L_D}N_{R_i}\widetilde{H}_{1}\xi 
    \right) 
    +\sum_{i=1}^{2}M_{R_{i}}N_{R_i}\overline{N_{R_i}^{C}}+ \mathrm{H.c.}
	\label{Lynu}
	\end{equation}
	with $\tilde{H}_a=i\sigma_2 H^*_a$. 
	We have defined the $\mathcal{Q}_{6}$ doublets as, $Q_{L_D}\equiv (Q_{L_1},Q_{L_2})^T$, $u_{R_D}\equiv (u_{R_1},{u}_{R_2})^T$, $d_{R_D}\equiv ({d}_{R_1},{d}_{R_2})^T$, $L_{L_D}\equiv (L_{L_1},L_{L_2})^T$, ${\ell}_{R_D}\equiv ({\ell }_{R_1},{\ell}_{R_2})^T$, $T_{L,R}\equiv (T_{L_1,R_1},T_{L_2,R_2})^T$
	and $B_{L,R}\equiv (B_{L_1,R_1},B_{L_2,R_2})^T$. Using the multiplication rules of $\mathcal{Q}_{6}$ given in Appendix~\ref{app}, the above Yukawa interactions can be rewritten as follows
	\begin{eqnarray}
	\mathcal{L}_{u}&=&
	 y_{u3} \overline{Q}_{L_3}u_{R_3}\widetilde{H}_{1}
    +y_{T}
    \left( \overline{Q}_{L_1}T_{R_1}-\overline{Q}_{L_2}T_{R_2}\right)\widetilde{H}_{1}
	+y_{T1}\left( \overline{T}_{L_1}T_{R_2}-\overline{T}_{L_2}T_{R_1}\right)\sigma_{1}\notag\\
	&+&y_{T2}
	\left( \overline{T}_{L_1}u_{R_1}-\overline{T}_{L_2}u_{R_2}\right)\sigma _{2} 
	+y_{T3}\left( \overline{T}_{L_1}\xi_{1}-\overline{T}_{L_2}\xi _{2}\right) u_{R_3} 
	+y_{T4}\left( \overline{T}_{L_1}T_{R_1}\xi_{2}-\overline{T}_{L2}T_{R_2}\xi _{1}\right) 
	\label{Lyu2}
	\end{eqnarray}
	 
	 \begin{eqnarray}
	  \mathcal{L}_{d}&=&
	 y_{d3}\overline{Q}_{L_3}d_{R_3}H_{2}
    +y_{B}\left( \overline{Q}_{L_1}B_{R_1}-\overline{Q}_{L_2} B_{R_2}\right)H_{2}
    +y_{B1}\left( \overline{B}_{L_1}B_{R_2}-\overline{B}_{L_2}B_{R_1}\right)\sigma _{1}\notag\\ 
	&+&y_{B2}\left( \overline{B}_{L_1}d_{R_1}-\overline{B}_{L_2}d_{R_2}\right)\sigma _{2}
	+y_{B3}\left( \overline{B}_{L_1}\xi _{1}{-}\overline{B}_{L_2}\xi _{2}\right) d_{R_3} 
	+y_{B4}\left( \overline{B}_{L_1}B_{R_1}\xi _{2}-\overline{B}_{L_2}B_{R_1}\xi _{1}\right) 
	+\mathrm{H.c.},  \label{Lyd2}
	\end{eqnarray}
	
	\begin{eqnarray}
	\mathcal{L}_{\ell} &=&
	y_{\ell3}\overline{L}_{L_3}\ell _{R_3}H_{2}
	+y_{E}\left( \overline{L}_{L_1}E_{R_1}-\overline{L}_{L_2}E_{R_2}\right) H_{2}
	+y_{E1}\left( \overline{E}_{L_1}E_{R_2}-\overline{E}_{L_2}E_{R_1}\right)\sigma _{1} \notag\\
	&+&y_{E2}\left( \overline{E}_{L_1}\ell _{R_1}-\overline{E}_{L_2}\ell _{R_2}\right)\sigma _{2}  
	+ y_{E3}\left( \overline{E}_{L_1}\xi_{1}{-}\overline{E}_{L_2}\xi _{2}\right)
	\ell_{R_3}
	+y_{E4}\left( \overline{E}_{L_1}E_{R_1}\xi _{2}-\overline{E}_{L_2}E_{R_2}\xi _{1}\right) + \mathrm{H.c.}, \label{Lyl2}
		\end{eqnarray}

	\begin{equation}
	\mathcal{L}_{\nu}=\sum_{i=1}^{2}\frac{1}{\Lambda }\left[ 
	y_{\nu _{i}}\overline{L}_{L_3}N_{R_i}\widetilde{H}_{1}\sigma_{1}
	+y'_{\nu _{i}}\left( \overline{L}_{L_1}N_{R_i}\widetilde{H}_{1}\xi _{1}-\overline{L}_{L_2}N_{R_i}\widetilde{H}_{1}\xi _{2}\right) %
	\right] +\sum_{i=1}^{2}M_{R_{i}}N_{R_i}\overline{N_{R_i}^{C}}+\mathrm{H.c.}
	\label{Lynu2}
	\end{equation}
	
	In addition, the invariant scalar potential reads
	\begin{equation}
	V=V_{\text{2HDM}}+V(\text{$H_{1}$,$H_{2}$,flavons})
	\label{ec:pot1}
	\end{equation}
	where the first term corresponds to the 2HDM potential 
	The second term
	in eq.~(\ref{ec:pot1}) has the contributions from the flavons fields, $\sigma_1$, $\sigma_2$ and $\xi$, i.e. the interactions among them and with the $\mathrm{SU(2)}$ scalar doublets. Explicitly,
	\begin{eqnarray}
	V_{\text{2HDM}}&=& -\mu _{1}^{2}\left( H_{1}^{\dagger }H_{1}\right) -\mu _{2}^{2}\left(
	H_{2}^{\dagger}H_{2}\right)
	+\frac{\lambda_{1}}{2}\left(H_{1}^{\dagger }H_{1}\right) ^{2} \notag\\
	&+& \frac{\lambda _{2}}{2}\left(H_{2}^{\dagger }H_{2}\right) ^{2}+\lambda _{3}\left(
	H_{1}^{\dagger }H_{1}\right) \left( H_{2}^{\dagger }H_{2}\right)
	+\lambda_{4}\left(H_{1}^{\dagger }H_{2}\right) \left(H_{2}^{\dagger
	}H_{1}\right) +\frac{\lambda _{5}}{2}\left[\left(H_{1}^{\dagger
	}H_{2}\right)^2+\mathrm{H.c.}\right]
	\label{ec:pot2hdm}
	\end{eqnarray}
	where $H_i=\left(\Phi^{+}_{i}, \Phi^{0}_{i}\right)^{T}$ with $i=1,2$ and
	\begin{eqnarray}  \label{eq:potential}
	V(\text{$H_{1}$,$H_{2}$,flavons}) &=&
	-\mu _{3}^{2}\sigma_1^{*}\sigma_1
	-\mu_{4}^{2}\sigma_2^{*}\sigma_2 
	-\mu_{5}^{2}\xi_1^{*}\xi_1
	-\mu_{6}^2\xi_2^{*}\xi_2
	-\mu_{7}^{2}(\xi_1^{*}\xi_2+\mathrm{H.c.}) 
	-\mu _{8} \left(\sigma_1 H_{1}^{\dagger }H_{2}+ \mathrm{H.c.} \right)
	+\lambda_{6} \left(H_{1}^{\dagger }H_{1}\right) (\sigma_1 ^{*}\sigma_1)\notag\\
	&+&\lambda_{7}\left(H_{2}^{\dagger }H_{2}\right) (\sigma_1 ^{*}\sigma_1)
	+\lambda _{8}\left(H_{1}^{\dagger }H_{1}\right) (\sigma_2 ^{*}\sigma_2)
	+\lambda_{9}\left(H_{2}^{\dagger }H_{2}\right) (\sigma_2 ^{*}\sigma_2) 	
	+\frac{\lambda _{10}}{2}(\sigma_1 ^{*}\sigma_1)^{2}
	+\frac{\lambda _{11}}{2}(\sigma_2 ^{*}\sigma_2)^2 \notag\\
	&+&\lambda_{12}(\sigma_1 ^{*}\sigma_1)(\sigma_2^{*}\sigma_2) 
	+\lambda^{^{\prime }}_{12}(\sigma_1 ^{*}\sigma_2)(\sigma_2^{*}\sigma_1) 
	+\lambda^{^{\prime \prime }}_{12}\left[(\sigma_1^{*}\sigma_2)^2+\mathrm{H.c.}\right] 
	+\frac{\lambda_{13}}{2}\left( \xi^*\xi \right)_{\mathbf{1}_{--}}\left(\xi^*\xi \right)_{\mathbf{1}_{--}},
	\end{eqnarray}

    The terms $\mu_{5,6,7}$ in the last equation break softly the $\mathcal{Q}_6$ symmetry\footnote{$\xi$ does not mix with the other scalars because of the $\mathcal{Q}_6$ symmetry. We have, for real $\xi$, $(\xi^2)_{++}=\xi_1 \xi_2 - \xi_2 \xi_1=0$ and, if complex, $(\xi^*\xi)_{++}+\mathrm{H.c.}=0$.} and prevent the appearance of either Goldstone or tachyonic fields. Since the flavons are real fields, they have no complex charge assignment,  the $\lambda_{12}$, $%
	\lambda^{\prime }_{12}$ and $\lambda^{\prime \prime }_{12}$ terms are equivalent. Then,
    eq.~(\ref{eq:potential}) can be rewritten (discarding redundant terms) as follows,
	\begin{eqnarray}
	V(\text{$H_{1}$,$H_{2}$,flavons}) &=&
	-\mu _{3}^{2}\sigma_1^{2}-\mu_{4}^{2}\sigma_2^{2} -\mu_{5}^{2}\xi_1^{2}-\mu_{6}^2\xi_2^{2}-\mu_{7}^{2}\xi_1\xi_2
	-\mu _{8} \left(\sigma_1 H_{1}^{\dagger }H_{2}+ \mathrm{H.c.} \right)
	+\lambda_6 \left(H_{1}^{\dagger }H_{1}\right) \sigma_1 ^{2}
	+\lambda_{7}\left(H_{2}^{\dagger }H_{2}\right) \sigma_1 ^{2} \notag \\
	&+&\lambda_{8}\left(H_{1}^{\dagger }H_{1}\right)\sigma_2 ^{2}
	+\lambda_{9}\left(H_{2}^{\dagger }H_{1}\right)\sigma_2 ^{2} 
	+\frac{\lambda_{10}}{2}\sigma_1^{4}
	+\frac{\lambda _{11}}{2}\sigma_2^4 
	+\lambda_{12}\sigma_1 ^{2}\sigma_2 ^{2}
	+\frac{\lambda _{13}}{2}\left(2 \xi_1\xi_2 \right)^2,
	\label{ec:potflavon}
	\end{eqnarray}
	
	All these scalars contribute to the symmetry breaking, they get a no-null vev,
	and they are shifted as follows
	\begin{equation}
	\Phi_i^0=\frac{1}{\sqrt{2}}\left(v_i+\varphi_{R_i}+i \varphi_{I_i}\right),
	\sigma_i=\frac{1}{\sqrt{2}}\left(v_{\sigma_i}+\sigma_{R_i}\right)\ \ \text{and%
	} \ \ \xi_i=\frac{1}{\sqrt{2}}\left(v_{\xi_i}+\xi_{R_i}\right),
	\end{equation}
	where $i=1,2$, the $SU(2)$ scalar vevs satisfy $v^2_1 +v^2_2 = v_{\rm EW}^2$ and $v_{\rm EW}\equiv246\,\text{GeV}$. 
	
	\subsection*{Fermion mass spectrum}
	\label{sec:fermionmasses}
	
	After the spontaneous breaking of 
	the $\mathrm{SU(3)_C\times SU(2)_L \times U(1)_Y \times \mathcal{Q}_{6}\times \mathcal{Z}_{2}}$ symmetry,
	using eqs.(\ref{Lyu}), (\ref{Lyd}) and (\ref{Lyl}), we get $5\times5$ fermion mass matrices,
	\begin{eqnarray}
	M_{f}&=&\frac{1}{\sqrt{2}}\left( 
	\begin{array}{ccccc}
	0 & 0 & 0 & y_{F} v_{H_{1}} & 0 \\ 
	0 & 0 & 0 & 0 & -y_{F}v_{H_{1}} \\ 
	0 & 0 & y_{f3}v_{H_{1}} & 0 & 0 \\ 
	y_{F2}v_{\sigma _{2}} & 0 & y_{F3}v_{\xi _{1}} & y_{F4}v_{\xi _{2}} & y_{F1}v_{\sigma _{1}} \\ 
	0 & -y_{F2}v_{\sigma _{2}} & -y_{F3}v_{\xi _{2}} & -y_{F1}v_{\sigma _{1}} & 
	-y_{F4}v_{\xi _{1}}%
	\end{array}%
	\right) =\left( 
	\begin{array}{cc}
	C_{f}& A_{f}\\ 
	B_{f}& M_{F}%
	\end{array}%
	\right) ,   \label{MF1}
	\end{eqnarray}
	where the sub-indices $f=u,d,\ell$ and $F=T,B,E$. The block matrices, in the previous equation, are defined as
	\begin{equation}
	C_{f}=\left( 
	\begin{array}{ccc}
	0 & 0 & 0 \\ 
	0 & 0 & 0 \\ 
	0 & 0 & M_{f_{33}}%
	\end{array}%
	\right), \ \
	A_{f}=\left( 
	\begin{array}{cc}
	M_{f_{14}} & 0 \\ 
	0 & M_{f_{25}} \\ 
	0 & 0%
	\end{array}%
	\right),  \ \
	 B_{f}=\left( 
	\begin{array}{ccc}
	M_{f_{41}} & 0 & M_{f_{43}} \\ 
	0 & M_{f_{52}} & M_{f_{53}}%
	\end{array}%
	\right), 
	\label{Mblocks}
	\end{equation}
    and
    \begin{equation}
	M_{F} =\left( 
	\begin{array}{cc}
	M_{f_{44}} & M_{f_{45}} \\ 
	M_{f_{54}} & M_{f_{55}}%
	\end{array}%
	\right) 
	\label{Mblocks2}
    \end{equation}

    The mass matrices in eq.(\ref{MF1})
are diagonalized via the following bi-unitary transformation:
\begin{equation}
\left(U_{L}^{f}\right)^{\dagger }M_{f}U_{R}^{f}=\text{diag}\left( m_{f_1},m_{f_2},m_{f_3},m_{F_1},m_{F_2}\right),
\label{matrixdiagonalization}
\end{equation}
where $f=u,d,\ell$ and $F=T,B,E$. 

    On can notice, from eq.(\ref{Mblocks}), that only the third fermion family 
    gets the mass through the Yukawa interaction with one of the two Higgs doublets.
    That is, the top quark gets tree-level mass from its Yukawa interaction with $H_{1}$, 
    whereas the bottom quark and tau lepton obtain their masses from their Yukawa
	interactions with the second Higgs doublet $H_{2}$. We will assume that 
	the symmetry breaking and the masses of the VLFs 
	are around the TeV scale. Thus, the mass matrices for the SM charged fermions, 
	resulting from a seesaw-like (or FN-like) mechanism, are given by
	\begin{eqnarray}
	\widetilde{M}_{f} &=&C_{f}-A_{f}M_{F}^{-1}B_{f}=
	\frac{M_{f_{14}}}{M_{f_{45}}^{2}+M_{f_{55}}M_{f_{44}}}\left( 
	\begin{array}{ccc}
	-M_{f_{55}} M_{f_{41}} & -M_{f_{45}} M_{f_{41}} & - M_{f_{55}}M_{f_{43}}-M_{f_{45}}M_{f_{53}} \\ 
	 M_{f_{45}} M_{f_{41}} & -M_{f_{44}} M_{f_{41}} &   M_{f_{45}}M_{f_{43}}+M_{f_{44}}M_{f_{53}} \\ 
	0 & 0 & M_{f_{33}}%
	\end{array}%
	\right)
	\end{eqnarray}
	where $f=u,d,\ell$, $F=T,B,E$ and we have used $M_{f_{25}}=-M_{f_{14}}$, $M_{f_{54}}=-M_{f_{45}}$,
	$M_{f_{52}}=-M_{f_{41}}$.

	Furthermore, from the neutrino Yukawa interactions, we obtain a $5\times5$ neutrino mass 
	matrix given by 
	\begin{equation}
	M_{\nu }=\left( 
	\begin{array}{cc}
	0_{3\times 3} & M_{D_\nu} \\ 
	M_{D_\nu}^{T} & M_{R}%
	\end{array}%
	\right) ,
	\end{equation}
	where $M_{D_\nu}$ is the Dirac mass matrix and $M_{R}$ is the Majorana mass matrix for 
	RH neutrinos. These matrices are
	\begin{equation}
	M_{D_\nu}=\left( 
	\begin{array}{cc}
	A_{\nu } & C_{\nu } \\ 
	-\tilde{A}_{\nu } & -\tilde{C}_{\nu } \\ 
	B_{\nu } & D_{\nu }%
	\end{array}%
	\right) ,\quad \quad M_{R}=\left( 
	\begin{array}{cc}
	M_{R_{1}} & 0 \\ 
	0 & M_{R_{2}}%
	\end{array}%
	\right),  \label{Mnu}
	\end{equation}
	with $A_\nu = y'_{\nu _{1}}v_1 v_{\xi_1}/2\Lambda$, $\tilde{A}_\nu = y'_{\nu _{1}}v_1 v_{\xi_2}/2\Lambda$, $C_\nu = y'_{\nu _{2}}v_1 v_{\xi_1}/2\Lambda$, $\tilde{C}_\nu = y'_{\nu_{2}}v_1 v_{\xi_2}/2\Lambda$, $B_\nu = y_{\nu _{1}}v_1 v_{\sigma_1}/2\Lambda$ and $D_\nu = y_{\nu _{2}}v_1 v_{\sigma_1}/2\Lambda$.
	Assuming that the right handed Majorana neutrinos have masses much larger
	than the electroweak symmetry breaking scale $v_{\rm EW}=246$ GeV, the type I seesaw
	mechanism can be implemented to generate the tiny masses of the light active
	neutrinos. The resulting mass matrix for light active neutrinos takes the
	form
	\begin{equation}\label{Mnulow}
	\begin{aligned}
 	&\widetilde{M}_{\nu } = M_{D_\nu}M_{R}^{-1}M_{D_\nu}^{T}
	\end{aligned}
	\end{equation}
	Then, the light active neutrino masses are given by
	\begin{equation}
	\label{eq:neutrino_masses}
	\begin{aligned}
    m_{\nu_1} = 0, \quad \quad m_{\nu_2,\nu_3} = \frac{\kappa \pm \kappa'^2}{2M_{R_1} M_{R_2}},
	\end{aligned}    
	\end{equation}
	where we have defined 
    \begin{equation}\label{eq:kappas_defs}
    \begin{aligned}
    &\kappa = \tilde{C}_\nu^2 M_{R_1} + C_\nu^2 M_{R_1} + D_\nu^2 M_{R_1} + \tilde{A}_\nu^2 M_{R_2} + A_\nu^2 M_{R_2} + B_\nu^2 M_{R_2}, \\
    &\kappa'^2 = -\tilde{C}_\nu^2 M_{R_1} - C_\nu^2 M_{R_1} - D_\nu^2 M_{R_1} - \tilde{A}_\nu^2 M_{R_2} - A_\nu^2 M_{R_2} - B_\nu^2 M_{R_2}^2)^2 - \\
    &\hspace{2.4em}- 4A_\nu^2 \tilde{C}_\nu^2 M_{R_1} M_{R_2} + 4B_\nu^2 \tilde{C}_\nu^2 M_{R_1} M_{R_2} - 8 \tilde{A}_\nu A_\nu \tilde{C}_\nu C_\nu M_{R_1} M_{R_2} + \\
    &\hspace{2.4em}+ 4 \tilde{A_\nu}^2 C_\nu^2 M_{R_1} M_{R_2} + 4 B_\nu^2 C_\nu^2 M_{R_1} M_{R_2} - 8 \tilde{A}_\nu B_\nu \tilde{C}_\nu D_\nu M_{R_1} M_{R_2} + \\
    &\hspace{2.4em}+ 8 A_\nu B_\nu C_\nu D_\nu M_{R_1} M_{R_2} + 4 \tilde{A}_\nu^2 D_\nu^2 M_{R_1} M_{R_2} + A_\nu^2 D_\nu^2 M_{R_1} M_{R_2}.
    \end{aligned}
    \end{equation}

	\subsection*{Scalar mass spectrum}
	
	Using eqs.~(\ref{ec:pot2hdm}) and (\ref{ec:potflavon}) and solving the tadpole equations, 
	the CP-even squared mass matrix becomes, 
	\begin{eqnarray}
	M^2_{\text{CP-even}}= \left( 
	\begin{array}{cc}
	[M^2]_{4\times4} & 0_{4\times2} \\ 
	0_{2\times4} & [\bar{M}^2]_{2\times2}%
	\end{array}
	\right)
	\end{eqnarray}
	where 
	\begin{eqnarray}  \label{eq:m013}
	M^2= \left( 
	\begin{array}{cccc}
	\lambda _1 v_1^2 & \left(\lambda _3+\lambda _4+\lambda _5\right) v_1 v_2 & 
	\lambda _6 v_1 v_{\sigma_1} & \lambda _8 v_1 v_{\sigma_2} \\ 
	\left(\lambda _3+\lambda _4+\lambda _5\right) v_1 v_2 & \lambda _2 v_2^2 & 
	\lambda _7 v_2 v_{\sigma_1} & \lambda _9 v_2 v_{\sigma_2} \\ 
	\lambda _6 v_1 v_{\sigma_1} & \lambda _7 v_2 v_{\sigma_1} & \lambda _{10}
	v_{\sigma_1}^2 & \lambda _{12} v_{\sigma_1} v_{\sigma_2} \\ 
	\lambda _8 v_1 v_{\sigma_2} & \lambda _9 v_2 v_{\sigma_2} & \lambda _{12}
	v_{\sigma_1} v_{\sigma_2} & \lambda _{11} v_{\sigma_2}^2 \\
	\end{array}
	\right)
	\end{eqnarray}
	and 
	\begin{eqnarray}  \label{eq:m022}
	\bar{M}^2= \left( 
	\begin{array}{cc}
	\frac{\mu_7^2 v_{\xi_2}}{2 v_{\xi_1}} & 2 \lambda _{13} v_{\xi_1} v_{\xi_2}-%
	\frac{\mu_7^2}{2} \\ 
	2 \lambda _{13} v_{\xi_1} v_{\xi_2}-\frac{\mu_7^2}{2} & \frac{\mu_7^2
		v_{\xi_1}}{2 v_{\xi_2}}\\
	\end{array}
	\right).
	\end{eqnarray}
	From the last equation we find that $\left\langle \xi_i\right\rangle\neq0$. Notice that
	the flavon fields get decoupled when $\left\langle \sigma_i\right\rangle \gg v_{\rm EW}$ 
	or simply if one takes $\lambda_{6,7,8,9} \ll 1$. In this case, the CP-even parts of the 
	$\mathrm{SU(2)}$ scalar doublets do not mix with the scalar singlets and their masses are obtained
	by diagonalizing the matrix
	\begin{eqnarray}  \label{eq:m012}
	M^2_{\text{2DHM}}\sim \left( 
	\begin{array}{cc}
	\lambda _1 v_1^2 & -\mu^2_{12}+\left(\lambda _3+\lambda _4+\lambda _5\right)
	v_1 v_2  \\ 
	-\mu^2_{12}+\left(\lambda _3+\lambda _4+\lambda _5\right) v_1 v_2 & \lambda
	_2 v_2^2 \\ 
	\end{array}
	\right)
	\end{eqnarray}
	The mass of the CP-odd and charged scalar are, respectively,
	\begin{equation}  \label{eq:m2A2}
	m_A^2=\frac{\mu_{12}^2}{v_1 v_2}-\lambda _5 \left(v_1^2+v_2^2\right),
	\end{equation}
	and 
	\begin{equation}  \label{eq:m2ch2}
	m_{H^\pm}^2=\frac{\mu_{12}^2}{v_1 v_2}-\frac{\left(\lambda
		_4+\lambda_5\right)}{2}\left(v_1^2+v_2^2\right)
	\end{equation}

Before concluding this section, let us briefly mention
that even though the $SU(2)$ scalar $H_1$ is only coupled to the SM up-type sector and $H_2$ to the down-type sector the Weinberg-Glasgow-Paschos theorem \cite{Glashow:1976nt,Paschos:1976ay} does not apply in this case, i.e. FCNCs might appear at tree-level due to the mixing between SM and vector like fermions. However,
we expect these FCNCs to be under control since they will be proportional to the small mixing angles\footnote{ For instance, we have numerically checked that the mixing angles between SM and vector like quarks are at most of the order of $10^{-3}$ which are sufficiently small to suppress FCNCs induced by these mixings.} and further suppressed by the square of the heavy non-SM scalar masses. A thorough analysis on this regard is out of the scope of this paper.

	\section{Muon anomalous magnetic moment\label{section:Muon_g2}}
	
	In this section we start discussing the implications of our model for the muon anomalous magnetic moment. It is worth mentioning that the Yukawa interactions, $-y_E\overline{L}_{L_2}E_{R_2} H_{2}$ and 
	$-y_{E2}\overline{E}_{L_2}\ell_{R_2} \sigma_{2}$ in eq.~(\ref{Lyl2}) as well as the quartic scalar interaction $\lambda _{9}( H_{2}^{\dagger }H_{2}) (\sigma _{2}^{\ast }\sigma
	_{2})$ in eq.~(\ref{ec:potflavon}), provide the dominant contributions to the muon anomalous magnetic moment. These contributions to $(g-2)_{\mu}$ are one-loop diagrams which involve the exchange of a electrically neutral CP-even scalar and the VLL $E_{2}$. 
	To simplify our analysis, we consider a benchmark close	to the decoupling limit
	where $\varphi_{R_2}$ and $\sigma_{R_2}$ are mainly
	composed of two orthogonal combinations involving two heavy CP-even $S_{1}^{0}$ 
	and $S_{2}^{0}$ physical scalar fields. Therefore, the muon anomalous magnetic moment takes the form:
	\begin{equation}
	\Delta a_{\mu}\simeq \frac{y_{E}y_{E2}m_{\mu }^{2}}{8\pi ^{2}}\left[ J\left(
	m_{E_{2}},m_{S_{1}^{0}}\right) -J\left( m_{E_{2}},m_{S_{2}^{0}}\right) %
	\right] \sin \theta \cos \theta ,
	\end{equation}
	where $S_{1}^{0}\simeq \cos \theta \sigma_{R_2} +\sin \theta \varphi_{R_2}$ , 
	$S_{2}^{0}\simeq -\sin \theta \sigma_{R_2} +\cos \theta \varphi_{R_2}$, and $%
	m_{E_{2}}$ is the mass of the VLL $E_{2}$. Furthermore, the
	loop $J\left( m_{E},m_{S}\right) $ function has the following form \cite{Diaz:2002uk,Jegerlehner:2009ry,Kelso:2014qka,Lindner:2016bgg}
	\begin{equation}
	J\left( m_{E},m_{S}\right) =\int_{0}^{1} dx\frac{x^{2}\left( 1-x+\frac{m_{E}}{%
			m_{\mu }}\right) }{m_{\mu }^{2}x^{2}+\left( m_{E}^{2}-m_{\mu }^{2}\right)
		x+m_{S}^{2}\left( 1-x\right) }.
	\end{equation}
 It is worth mentioning that in this model there exit other BSM contributions to the muon anomalous magnetic moment 
		but they turn out to be subleading. For instance, the loop contributions mediated by heavy neutrinos and the $W$ gauge boson are strongly suppressed by the quadratic power of both the very tiny active-sterile neutrino mixing angle and the small effective Dirac neutrino Yukawa coupling. Notice that the smallness of this coupling is due to the neutrino Yukawa interactions are dimension-5, Eq. (\ref{Lynu2}). Therefore, Dirac neutrino Yukawas also suppress one-loop diagrams where neutrinos and electrically charged Higgs are exchanged. Furthermore, following Refs. \cite{Raby:2017igl,CarcamoHernandez:2019ydc,Dermisek:2020cod}, one can estimate that the $\left(g-2\right)_{\mu}$ contribution that involve mediation of the $Z$ gauge boson and heavy vector like leptons turns out be $\Delta a^{(Z)}_{\mu}\sim\frac{m_{\mu}m_E}{8\pi^{2}m_Z^{2}}\theta^2G_{loop}$. Hence, $\Delta a^{(Z)}_{\mu}\sim\mathcal{O}\left(10^{-11}\right)$, for $200$ GeV masses of charged vector like leptons and SM-heavy vector like lepton mixing angle satisfying $\theta\sim\mathcal{O}\left(10^{-3}\right)$.

\begin{figure}[h]
		\centering
		\captionsetup{justification=raggedright,singlelinecheck=false}
		\hspace{-6em}
		\includegraphics[width=0.75\textwidth]{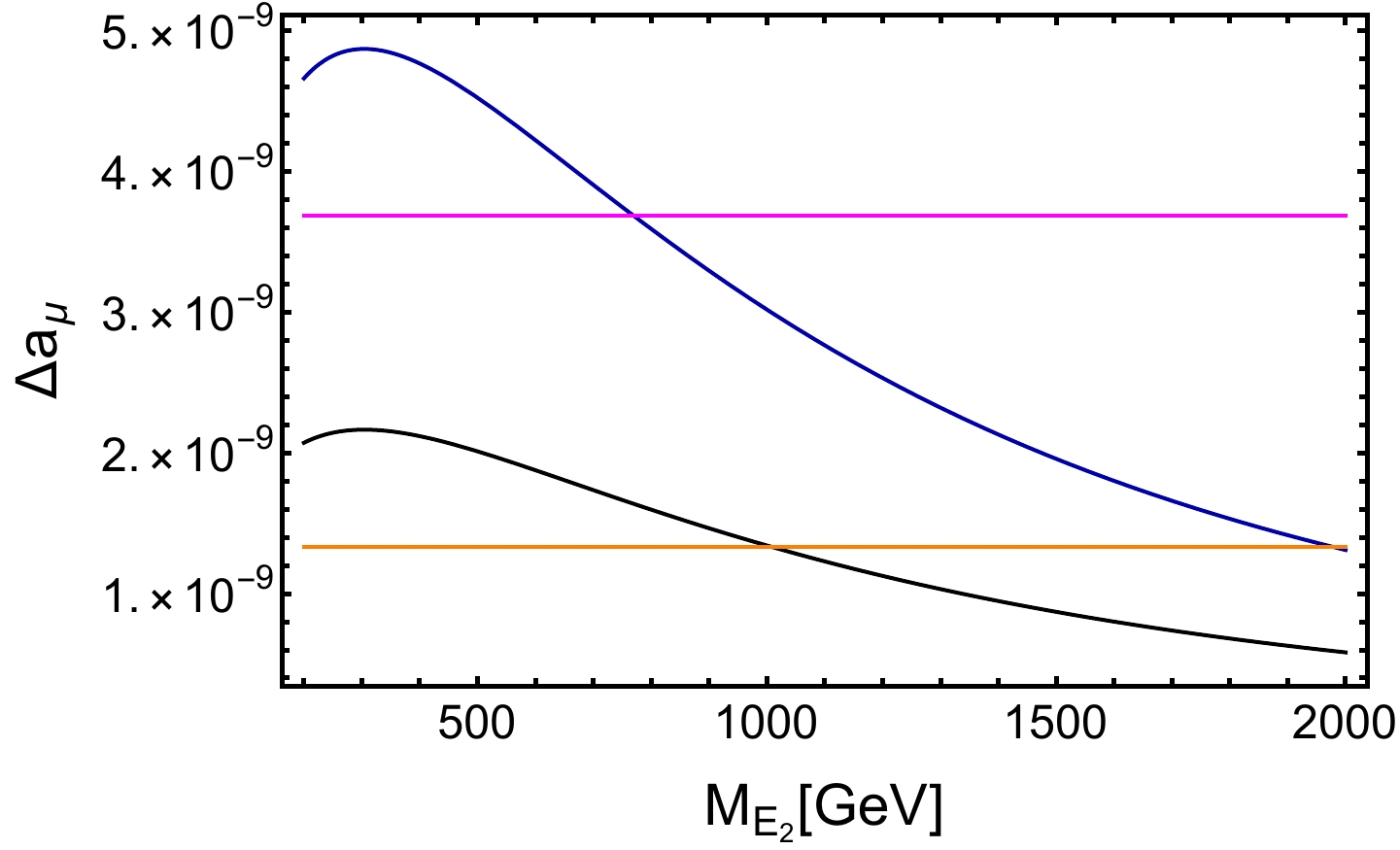}
		\caption{Muon anomalous magnetic moment as a function of the charged exotic lepton mass $M_{E_{2}}$. The black and blue curves correspond to $y_{E}=y_{E2}=0.2$ and $y_{E}=y_{E2}=0.3$, respectively. The horizontal magenta and orange lines correspond to the $2\sigma$ upper and lower bounds for the muon anomalous magnetic moment, respectively.}
		\label{gminus2muonvsmE}
	\end{figure}

Fig.~\ref{gminus2muonvsmE} shows the muon anomalous magnetic moment as a function of the VLL mass $M_{E_{2}}$. The solid horizontal lines correspond to the current upper and lower experimental bounds for the muon anomalous magnetic moment which are set by \cite{Abi:2021gix} 
\begin{equation}
\label{eq:exp_values_g2}
\begin{aligned}
&(\Delta a_{\mu })_{\exp } =\left( 2.51\pm 0.59\right) \times
10^{-9}.
\end{aligned}
\end{equation}
In our numerical analysis we have considered the benchmark point with $\theta =\pi/4$, $M_{S_{1}}=1.5$ TeV and $M_{S_{2}}=2$ TeV. Furthermore, we have set $y_{E}=y_{E2}=0.2$ and $y_{E}=y_{E2}=0.3$ for the black and blue curves, respectively. The mass of the charged exotic lepton $E_{2}$ has been varied in the range $0.2$ TeV$\leqslant M_{E_{2}}\leqslant $ $2$ TeV. Fig.~\ref{gminus2muonvsmE} shows that our model can successfully accommodate the current experimental anomaly $\Delta a_{\mu }$ within the considered mass range. One can see that $\Delta a_{\mu }$ requires vector-like states below the TeV scale. For this reason, in what follows, we will analyze how likely one can observe these VLLs at the upcoming LHC searches.
	
\section{Exotic fermionic signatures: Analysis setup}\label{section:VLLs_collider}
		
As it was shown in Sect.~\ref{sec:model}, two generations of VLLs and four VLQs are present in the model. For the VLQs, two are of the up-type whereas the other two are of the down-type. Such states have masses at the TeV scale, within the range of future collider runs at the LHC. In this section, we focus on a discussion of potential signatures characteristic of these particles as well as on the numerical techniques that we propose to probe them. Such analysis will be boosted via the implementation of neural networks (NNs), whose hyperparameters are optimized through the use of genetic algorithms, based on previous work by some of the authors \cite{Freitas:2020ttd}. We focus in pair-production topologies for both VLQs and VLLs, while single-production is considered only for the VLLs.

The model, at the Lagrangian level, is implemented in \texttt{SARAH} \cite{Staub:2013tta}, from which we generate the relevant \texttt{UFO} \cite{Degrande:2011ua} python codes that interface with Monte-Carlo simulators. In particular, we employ \texttt{MadGraph} (MG5) \cite{Alwall:2014hca} for simulation of particle collisions at parton-level for both signal and background topologies. We add hadronization and showering effects with \texttt{Pythia8} \cite{Sjostrand:2014zea} and \texttt{Delphes} \cite{deFavereau:2013fsa} for fast detector simulation. Angular and kinematic distributions are extracted from this last step, with the help of \texttt{ROOT} \cite{Brun:1997pa}, which are used as inputs in the NNs, for signal/background separation and computation of statistical significance. All parton-level events are generated for proton-proton collisions at the LHC, for a centre-of-mass energy $\sqrt{s} = 14$ TeV and for the parton-distribution function nn23lo1, which automatically fixes the strong coupling, $\alpha_s$, and its evolution. We generate 250 000 events for each individual topology (background and signal). We consider the MLM matching scheme \cite{Hoche:2006ph} for topologies with at least two jets as final states. 

VLLs have long been motivated by various SM extensions and Grand Unified Theory (GUT) frameworks (see, e.g. \cite{Raby:2017igl,Garcia:2015sfa,Bhattacherjee:2017cxh}), and like it was shown in Sec.~\ref{section:Muon_g2}, are important in addressing the muon $(g-2)$ anomaly, whose relevance has recently come to the forefront of new physics explorations \cite{Abi:2021gix}\footnote{ For further constraints on VLLs we refer the reader to \cite{Crivellin:2020ebi}.}. Despite the strong theoretical motivations, very limited collider searches have been performed so far, with the most stringent constraints coming from CMS \cite{Sirunyan:2019ofn}, for doublet VLLs that strongly couple to the tau lepton. Older searches at LEP \cite{Achard:2001qw} constrain these exotic states to be heavier than $101.2~\mathrm{GeV}$. Therefore, there is still a plenty of parameter space left to be explored and as such, phenomenological studies such as these may help pin-point regions of the model parameter space to look for at collider experiments. 

\begin{figure*}[h!]
			\centering
			\captionsetup{justification=raggedright}
			\subfloat[]{{\includegraphics[width=0.28\textwidth]{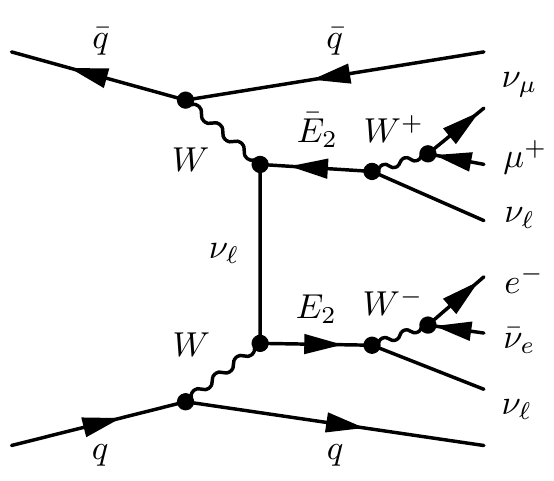} }} 
			\subfloat[]{{\includegraphics[width=0.28\textwidth]{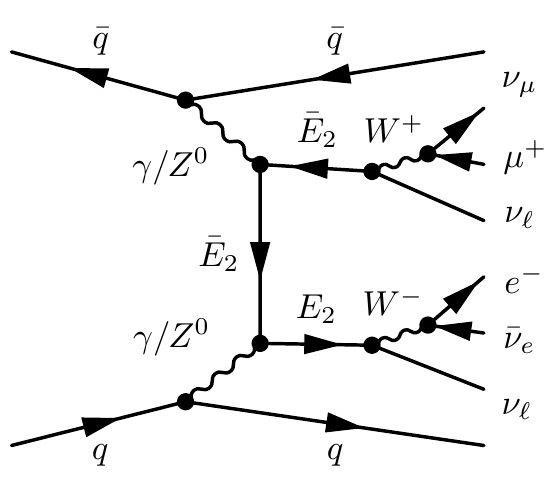} }} \\
			\caption{Leading-order Feynman diagrams for the VBF topologies. Original quarks from colliding protons are indicated as $q$ and $\bar{q}$, while $E_2$ represents the lightest VLL. Besides the forward jets, we have purely leptonic final states originating from $W^\pm$ decays, with one anti-muon, $\mu^+$, and an electron $e^-$, as well as their associated neutrino. $\nu_\ell$ are the SM neutrinos.
			\label{fig:VBF-events}}
		\end{figure*}

\begin{figure}[h!]
			\centering
			\captionsetup{justification=raggedright}
			\includegraphics[width=0.34\textwidth]{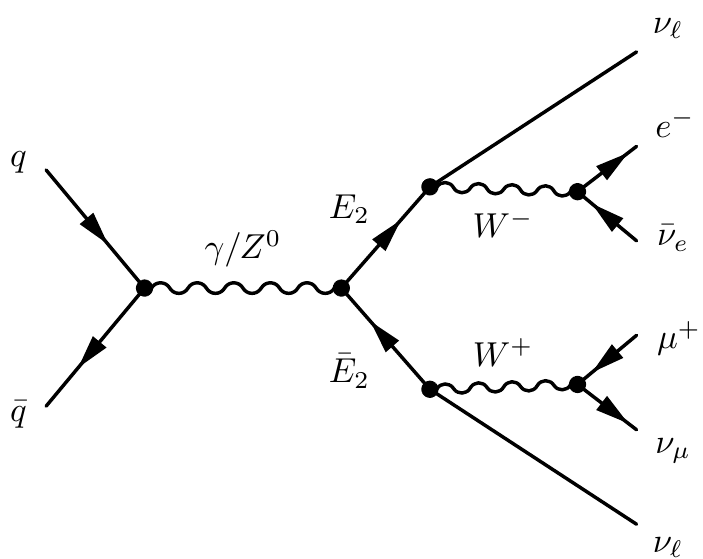}
			\caption{Leading-order Feynman diagram for the ZA topologies. The same nomenclature as in Fig.~\ref{fig:VBF-events} applies here.}
			\label{fig:ZA-events}
		\end{figure}

\begin{figure}[h!]
			\centering
			\captionsetup{justification=raggedright}
			\includegraphics[width=0.34\textwidth]{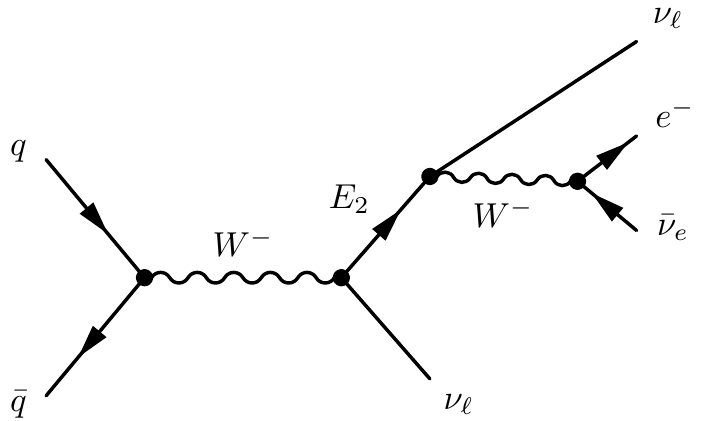}
			\caption{Leading-order Feynman diagrams for the VLBSM topologies. The same nomenclature as in Fig.~\ref{fig:VBF-events} applies here.}
			\label{fig:VLBSM-events}
		\end{figure}

For the VLLs' search, we consider topologies identical to some of those studied in our previous work \cite{Freitas:2020ttd}. These include pair-production in the t-channel via vector-boson fusion (VBF) processes (see Fig.~\ref{fig:VBF-events}), characterized by two light jets in the forward region originating from colliding protons. We also include contributions from pair production via the exchange of a virtual photon or a $Z^0$ boson, which we name as ``ZA'' in what follows (see Fig.~\ref{fig:ZA-events}). Both topologies are characterized by having two leptons and a large missing transverse energy (MET) as final states. The single-production diagram is characterized by a single lepton and a large MET in the final state (see Fig.~\ref{fig:VLBSM-events}), which we dub as the ``VLBSM topology''.
For these processes, we consider the main irreducible backgrounds as follows: 
\begin{enumerate}
			\item  For ZA topologies, main backgrounds include top quark pair production, $t\bar{t}$, with two b-jets and fully leptonic decays for the W bosons. $t\bar{t}$ plus $Z^0$ production is also considered, with the $Z^0$ decaying in the fully invisible channel (two neutrinos) or into two leptons;
			\item For VBF topologies, diboson $W^+W^-$ production is considered, with subsequent decay into leptons. We also take into account $t\bar{t}$ pair production plus one or two jets, that is, the top decays into leptons and is accompanied by one or two light jets;
			\item For VLBSM topologies, we consider all production channels with a single lepton in the final state, that is, $p p \rightarrow \ell \nu_\ell$, with up to two light jets.
		\end{enumerate}
To maximize the signal region and to reduce the main irreducible backgrounds, specific kinematic cuts are imposed in \texttt{ROOT}. In particular, we consider
\begin{enumerate}
			\item For VBF and ZA topologies, we require at least two lepton candidates with an opposite flavour and opposite charge. In particular, one anti-muon originating from $W^+$ decays and one electron candidate originating from $W^-$. For VLBSM, we require only one lepton, an electron.
			\item Generic to all topologies, we impose kinematic constraints for the final charged lepton states with $p_{T} > 25$ GeV and $|\eta| \leq 2.5$. A minimum MET is also considered with $\mathrm{MET} > 15$ GeV.
			\item For jet reconstruction, we use the Cambridge/Aachen algorithm \cite{CMS:2009lxa} with cone radius $\Delta R = 1.0$, with kinematic constraints $p_T > 35$ GeV and $|\eta| \leq 2.5$. For jets originating from bottom quarks, we consider tight-working points with 90\% b-tagging efficiency.
\end{enumerate}

The reconstruction procedure via the use of invisible particles in the final states follows the same approach as thoroughly described in \cite{Freitas:2020ttd}. The dimension-full and dimension-less variables are extracted from the final states to train our Deep Learning models, for different reference frames, including the laboratory frame, the $W$ bosons frame and $\bar{E}_2E_2$ frame. All chosen observables for double production topologies are shown in Table~\ref{tab:vars_Zp}, while in Table~\ref{tab:vars_Zp_VLBSm} we present the variables used for the VLBSM topology.
		\begin{table*}[ht!]    
			\centering
			\captionsetup{justification=raggedright,singlelinecheck=false}
			\resizebox{0.75\textwidth}{!}{\begin{tabular}{|c|c|c|c|}
					\toprule
					\hline
					& Dimension-full & \multicolumn{2}{|c|}{Dimensionless} \\
					\hline
					\hline
					\midrule
					\makecell{Lab. \\
						frame}  & \makecell{$p_T(e^-)$, $p_T(\mu^+)$,$p_T(E_2)$ \\
						$p_{T}(\bar{E}_2)$, $M({E_2})$, $M(\bar{E}_2)$ \\
						$M_T(W^-)$, $M_T(W^+)$, $p_T(W^+)$, \\
						$p_T(W^-)$, MET} & 
					\makecell{
						$\cos(\theta_{\bar{\nu}_e e})$, $\cos(\theta_{\bar{\nu}_\mu \mu^+})$, \\
						$\cos(\theta_{W^- W^+})$, \\ $\cos(\Delta \phi)$,
						$\cos(\Delta \theta)$, \\ $\eta(e^-)$, $\eta({\mu^+})$, $\eta({E_2})$, $\eta(\bar{E}_2)$ \\ $\eta(W^+)$, $\eta(W^-)$} & 
					\makecell{$\Delta R(e, \bar{\nu_e})$, $\Delta R(\mu^+, \nu_{\mu^+})$}\\
					\hline
					\makecell{\\[-0.5em]$W^-$ \\
						frame} & \makecell{$p_T(e^-)$, $p_T(E_2)$}& \makecell{
						$\cos(\theta_{\bar{\nu}_e e})$, \\
						$\eta(e^-)$, $\eta({E_2})$}  & 
					\makecell{}\\
					\hline
					\makecell{\\[-0.5em]$W^+$ \\
						frame} &\makecell{$p_T(\mu^+)$, $p_T(\bar{E}_2)$} & \makecell{
						$\cos(\theta_{\nu_\mu \mu^+})$, \\
						$\eta(\mu^+)$, $\eta(\bar{E}_2)$} & 
					\makecell{}\\           
					\hline
					\makecell{\\[-0.5em] $E_2\bar{E_2}$ \\
						frame} &\makecell{} & \makecell{
						$\cos(\Delta \Phi)$,
						$\cos(\Delta \Theta)$} & 
					\makecell{}\\
					\hline
					\hline
			\end{tabular}}
			\caption{Angular and kinematic observables selected for study of the pair-production topologies, for different frames of reference, the laboratory frame (in the first row), $W^+$ and $W^-$ frames (in the second and third row, respectively) and the vector-like frame $E_2\bar{E}_2$ (last row). $\theta_{i,j}$ denotes the angle between different particles, either in the final state or reconstructed ones. In the $E_2\bar{E_2}$ frame, the angles $\Delta\Phi$ and $\Delta\Theta$ correspond to the azimuthal and polar angles formed by the decay plane of the two $W$ bosons (see \cite{Freitas:2020ttd}).}
			\label{tab:vars_Zp}
		\end{table*}
		\begin{table*}[ht!]    
			\centering
			\captionsetup{justification=raggedright,singlelinecheck=false}
			\begin{tabular}{|c|c|c|}
				\toprule
				\hline
				& Dimension-full & {Dimensionless} \\
				\hline
				\hline
				\midrule
				\makecell{Lab. \\
					frame}  & \makecell{$p_T(e^-)$,$M_T(W^-)$, \\
					$p_T(W^-)$, MET} & 
				\makecell{
					$\cos(\theta_{e^-})$, $\cos(\theta_{\bar{\nu}_e e^-})$, \\
					$\cos(\theta_{W^-})$, $\eta(e^-)$, $\eta(W^-)$, $\phi(e^{-})$}  \\
				\hline
				\hline
			\end{tabular}
			\caption{Angular and kinematic observables selected to study the single production channel (VLBSM). We compute observables in the laboratory frame. The same nomenclature of Table~\ref{tab:vars_Zp} for angles applies here.}
			\label{tab:vars_Zp_VLBSm}
		\end{table*}
		
\begin{figure}[h!]
			\centering
			\captionsetup{justification=raggedright}
			\includegraphics[width=0.34\textwidth]{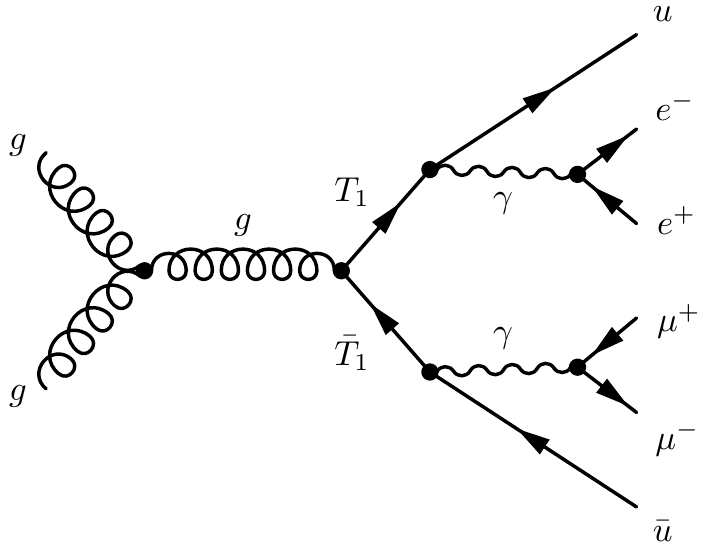}
			\caption{Leading-order Feynman diagram for the VLQ pair-production via gluon-gluon fusion. $T_1$ represents the lightest up-type VLQ and $u/\bar{u}$ indicate light up-type quarks (up or charm quarks). This diagram provides a larger contribution than that with $T_1 u Z^0$ vertices. We refer to Appendix \ref{app:feynman} for further details.}\label{fig:VLQ-events}
		\end{figure}

A similar analysis is also built for the VLQs. The prediction of VLQs is not exclusive to the model under consideration. In fact, the existence of such states has been predicted by a series of distinct models in previous literature, such as in $\mathrm{E_6}$-inspired string and GUT models \cite{Hebbar:2016gab,HEWETT1989193} or in other extensions to the SM \cite{Benbrik:2015fyz,Hernandez:2021uxx}. However, unlike VLLs, from an experimental point-of-view, there is also a good history of searches, in particular, at the LHC (see, for example, \cite{Aaboud:2018ifs,Aaboud:2018zpr,Sirunyan:2019sza}), where the current constraints on parameter space constrain the VLQ masses to be between 690 GeV up to 1.85 TeV (for current constraints, as of March 2021, see the summary plots in \cite{ATLAS_twiki_VLQs}). See also \cite{Roy:2020fqf,Araque:2015cna} where possible interpretations of current VLQ searches at the LHC are undertaken and \cite{Romao:2019dvs,Romao:2020ocr} for a discussion about Deep-Learning based methods that can be applied in direct VLQ searches. Naturally, the constraints are heavily dependent on considered assumptions, mainly when it comes to couplings to the SM states. In this regard, the majority of searches focus on dominant couplings with the top quark, where the primary decay channel $\text{VLQ} \rightarrow t(b)W$ is studied, characterized by a final b-jet and two charged leptons. Of course, such assumption is not the most general, as there is no reason for mixings with other SM quarks to not exist. For the purpose of this work, the proposed search topology is focused on a channel with light jets as final states, as seen in Fig.~\ref{fig:VLQ-events}.

Identical cuts to the VLL scenario are imposed, with two main differences. The more obvious one is that we now require at least four lepton candidates (an anti-muon/muon pair and a positron/electron pair) and at least two light jets. We also alter the minimum transverse momentum for the jet candidates, with now $p_T > 50$ GeV. The reason for such alteration comes from the fact that we plan to probe higher masses ($m > 1.8$ TeV), and therefore, the final jets emerging from VLQs will be highly boosted and energetic when compared to SM processes, helping to reduce the number of relevant backgrounds. Without any missing energy, both VLQs can be more easily reconstructed from the leptons and light jets. For irreducible backgrounds, we consider the same $t\bar{t} + Z^0$ background as for the ZA topology, with $Z^0$ decaying into two charged leptons. We also include all production channels with the same final states $p p \rightarrow e^+ e^- \mu^+\mu^- j j$, where $j$ is a light jet. Such a process includes the main diboson production backgrounds. Similarly, dimension-full and dimension-less variables from final and reconstructed states are used for NNs' training in three distinct reference frames: laboratory frame, $j_1 + \gamma$ frame and $j_2 + \gamma$ frame, where we define $j_1$ as the leading jet (greatest $p_T$), while $j_2$ is the sub-leading jet. All distributions are indicated in Table.~\ref{tab:vars_VLQ}
		\begin{table*}[ht!]    
			\centering
			\captionsetup{justification=raggedright,singlelinecheck=false}
			\resizebox{\textwidth}{!}{\begin{tabular}{|c|c|c|c|}
					\toprule
					\hline
					& Dimension-full & \multicolumn{2}{|c|}{Dimensionless} \\
					\hline
					\hline
					\midrule
					\makecell{Lab. \\
						frame}  & \makecell{$M(e^+,e^-)$, $M(\mu^+,\mu^-)$, $M(e^-,\mu^-)$, \\[0.2em] $M(j_1,j_2)$, $p_T(e^-)$, $p_T(e^+)$, $p_T(\mu^+)$, \\[0.2em] $p_T(\mu^-)$, $p_T(j_n)$, $M(e^+,e^-,j_n)$,\\[0.2em]  $M(\mu^+,\mu^-,j_n)$} & 
					\makecell{$\eta(e^-)$, $\phi(e^-)$, $\eta(e^+)$, $\phi(e^+)$, \\[0.2em] $\eta(\mu^-)$, $\phi(\mu^-)$, $\eta(\mu^+)$, $\phi(\mu^-)$, \\[0.2em] $\eta(j_n)$, $\phi(j_n)$, \\[0.2em] $\cos(\theta_{e^+ e^-})$, $\cos(\theta_{\mu^+ \mu^-})$ $\cos(\theta_{j_1 j_2})$\\[0.2em] $\cos(\theta_{e^- \mu^-})$, $\cos(\theta_{e^- \mu^+})$, $\cos(\theta_{e^- j_n})$\\[0.2em] $\cos(\theta_{e^+ j_n})$, $\cos(\theta_{\mu^- j_n})$, $\cos(\theta_{\mu^+ j_n})$, \\[0.2em] $\Delta\phi(e^+,e^-)$, $\Delta\phi(e^-,j_n)$, $\Delta\phi(e^+,j_n)$, \\[0.2em] $\Delta\phi(\mu^+,\mu^-)$, $\Delta\phi(\mu^-,j_n)$, $\Delta\phi(\mu^+,j_n)$, \\[0.2em] $\Delta\phi(e^-,\mu^-)$, $\Delta\phi(e^-,\mu^+)$, $\Delta\phi(e^+,\mu^-)$, \\[0.2em] $\Delta\phi(e^+,\mu^+)$} & 
					\makecell{$\Delta R(e^+, e^-)$, $\Delta R(e^-, j_n)$, $\Delta R(e^+, j_n)$ \\[0.2em] $\Delta R(\mu^+, \mu^-)$, $\Delta R(\mu^-, j_n)$, $\Delta R(\mu^+, j_n)$, \\[0.2em] $\Delta R(e^-,\mu^-)$, $\Delta R(e^-,\mu^+)$ , $\Delta R(e^+,\mu^-)$, \\[0.2em] $\Delta R(e^+,\mu^+)$}\\
					\hline
					\makecell{\\[-0.5em]$j_1 + \gamma$ \\
						frame} & \makecell{}& \makecell{
						$\cos(\theta_{\mu^- j_1})$, $\cos(\theta_{\mu^+ j_1})$, $\cos(\Delta\Phi_1)$}  & 
					\makecell{}\\
					\hline
					\makecell{\\[-0.5em]$j_2 + \gamma$ \\
						frame} &\makecell{} & \makecell{
						$\cos(\theta_{\mu^- j_2})$, $\cos(\theta_{\mu^+ j_2})$, $\cos(\Delta\Phi_2)$} & 
					\makecell{}\\
					\hline
					\hline
			\end{tabular}}
			\caption{Angular and kinematic observables selected for study of VLQ pair-production, for different frames of reference, the laboratory frame (in the first row), $j_1 + \gamma$ and $j_2 + \gamma$ frames (in the second and third row, respectively).  The same nomenclature of Table~\ref{tab:vars_Zp} for angles applies here. $\Delta\Phi_{1,2}$ corresponds to the azimuthal angle formed by the decay planes of the two virtual photons. To simplify notation, it is defined $j_n = j_1, j_2$.}
			\label{tab:vars_VLQ}
		\end{table*}
	
For both VLQs and VLLs, the main objective is to combine the kinematic distributions into multi-dimensional distributions that are then fed into the NN, whose job is to solve a classification task, that is, to distinguish between the background and signal events, so that we can evaluate the statistical significance. To obtain the most optimal results, one must take special care in the employed architecture, the number of layers and nodes, activation functions, etc. The selection of these parameters is often made based on arbitrary decision-making, by choosing the set of hyperparameters that have shown to guarantee the best results in earlier analyses. Such a method is often arbitrary and can not be generalised to other applications. 

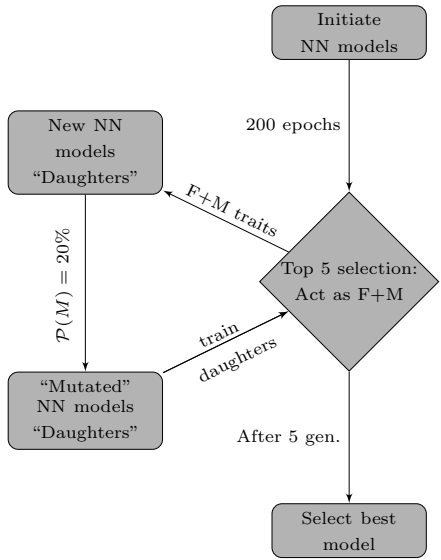
\begin{figure}[th!]
			\centering
			\tikzstyle{Rectangulo} = [draw, rectangle, fill=black!30, text width=6em, text centered, minimum height=2em]
			\tikzstyle{Diamante} = [draw, diamond, fill=black!30, text width=6em, text badly centered, inner sep=0pt]
			\tikzstyle{Linha} = [draw, -latex']
			\resizebox{0.32\textwidth}{!}{\begin{tikzpicture}[node distance = 1.5cm, auto]
				\node [Rectangulo, rounded corners] (step1) {\scriptsize Initiate NN models};
				\node [Diamante, below of=step1,node distance=3.5cm] (step2) {\scriptsize Top 5 selection: Act as F+M};
				\node [Rectangulo, rounded corners, below of=step2, node distance=3.5cm] (step3) {\scriptsize Select best model};
				\node [Rectangulo, rounded corners, above left of=step2, node distance=2.0cm, above left=0.4cm, left=0.8cm] (step4) {\scriptsize New NN models ``Daughters''};
				\node [Rectangulo, rounded corners, below left of=step2, node distance=2.0cm, below left=0.4cm, left=0.8cm] (step5) {\scriptsize ``Mutated'' NN models \\ ``Daughters''};
				\path [Linha] (step1) -- node [left] {\scriptsize 200 epochs} (step2);
				\path [Linha] (step2) -- node [left] {\scriptsize After 5 gen.} (step3);
				\path [Linha] (step2) -- node [above=0.50cm, right=-0.65cm, rotate=-26] {\scriptsize F+M traits} (step4);
				\path [Linha] (step4) -- node [left, rotate=90, below=-0.3cm, left=-0.9cm] {\scriptsize $\mathcal{P}(M) = 20\%$}(step5);
				\path [Linha] (step5) node [above=0.90cm , right=1.50cm, rotate=26] {\scriptsize train} -- (step2);
				\path [Linha] (step5) node [below=-0.40cm, right=1.50cm, rotate=26] {\scriptsize daughters} -- (step2);
				\end{tikzpicture}}
			\caption[]{Diagram representative of the iterations involved in the evolutive algorithm as used in this work.}
			\label{fig:EVO-algo}
		\end{figure}

For the purpose of this work, we implement the same techniques as in \cite{Freitas:2020ttd}, via a genetic algorithm, that chooses the parameters that best improve a given significance. A diagrammatic representation of the algorithm can be seen in Fig.~\ref{fig:EVO-algo}.  The entire process starts by providing a list of possible input parameters which the algorithm chooses from. From this list, the algorithm picks, in a random fashion, a series of parameters from which it can build an arbitrary number of NNs. Once all the networks have been constructed, we train them for fixed number of epochs. From the trained networks, we choose the top networks, that is, the ones that better maximize a given metric. It is from this point that the evolutionary part of the algorithm kicks in. The idea is inspired by the process of natural selection. From the best networks, we create Father-Mother pairs, where 50\% of the father's traits and 50\% of the mother's traits are used to construct new NNs, that we dub ``daughters''. We also impose a probability of mutation, $\mathcal{P}(M)$, meaning that after the daughters have been built, we also add a non-zero probability that each of the traits may change to another, leading to the creation of ``mutated daughters''. Then, we train these new daughter networks, and the loop repeats for a given number of generations. At the end, we select the one with the better performance, for a given metric, in our case, the Asimov significance.
		
We choose the same set of hyperparameters as in our previous work \cite{Freitas:2020ttd}, which we detail as follows: 
		\begin{itemize}
			\item Number of hidden layers: 1 to 5;
			\item Number of nodes per layers: 256, 512, 1024 or 2048;
			\item Initialisers: '\texttt{normal}', '\texttt{he normal}' and '\texttt{he uniform}';
			\item \texttt{L2} regulariser, with penalties $1\times 10^{-3}$, $1\times 10^{-5}$ or $1\times 10^{-7}$;
			\item Activation functions: '\texttt{ReLU}', '\texttt{eLU}', '\texttt{tanh}' and '\texttt{sigmoid}';
			\item Optimisers: '\texttt{Adam}', '\texttt{sgd}', '\texttt{AdaMax}' and '\texttt{NAdam}'. 
		\end{itemize}
The networks are built in \texttt{Keras} \cite{chollet2015keras} with \texttt{TensorFlow 2.0} \cite{Abadi:2016kic} as back-end. Some architectural considerations remain fixed during the evolution of the genetic algorithm. Namely, the final output layer works as a prediction layer where we funnel the data into a vector with dimensions $\text{dim}N = 1 + N_b$, with $N_b$ the number of backgrounds. Each entry corresponds to the probability of being a signal or background. For example, consider a vector $(S,B1,B2) = (0.98, 0.01, 0.01)$. The output of this form indicates that the network considers this event as a signal with probability of $\mathcal{P}_S = 0.98$ whereas backgrounds have a probability $\mathcal{P}_{B_{1,2}} = 0.01$. The inputs of the networks are also identical, with normalized and balanced distributions of kinematic data. The normalized data are important for training due to potentially high variability in numerical data structures. By normalizing the datasets, we mitigate numerical errors when computing gradients during back propagation, which in turn allows for faster learning and improved performance \cite{589532,LeCun2012}. The datasets are also balanced, as the imposition of cuts in the kinematics of final states reduces the number of entries for both background and signal classes. Such unbalanced nature may lead to over-fitting problems and poor generalization to validation datasets. We use \texttt{SMOTE} \cite{Chawla_2002} algorithm to balance the data, by oversampling the minority classes. We also employ a cyclic learning rate during the training, with the initial value of 0.01 and the maximum allowed value of 0.1. A fixed batch size of 32544 is considered.
		
In this work we are interested in models that provide us with the best significance. Ss such, we define the Asimov significance that can be modified to work as the loss function. In our case we define our loss as $1/(\mathcal{Z_A} + \epsilon)$, that we plan to minimize\footnote{This methodology was first proposed by Adam Elwood and Dirk Krücker in \cite{Elwood:2018qsr}.}, with
\begin{equation}
\label{eq:Asimov_sig}
	\mathcal{Z_A} = \Bigg[2\Bigg((s + b)\ln\Bigg(\frac{(s+b)(b+\sigma_b^2)}{b^2 + (s+b)\sigma_b^2}\Bigg) -\frac{b^2}{\sigma_b^2}\ln\Bigg(1+\frac{\sigma_b^2 s}{b(b+\sigma_b)}\Bigg)\Bigg)\Bigg]^{1/2},
\end{equation}
where $s$ is the number of signal events, $b$ the number of backgrounds and $\sigma_b^2$ is the variance of backgrounds events. Note that in the limit of large backgrounds, $\mathcal{Z}_A \approx s/\sqrt{b}$.

\section{Exotic fermionic signatures: Results}\label{section:Results}
		
In this section we discuss the results obtained for collider signatures of exotic fermions in the context of the $\mathcal{Q}_6$ flavored multiscalar model under consideration. Due to their distinct nature, the mass range that the LHC can probe for the model's VLLs differs from that of the VLQs. In particular, and based on the current exclusion bounds, we focus on the following two sets,
\begin{equation}\label{eq:mass_ranges}
	m_{E_2}\in [200,800] \hphantom{.}\mathrm{GeV}, \quad m_{T_1}\in [2.2,4] \hphantom{.}\mathrm{TeV},
\end{equation} 
where the mass range chosen for $m_{E_2}$ was based on the $\Delta a_{\mu }$ analysis performed above. Note that both the VLL and VLQ decay widths are automatically computed by \texttt{MadGraph} for each mass.

In all studied signal events the internal vertices are gauge-interacting, always involving couplings with the SM vector bosons ($Z^0$, $\gamma$, $W^\pm$ or $g$). These do indeed provide the dominant contributions and their strength is well known. However, the fermion-mixing effects must be considered and typically provide a suppression factor. This is what happens in flavor non-diagonal scenarios as is the case of the model that we study in this article. For the purpose of this work, we assume \textit{vector-to-chiral} fermion mixing of order $\mathcal{O}(10^{-2}-10^{-3})$ according to the structures specified in section~\ref{sec:model}, which in turn corresponds to the benchmark point $y_E = y_{E_2} =0.2$ shown in Fig.~\ref{gminus2muonvsmE}. For example, the lepton mixture is relevant when probing the $E_2\nu_\ell W$ vertex via an extended Pontecorvo-Maki-Nakagawa-Sakata (PMNS) mixing. In particular, we fix the PMNS matrix in such a way that the block that mixes the charged chiral leptons and SM-like neutrinos is phenomenologically consistent \cite{Zyla:2020zbs}. The same is done in the quark sector with an extended version of the Cabibbo-Kobayashi-Maskawa (CKM) matrix.

\begin{figure*}[h!]
			\centering
			\captionsetup{justification=raggedright}
			\subfloat[VLLs]{{\includegraphics[width=0.46\textwidth]{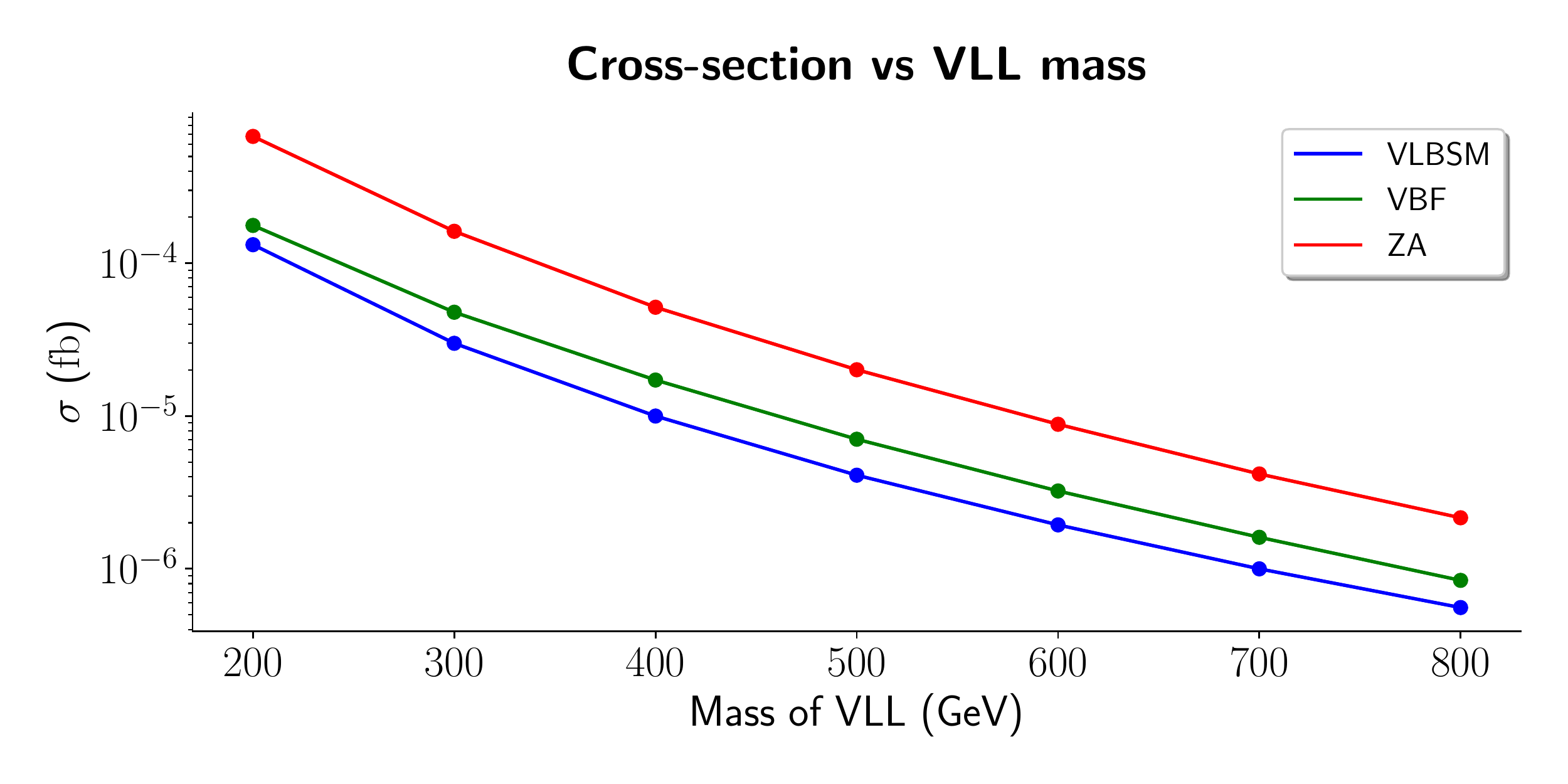} }} 
			\subfloat[VLQs]{{\includegraphics[width=0.46\textwidth]{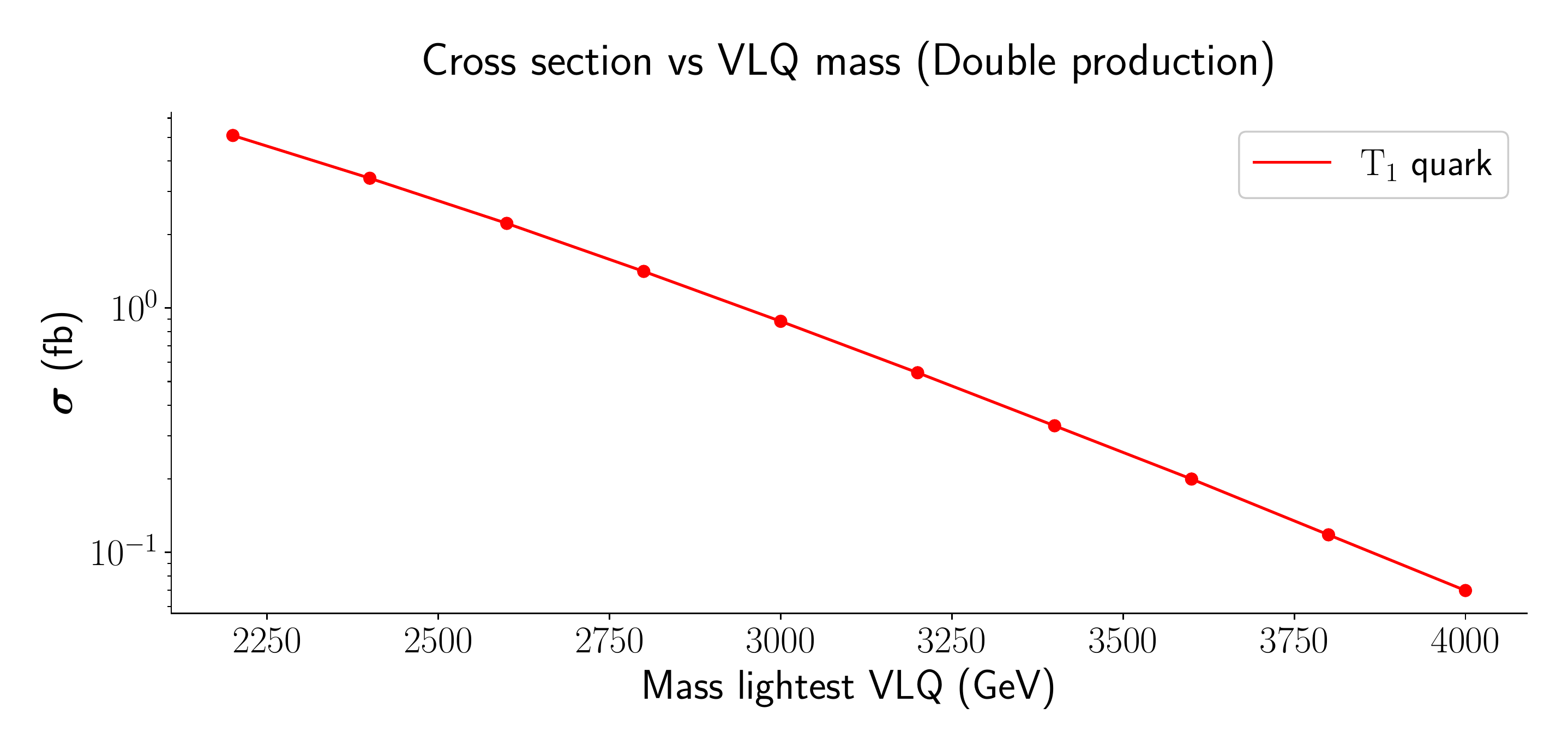} }} \\
			\caption{The production cross-section as a function of the VLLs mass (left panel) and the VLQ mass (right panel).
				\label{fig:Xsecs_VLLs_VLQs}}
		\end{figure*}

Using the cuts detailed in the previous section we can estimate the production cross-section for each signal/background topology. For the VLL channel, and fixing $m_{E_2} = 200~\mathrm{GeV}$ as well as the mixing matrix in eq.(\ref{matrixdiagonalization}):
\begin{equation}\label{eq:lepton_mixing}
U^e_L = \begin{bmatrix}
-0.999997 & 0.00122671 & -0.00196369 & -2.41219\times 10^{-6} & -9.55935\times 10^{-6} \\
0.00123261 & 0.999993 & -0.00300585 & -0.00200382 &  1.1783\times 10^{-8} \\
0.00195999 &-0.00300821 & -0.999994 & 0.0000294637 & 9.03987\times 10^{-8} \\  -1.17835\times 10^{-8} & 0.0080039 & 0.0000234357 & 0.999997 & 0.00123275 \\
9.55951\times 10^{-6} & 2.4701\times 10^{-6} & -4.27716\times 10^{-8} & 0.00123275 & -0.999999
\end{bmatrix}
\end{equation}
we have obtained,
		\begin{equation}\label{eq:production_xsec}\nonumber
		\begin{aligned}
		&\text{VLBSM signal:}\quad \sigma = 1.32\times 10^{-4} \hphantom{.}\mathrm{fb};\\
		&\text{ZA signal:}\quad \sigma = 6.77\times 10^{-4} \hphantom{.}\mathrm{fb};\\
		&\text{VBF signal:}\quad \sigma = 1.77\times 10^{-4} \hphantom{.}\mathrm{fb};\\
		&pp\rightarrow e^-\bar{\nu}_e\quad \sigma = 1.96\times 10^{6} \hphantom{.}\mathrm{fb}; \\
		&pp\rightarrow e^-\bar{\nu}_e (j,jj)\quad \sigma = 8.02\times 10^{5} \hphantom{.}\mathrm{fb};\\
		&t\bar{t}\quad \sigma = 1.21\times 10^{3} \hphantom{.}\mathrm{fb};\\
		&t\bar{t} (j,jj)\quad \sigma = 2.39\times 10^{3} \hphantom{.}\mathrm{fb};\\
		&W^+W^-\quad \sigma = 1.63\times 10^{2} \hphantom{.}\mathrm{fb};\\
		&t\bar{t}Z^0(e^-e^+)\quad \sigma = 0.18 \hphantom{.}\mathrm{fb};\\
		&t\bar{t}Z^0(\bar{\nu}_\ell\nu_\ell)\quad \sigma = 0.20 \hphantom{.}\mathrm{fb}
		\end{aligned}
		\end{equation}
and, as one notices, all backgrounds sit well above the expected cross sections for the signal events. On the other hand, VLQ pair-production reveals the opposite behaviour. Considering the scenario where $m_{T_1} = 2.2$ TeV and the mixing matrix in eq. (\ref{matrixdiagonalization}):
\begin{equation}\label{eq:quark_mixing}
U^u_L = \begin{bmatrix}
-1. & 0.000177353 & -0.000313111 & -4.77474\times10^{-8} & -9.09229\times10^{-6} \\
0.0003595 & -0.99997 & 0.005294 & -0.0034525 & 3.26836\times10^{-9} \\
-0.00001530 & 0.005294 & 0.999968 & 0.003161 & -1.07557\times10^{-7} \\ -9.09229\times10^{-6} & -9.64642\times10^{-7} & -6.00768\times10^{-7} & 0.00039356 & 1.\\
3.2693\times10^{-9} & -0.00026828 & -0.0016711 & 0.999995 & -0.0003935
\end{bmatrix},
\end{equation}
the cross sections for this analysis read as,
		\begin{equation}\label{eq:production_xsec_VLQs}\nonumber
		\begin{aligned}
		& \text{VLQ signal:}\quad \sigma = 5.09 \hphantom{.}\mathrm{fb};\\
		&pp\rightarrow e^-e^+\mu^+\mu^-j j\quad \sigma = 0.16\hphantom{.}\mathrm{fb}; \\
		&t\bar{t}Z^0(\mu^-\mu^-)\quad \sigma = 1.23\times 10^{-2} \hphantom{.}\mathrm{fb}\\
		\end{aligned}
		\end{equation}
where we note that for the $t\bar{t}Z^0(\mu^-\mu^-)$ background, the electron and positron originate from $W$ decays. As one can clearly see, the signal production cross-section sits above the main irreducible backgrounds. This fact remains true up until $m_{T_1} \sim 3.8$ TeV, after which the suppression coming from the VLQ mass and its decay width becomes large enough yielding an increasingly smaller cross-section. The noticeable difference between the VLQ and VLL sectors can be mainly attributed to the chosen collider and couplings present in the production channels. Proton-proton collisions heavily favour a pair-production of colored particles, which is further maximized via the strong coupling in the three gluon $ggg$ and $g\bar{T}_1T$ vertices, as opposed to the weak gauge coupling present in all VLL pair-production processes. We show in Fig.~\ref{fig:Xsecs_VLLs_VLQs} both the VLL and VLQ production cross-sections in terms of the exotic fermion masses for each of the studied processes.

As mentioned in the previous section, the main goal is to compute the statistical significance of a hypothetical discovery. As such, we need to extract the relevant kinematic information that helps separating the signal from background. Let us first dedicate our attention to VLL production focusing on a scenario where $m_{E_2} = 200~\mathrm{GeV}$. We show in Appendix~\ref{app2} the distributions in the laboratory frame in Figs.~\ref{fig:VBF_vars_LabFrame}, \ref{fig:ZA_vars_LabFrame} and \ref{fig:VLBSM_vars_LabFrame}. In these figures, the angular distributions in $\bar{E}_2E_2$ frame are also presented. The remaining distributions, in the $W$ frame of reference, are shown in Fig.~\ref{fig:Vars_BoostedFrame}. It is common to all topologies that the angular distributions are dominated by the imposed backgrounds, since the signal events do not have major qualitative differences that help separating different $\cos(\theta)$ distributions. On the other hand, $\Delta R$ distributions are particularly interesting in this regard, as signal topologies typically involve a peak around $\Delta R \sim 1$, as opposed to the backgrounds, with a flatter structure and a peak at higher values, $\Delta R \sim 1-2$. Kinematic information, such as pseudo-rapidity, offers additional discrimination as the signal events are characterized by a strong peak at $\eta=0$, while backgrounds typically either possess a double-peak structure (see $\eta$ distributions for reconstructed particles such as $E_2$ and $W^+$ in Fig.~\ref{fig:VBF_vars_LabFrame}), or a more uniform distribution over the allowed pseudo-rapidity range (see $\eta$ plots for the electron and the anti-muon in Fig.~\ref{fig:VBF_vars_LabFrame}). This is true for VBF topologies, while for ZA and VLBSM events it is not (see Figs.~\ref{fig:ZA_vars_LabFrame} and \ref{fig:VLBSM_vars_LabFrame}). For instance, pseudo-rapidity distributions for ZA and VLBSM typically follow the same characteristics as the backgrounds. Therefore, they do not offer the same discriminating power. Transverse momentum distributions, for both topologies, dot not provide a greater discriminating power either. However, MET distributions supply some key differences such as longer tails at large MET values, i.e.~$\mathrm{MET} > 100~\mathrm{GeV}$. In fact, MET distributions are rather relevant, as this observable is representative of the signal events that we propose, in particular for the ZA/VBF cases where the final state contains four neutrinos. For the case of VLBSM topologies this no longer applies. Even though the final state contains three neutrinos, MET distributions have the same shape as the SM backgrounds (see middle panel in the second row of Fig.~\ref{fig:VLBSM_vars_LabFrame}).

Turning our attention to the quark sector, all relevant distributions are shown in Fig.~\ref{fig:VLQ_vars} where we fix the VLQ mass as $m_{T_1}=2.2~\mathrm{TeV}$. For VLQ production, a substantial amount of variables were used. As such, not all of them, but only a representative subset is shown. As opposed to the VLLs, $\cos(\theta)$ distributions do offer discriminating power over backgrounds. For the pair of particles $(e^+,j_1)$, $(e^-,j_2)$ and $(\mu^+,j_2)$, the cosine distributions preferably peak at $\cos(\theta) = -1$, i.e.~at $\theta = \pi$, for signal events, implying that the outgoing particles are produced back-to-back. This contrasts with the studied backgrounds where a significant portion of events feature highly collinear particles, i.e.~$\cos(\theta) = 1 \Leftrightarrow \theta = 0$. Other angular variables such as $\Delta \phi$ offer further distinction with a typical double-peak structure of signal events at $|\Delta\phi| \geq 2$, whereas the corresponding backgrounds tend to populate near $\Delta \phi = 0$. The transverse momentum distributions of the final leptons are also relevant as one notices from the bottom rows of Fig.~\ref{fig:VLQ_vars}. In particular, signal events tend to populate regions of phase-space with larger momentum ($p_T > 300$ GeV), in stark contrast with the SM backgrounds, which preferably populate regions of much lower $p_T$ values.

A proper analysis requires a combination of the various kinematical variables into a single multi-dimensional distribution that the NN uses as an input. This allows one to find the regions of the parameter space that better enhance the signal region while minimizing the background effects. More importantly, it also provides us with the ability to compute the statistical significance as well as determining which mass regions can be excluded within the studied model. For a complete analysis we calculate the significance focusing on three distinct statistical metrics, each one being more conservative than the other, in order to provide the most rigorous and realistic scenarios for further investigation. 

We then consider the following
		\begin{itemize}
			\item $\mathcal{Z}_A:$ The Asimov significance, as it is defined in Eq.~\eqref{eq:Asimov_sig}, assuming 1\% systematic errors;
			\item $\mathcal{Z}(<1\%):$ An adapted version of the Asimov significance, as it is defined in Eq.~\eqref{eq:Asimov_sig}. For this case we assume a much lower systematic uncertainty, in particular $10^{-3}$. As such, this is the most lenient metric and typically offers the highest values for the significance;
			\item $s/\sqrt{s+b}:$ A more traditional metric, which is the limiting scenario of the Asimov significance when $s \ll b$.
		\end{itemize}
With this in mind, we apply the genetic algorithm as described in the previous section and calculate the aforementioned significance metrics in terms of the score that the NN gives to each event. For the scenarios that have so far been discussed, $m_{E_2} = 200$ GeV and $m_{T_1} = 2.2$ TeV, our results for the VLLs significance in each channel can be seen in Fig.~\ref{fig:ACC-Sig-plots} of Appendix~\ref{app:sig_plots_DL}. Notice that the significance is calculated under an assumption of the high-luminosity LHC, with an integrated luminosity of $\mathcal{L} = 3000$ $\mathrm{fb^{-1}}$.

Let us first focus on the VLL sector. Provided that each topology is an independent event we can safely combine their individual significances as
        \begin{equation}\label{eq:combined_significance}
		\sigma_C = \sigma_{\mathrm{VBF}} + \sigma_{\mathrm{VLBSM}} + \sigma_{\mathrm{ZA}}.
		\end{equation}
With this in mind, we notice that the highest combined significance is obtained for the $\mathcal{Z}(<1\%)$ metric, in particular, with $\sigma_C = 3.78\sigma$. The dominant contributions to this value are those of the VLBSM and VBF channels with $\sigma_{\mathrm{VLBSM}} = 1.76\sigma$ and $\sigma_{\mathrm{VBF}} = 1.03\sigma$, respectively. Equally interesting is the $s\sqrt{s+b}$ metric, for which the combined significance is slightly smaller, $\sigma_C = 2.56\sigma$, with the VLBSM topology providing the highest contribution, with $1.13\sigma$. While these results do not permit a confident exclusion at this mass range, however it is sufficiently large to merit further inspection if an hypothetical anomaly appears at the LHC experiments. If this turns out to be the case, we argue here that, if the next generation of colliders beyond the LHC are still far from becoming operational, a continuation of the high-luminosity program must be put on the table as we further discuss below.

As expected, the most conservative metric, $\mathcal{Z}_A$, shows the lowest significance values with $\sigma_C = 0.032\sigma$. Therefore, within the context of the model under consideration, we can not safely exclude VLLs up to masses of $200~\mathrm{GeV}$. In fact, singlet VLLs do not couple directly to $W$ bosons as it happens for doublet VLLs (see \cite{Freitas:2020ttd} for a previous study.) Instead, such interactions with $W$ bosons relevant for the three signal processes under consideration, are indirectly induced via off-diagonal Yukawa interactions becoming suppressed by a factor of the mass of the VLL itself \cite{Kumar:2015tna}. This implies that the larger the VLL mass, the smaller the interaction strength with $W$ bosons, thus, the smaller the cross section and significance.

\begin{figure*}[ht!]
			\captionsetup{justification=raggedright}
			\subfloat[$\mathcal{Z}_A$ significance]{{\includegraphics[width=0.50\textwidth]{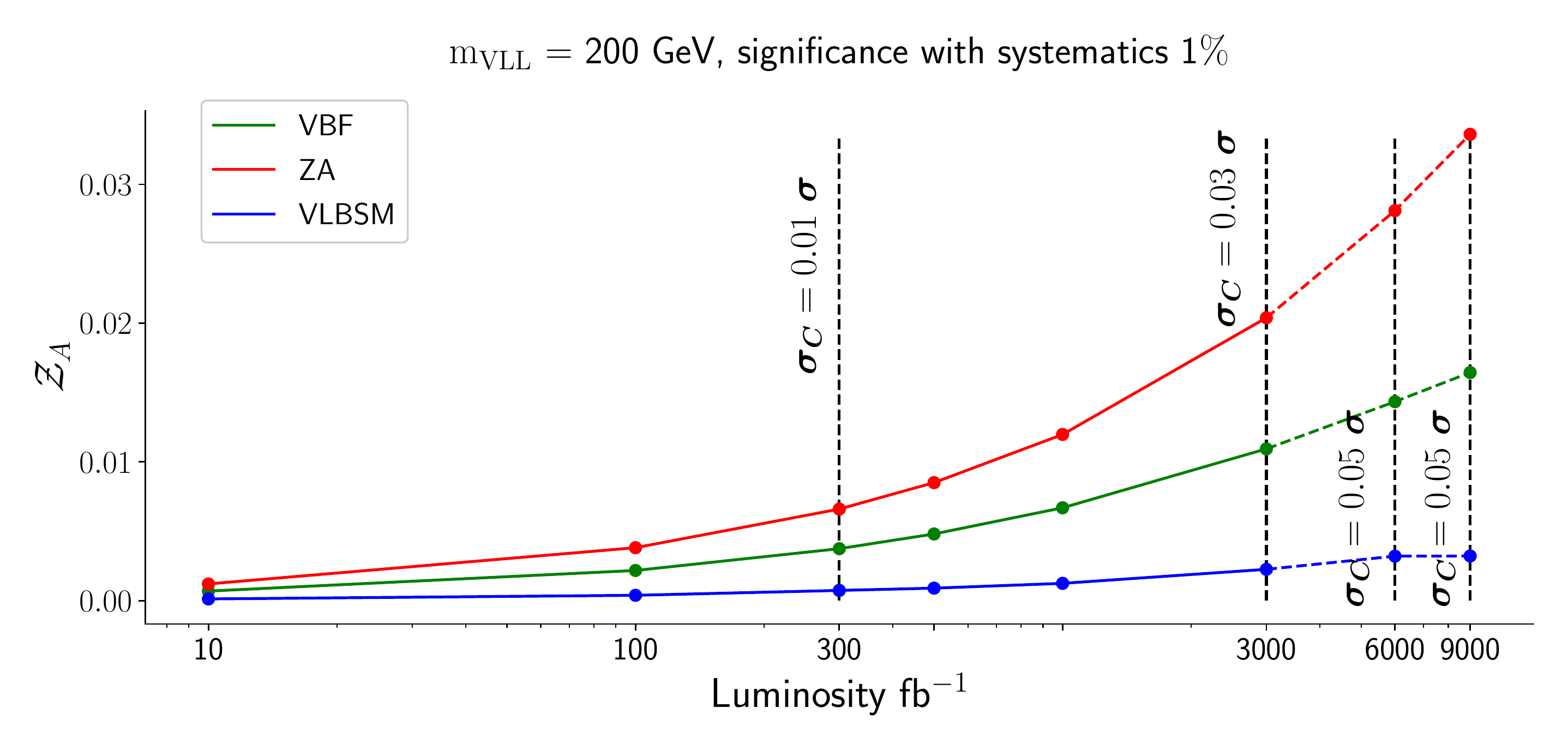} }} 
			\subfloat[$\mathcal{Z}(<1\%)$ significance]{{\includegraphics[width=0.50\textwidth]{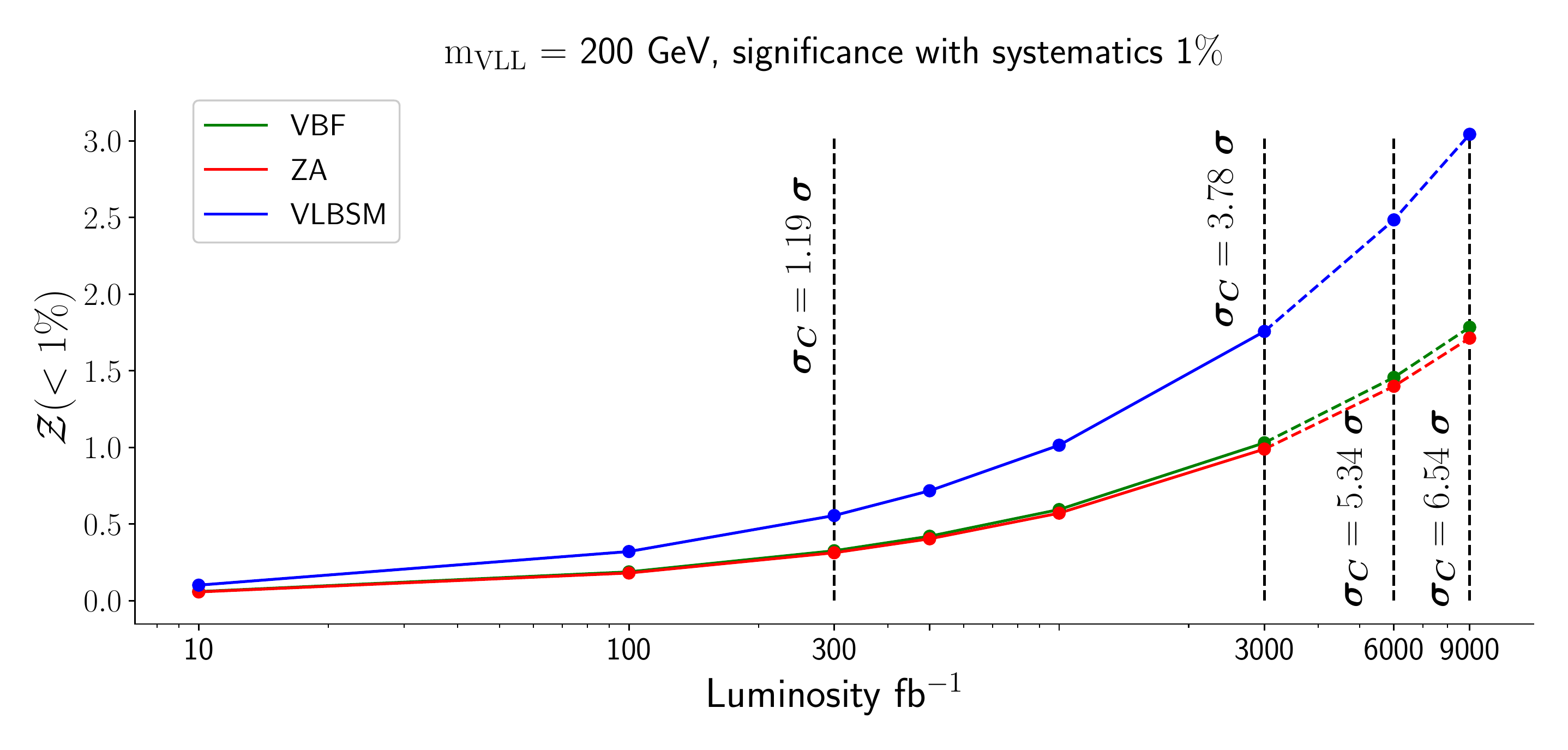} }} \\
			\subfloat[$s/\sqrt{s+b}$]{{\includegraphics[width=0.50\textwidth]{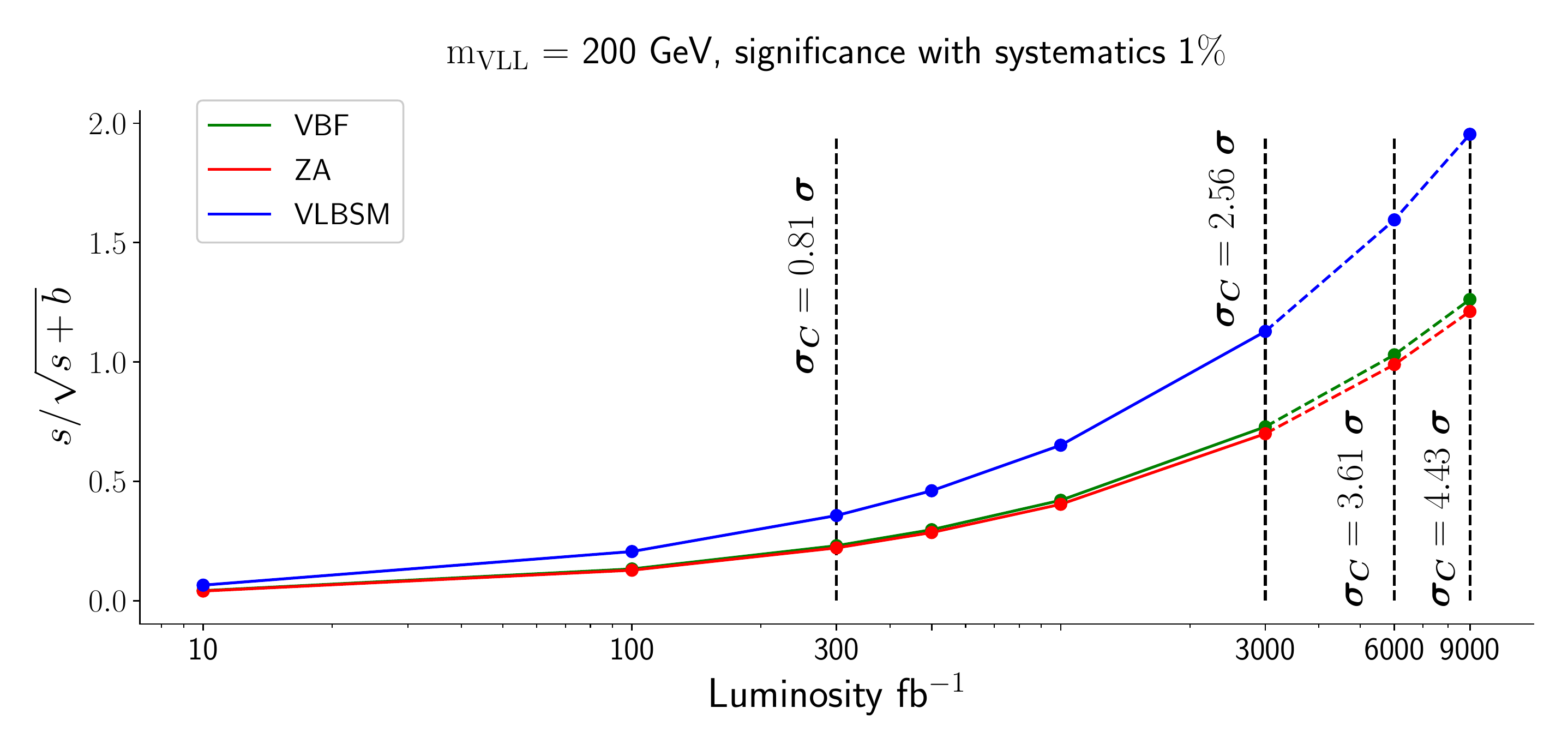} }}\\
			\caption{Statistical significance as a function of the integrated luminosity in $\mathrm{fb^{-1}}$ for the three different statistics adopted in the current analysis and for a fixed VLL mass, $m_{E_2} = 200$ GeV. The $x$ axis is on logarithmic scale. In (a) we showcase the Asimov significance, in (b) -- the adapted Asimov significance, and in (c) -- the $s/\sqrt{s+b}$ metric. The colours represent the distinct signal processes under consideration. In particular, the green curve is representative of VBF events, the red curve indicates ZA topologies, while the blue curve refers to VLBSM single production events. The dashed curves indicate that the considered values of the luminosity are beyond the LHC operation regime.
				\label{fig:Sig_plots_luminosity}}
		\end{figure*}

We show in Fig.~\ref{fig:Sig_plots_luminosity} the dependence of the significance as a function of the luminosity for a VLL mass of $200~\mathrm{GeV}$.
In particular, we see that by the end of the LHC program, i.e.~$\mathcal{L} = 3000~\mathrm{fb^{-1}}$, a combined significance of $\sigma_C = 3.78\sigma$ can be achieved, for the $\mathcal{Z}(<1\%)$ metric. Alternatively, if the standard $s\sqrt{s+b}$ measure is considered, a $\sigma_C = 2.56\sigma$ anomaly could be observed. In either scenario, we argue that, if a new generation of colliders that will succeed the LHC remains decades away from the beginning of operations, such an anomaly (or any other excess) may justify a continuation of the high-luminosity runs. 

As an example, we show in all panels of Fig.~\ref{fig:Sig_plots_luminosity} the continuation of the significance curves for higher luminosities choosing $\mathcal{L} = 6000~\mathrm{fb^{-1}}$ and $\mathcal{L} = 9000~\mathrm{fb^{-1}}$ as merely indicative values. In particular, we see that a signal confirmation or exclusion would be realizable with $\sigma_C = 6.54\sigma$ for the $\mathcal{Z}(<1\%)$ metric, at $\mathcal{L} = 9000~\mathrm{fb^{-1}}$, and $\sigma_C = 5.34\sigma$ at $\mathcal{L} = 6000~\mathrm{fb^{-1}}$. Note that we use dashed lines in the continuation of the significance curves for the region beyond $\mathcal{L} = 3000~\mathrm{fb}^{-1}$ to indicate that such a regime is beyond the planned LHC operation program.
		\begin{table}[h!]
			\centering
			\captionsetup{justification=raggedright,singlelinecheck=false}
			\resizebox{0.9\textwidth}{!}{\begin{tabular}{c||cccc||cccc||cccc||}
					\multirow{2}{*}{Mass of VLL} & \multicolumn{4}{c||}{$s/\sqrt{s+b}$}                                               & \multicolumn{4}{c||}{$\mathcal{Z}(<1\%)$}                              & \multicolumn{4}{c||}{$\mathcal{Z}_{A}$}                                            \\ \cline{2-13} 
					& ZA                          & VBF                         & VLBSM &$\sigma_\mathrm{C}$               & ZA                           & VBF                         & VLBSM  &$\sigma_\mathrm{C}$ & ZA                          & VBF                          & VLBSM     &$\sigma_\mathrm{C}$           \\ \hline
					$200$ GeV                      & \multicolumn{1}{c|}{$0.70$} & \multicolumn{1}{c|}{$0.73$} & \multicolumn{1}{c|}{$1.13$}              &\multicolumn{1}{c||}{\textbf{2.56}} &\multicolumn{1}{c|}{$0.99$} & \multicolumn{1}{c|}{$1.03$} & \multicolumn{1}{c|}{$1.76$} &\multicolumn{1}{c||}{\textbf{3.78}}& \multicolumn{1}{c|}{$0.02$} & \multicolumn{1}{c|}{$0.011$} & \multicolumn{1}{c|}{$0.0022$} & \multicolumn{1}{c||}{\textbf{0.033}}              \\
					$300$ GeV                      & \multicolumn{1}{c|}{$0.37$} & \multicolumn{1}{c|}{$0.38$} & \multicolumn{1}{c|}{$0.59$} &\multicolumn{1}{c||}{\textbf{1.34}}& \multicolumn{1}{c|}{$0.57$} & \multicolumn{1}{c|}{$0.54$} & \multicolumn{1}{c|}{$0.91$}  &\multicolumn{1}{c||}{\textbf{2.02}}& \multicolumn{1}{c|}{$0.018$} & \multicolumn{1}{c|}{$3.08\times 10^{-5}$}  & \multicolumn{1}{c|}{$0.0012$} &\multicolumn{1}{c||}{\textbf{0.019}}\\
					$400$ GeV                      & \multicolumn{1}{c|}{$0.25$} & \multicolumn{1}{c|}{$0.23$} & \multicolumn{1}{c|}{$0.38$}
					&\multicolumn{1}{c||}{\textbf{0.86}}&               \multicolumn{1}{c|}{$0.36$}  & \multicolumn{1}{c|}{$0.32$} & \multicolumn{1}{c|}{$0.55$}
					&\multicolumn{1}{c||}{\textbf{1.23}}  & \multicolumn{1}{c|}{$0.0086$} & \multicolumn{1}{c|}{$0.0077$} & \multicolumn{1}{c|}{$0.0022$} &\multicolumn{1}{c||}{\textbf{0.019}}\\ 
					$500$ GeV                      & \multicolumn{1}{c|}{$0.19$} & \multicolumn{1}{c|}{$0.15$} & \multicolumn{1}{c|}{$0.30$} & \multicolumn{1}{c||}{\textbf{0.64}}  & \multicolumn{1}{c|}{$0.30$} & \multicolumn{1}{c|}{$0.21$} & \multicolumn{1}{c|}{$0.43$} & \multicolumn{1}{c||}{\textbf{0.94}}& \multicolumn{1}{c|}{$0.0079$} & \multicolumn{1}{c|}{$0.0037$} & \multicolumn{1}{c|}{$0.0020$}& \multicolumn{1}{c||}{\textbf{0.014}}    \\ 
					$600$ GeV                      & \multicolumn{1}{c|}{$0.15$} & \multicolumn{1}{c|}{$0.13$} & \multicolumn{1}{c|}{$0.19$} & \multicolumn{1}{c||}{\textbf{0.47}}             & \multicolumn{1}{c|}{$0.20$} & \multicolumn{1}{c|}{$0.15$} & \multicolumn{1}{c|}{$0.32$} & \multicolumn{1}{c||}{\textbf{0.67}}& \multicolumn{1}{c|}{$0.00023$} & \multicolumn{1}{c|}{$0.0022$} & \multicolumn{1}{c|}{$0.0013$} & \multicolumn{1}{c||}{\textbf{0.0037}}\\ 
					$700$ GeV                      & \multicolumn{1}{c|}{$0.0024$} & \multicolumn{1}{c|}{$0.069$} & \multicolumn{1}{c|}{$0.095$} & \multicolumn{1}{c||}{\textbf{0.17}}             & \multicolumn{1}{c|}{$0.091$} & \multicolumn{1}{c|}{$0.098$} & \multicolumn{1}{c|}{$0.11$} & \multicolumn{1}{c||}{\textbf{0.30}}& \multicolumn{1}{c|}{$5.40\times 10^{-5}$} & \multicolumn{1}{c|}{$0.0015$} & \multicolumn{1}{c|}{$0.0016$} & \multicolumn{1}{c||}{\textbf{0.0032}}              \\
					$800$ GeV                      & \multicolumn{1}{c|}{$0.0020$} & \multicolumn{1}{c|}{$3.8849\times 10^{-6}$} & \multicolumn{1}{c|}{$0.055$} & \multicolumn{1}{c||}{\textbf{0.057}}             & \multicolumn{1}{c|}{$0.0034$} & \multicolumn{1}{c|}{$0.071$} & \multicolumn{1}{c|}{$0.077$} &
					\multicolumn{1}{c||}{\textbf{0.15}} & \multicolumn{1}{c|}{$2.78\times 10^{-5}$} & \multicolumn{1}{c|}{$2.20 \times 10^{-5}$} & \multicolumn{1}{c|}{$0.0015$}   & \multicolumn{1}{c||}{\textbf{0.0015}}           \\                                           
			\end{tabular}}
			\caption{Signal significance for the lightest VLL. The computation follows an evolutive algorithm that maximizes the Asimov significance metric. All significances are computed for $\mathcal{L} = 3000$ fb$^{-1}$ with centre-of-mass energy of $\sqrt{s}=14$ TeV. $\sigma_C$ is the combined significance as defined in \eqref{eq:combined_significance}.}\label{tab:Evolve_Asimov_table}
		\end{table}
			\begin{table}[h!]
			\centering
			\captionsetup{justification=raggedright,singlelinecheck=false}
			\resizebox{0.9\textwidth}{!}{\begin{tabular}{c||cccc||cccc||cccc||}
					\multirow{2}{*}{Mass of VLL} & \multicolumn{4}{c||}{$s/\sqrt{s+b}$}                                               & \multicolumn{4}{c||}{$\mathcal{Z}(<1\%)$}                              & \multicolumn{4}{c||}{$\mathcal{Z}_{A}$}                                            \\ \cline{2-13} 
					& ZA                          & VBF                         & VLBSM &$\sigma_\mathrm{C}$               & ZA                           & VBF                         & VLBSM  &$\sigma_\mathrm{C}$ & ZA                          & VBF                          & VLBSM     &$\sigma_\mathrm{C}$           \\ \hline
					$200$ GeV                      & \multicolumn{1}{c|}{$1.22$} & \multicolumn{1}{c|}{$1.26$} & \multicolumn{1}{c|}{$1.95$}              &\multicolumn{1}{c||}{\textbf{4.43}} &\multicolumn{1}{c|}{$1.71$} & \multicolumn{1}{c|}{$1.79$} & \multicolumn{1}{c|}{$3.04$} &\multicolumn{1}{c||}{\textbf{6.54}}& \multicolumn{1}{c|}{$0.034$} & \multicolumn{1}{c|}{$0.016$} & \multicolumn{1}{c|}{$0.0032$} & \multicolumn{1}{c||}{\textbf{0.053}}              \\
					$300$ GeV                      & \multicolumn{1}{c|}{$0.63$} & \multicolumn{1}{c|}{$0.66$} & \multicolumn{1}{c|}{$1.03$} &\multicolumn{1}{c||}{\textbf{2.32}}& \multicolumn{1}{c|}{$0.99$} & \multicolumn{1}{c|}{$0.93$} & \multicolumn{1}{c|}{$1.57$}  &\multicolumn{1}{c||}{\textbf{3.49}}& \multicolumn{1}{c|}{$0.030$} & \multicolumn{1}{c|}{$3.73 \times 10^{-5}$}  & \multicolumn{1}{c|}{$0.0025$} &\multicolumn{1}{c||}{\textbf{0.033}}\\
					$400$ GeV                      & \multicolumn{1}{c|}{$0.44$} & \multicolumn{1}{c|}{$0.39$} & \multicolumn{1}{c|}{$0.65$}
					&\multicolumn{1}{c||}{\textbf{1.48}}&               \multicolumn{1}{c|}{$0.62$}  & \multicolumn{1}{c|}{$0.56$} & \multicolumn{1}{c|}{$0.95$}
					&\multicolumn{1}{c||}{\textbf{2.13}}  & \multicolumn{1}{c|}{$0.015$} & \multicolumn{1}{c|}{$0.013$} & \multicolumn{1}{c|}{$0.0036$} &\multicolumn{1}{c||}{\textbf{0.032}}\\ 
					$500$ GeV                      & \multicolumn{1}{c|}{$0.34$} & \multicolumn{1}{c|}{$0.25$} & \multicolumn{1}{c|}{$0.52$} & \multicolumn{1}{c||}{\textbf{1.11}}  & \multicolumn{1}{c|}{$0.35$} & \multicolumn{1}{c|}{$0.36$} & \multicolumn{1}{c|}{$0.52$} & \multicolumn{1}{c||}{\textbf{1.13}}& \multicolumn{1}{c|}{$0.0079$} & \multicolumn{1}{c|}{$0.0065$} & \multicolumn{1}{c|}{$0.0035$}& \multicolumn{1}{c||}{\textbf{0.0179}}    \\ 
					$600$ GeV                      & \multicolumn{1}{c|}{$0.23$} & \multicolumn{1}{c|}{$0.20$} & \multicolumn{1}{c|}{$0.33$} & \multicolumn{1}{c||}{\textbf{0.76}}             & \multicolumn{1}{c|}{$0.23$} & \multicolumn{1}{c|}{$0.29$} & \multicolumn{1}{c|}{$0.40$} & \multicolumn{1}{c||}{\textbf{0.92}}& \multicolumn{1}{c|}{$0.0039$} & \multicolumn{1}{c|}{$0.0064$} & \multicolumn{1}{c|}{$0.0037$} & \multicolumn{1}{c||}{\textbf{0.014}}\\ 
					$700$ GeV                      & \multicolumn{1}{c|}{$0.10$} & \multicolumn{1}{c|}{$0.12$} & \multicolumn{1}{c|}{$0.20$} & \multicolumn{1}{c||}{\textbf{0.42}}             & \multicolumn{1}{c|}{$0.13$} & \multicolumn{1}{c|}{$0.17$} & \multicolumn{1}{c|}{$0.23$} & \multicolumn{1}{c||}{\textbf{0.53}}& \multicolumn{1}{c|}{$0.00027$} & \multicolumn{1}{c|}{$0.0026$} & \multicolumn{1}{c|}{$0.0030$} & \multicolumn{1}{c||}{\textbf{0.00587}}              \\
					$800$ GeV                      & \multicolumn{1}{c|}{$0.08$} & \multicolumn{1}{c|}{$6.73\times 10^{-6}$} & \multicolumn{1}{c|}{$0.095$} & \multicolumn{1}{c||}{\textbf{0.18}}             & \multicolumn{1}{c|}{$0.07$} & \multicolumn{1}{c|}{$0.12$} & \multicolumn{1}{c|}{$0.13$} &
					\multicolumn{1}{c||}{\textbf{0.32}} & \multicolumn{1}{c|}{$0.00015$} & \multicolumn{1}{c|}{$3.72\times 10^{-5}$} & \multicolumn{1}{c|}{$0.0028$}   & \multicolumn{1}{c||}{\textbf{0.0030}}           \\                                         
			\end{tabular}}
			\caption{Signal significance for the lightest VLL. The computation follows an evolutive algorithm that maximizes the Asimov significance metric. All significances are computed for $\mathcal{L} = 9000$ fb$^{-1}$ with centre-of-mass energy of $\sqrt{s}=14$ TeV. $\sigma_C$ is the combined significance as defined in \eqref{eq:combined_significance}.}\label{tab:Evolve_Asimov_table_2}
		\end{table}
		\begin{table}[h!]
			\centering
			\captionsetup{justification=raggedright,singlelinecheck=false}
			\resizebox{0.9\textwidth}{!}{\begin{tabular}{c||cccc||cccc||cccc||}
					\multirow{2}{*}{Mass of VLL} & \multicolumn{4}{c||}{$s/\sqrt{s+b}$}                                               & \multicolumn{4}{c||}{$\mathcal{Z}(<1\%)$}                              & \multicolumn{4}{c||}{$\mathcal{Z}_{A}$}                                            \\ \cline{2-13} 
					& ZA                          & VBF                         & VLBSM &$\sigma_\mathrm{C}$               & ZA                           & VBF                         & VLBSM  &$\sigma_\mathrm{C}$ & ZA                          & VBF                          & VLBSM     &$\sigma_\mathrm{C}$           \\ \hline
					$500$ GeV, $\sqrt{s}=28$ TeV                      & \multicolumn{1}{c|}{$0.49$} & \multicolumn{1}{c|}{$0.35$} & \multicolumn{1}{c|}{$0.36$}              &\multicolumn{1}{c||}{\textbf{1.20}} &\multicolumn{1}{c|}{$0.70$} & \multicolumn{1}{c|}{$0.50$} & \multicolumn{1}{c|}{$0.72$} &\multicolumn{1}{c||}{\textbf{1.92}}& \multicolumn{1}{c|}{$0.032$} & \multicolumn{1}{c|}{$0.022$} & \multicolumn{1}{c|}{$0.0022$} & \multicolumn{1}{c||}{\textbf{0.0562}}              
					\\    
					$500$ GeV, $\sqrt{s}=14$ TeV                      & \multicolumn{1}{c|}{$0.19$} & \multicolumn{1}{c|}{$0.15$} & \multicolumn{1}{c|}{$0.30$} & \multicolumn{1}{c||}{\textbf{0.86}}  & \multicolumn{1}{c|}{$0.36$} & \multicolumn{1}{c|}{$0.32$} & \multicolumn{1}{c|}{$0.55$} & \multicolumn{1}{c||}{\textbf{1.23}}& \multicolumn{1}{c|}{$0.0086$} & \multicolumn{1}{c|}{$0.0077$} & \multicolumn{1}{c|}{$0.0022$}& \multicolumn{1}{c||}{\textbf{0.019}}    \\                         
			\end{tabular}}
			\caption{Signal significance for the lightest VLL. The computation follows an evolutive algorithm that maximizes the Asimov significance metric. All significances are computed for $\mathcal{L} = 3000$ fb$^{-1}$ for a fixed mass $m_{E_2}=500$ GeV. In the first row, computations are performed for the centre-of-mass energy of $\sqrt{s}=28$ TeV, while the case of $\sqrt{s} = 14$ TeV is in the second row. $\sigma_C$ is the combined significance as defined in \eqref{eq:combined_significance}.}\label{tab:Evolve_Asimov_table_3}
		\end{table}
		
To further complement our analysis, we also perform a scan for different VLL masses, summarizing our results in Tab.~\ref{tab:Evolve_Asimov_table} for $3000$ $\mathrm{fb^{-1}}$ of integrated luminosity. For completeness, in particular, in order to understand how much we would gain in prolonging the LHC high-luminosity runs, we show in Tab.~\ref{tab:Evolve_Asimov_table_2} the same results but for an integrated luminosity of $9000$ $\mathrm{fb^{-1}}$.

First, in Table \ref{tab:Evolve_Asimov_table} we obtain significances of the order of $2\sigma$ up until a mass of 300 GeV, whereas, in Table \ref{tab:Evolve_Asimov_table_2}, the same significance would be achieved for a mass of 400 GeV. It is also interesting to observe what would be the effect of a collider at a higher centre-of-mass energy. For this scenario, we focus our attention at a particular point, $m_{E_2}=500$ GeV and a centre-of-mass energy $\sqrt{s} = 28$ TeV. The results are shown in Tab.~\ref{tab:Evolve_Asimov_table_3}. As expected, the discovery significance increases with the centre-of-mass energy. For example, combined significance of the $s/\sqrt{s+b}$ metric increases by a factor of $1.4$. Similarly, the other metrics also showcase improvements with an increase by factors of $1.6$ and $0.3$ for the $\mathcal{Z}(<1\%)$ and $\mathcal{Z}_A$ metrics, respectively. We also note that, for the $s/\sqrt{s+b}$ and $\mathcal{Z}_A$ measures of the VLBSM topology, small variations in statistical significances are obtained, with $\mathcal{Z}_A$ seeing no deviation. These small fluctuations might be associated with the stochastic nature of the tested NNs.
		
\begin{figure}[h!]
			\centering
			\captionsetup{justification=raggedright}
			\includegraphics[width=0.63\textwidth]{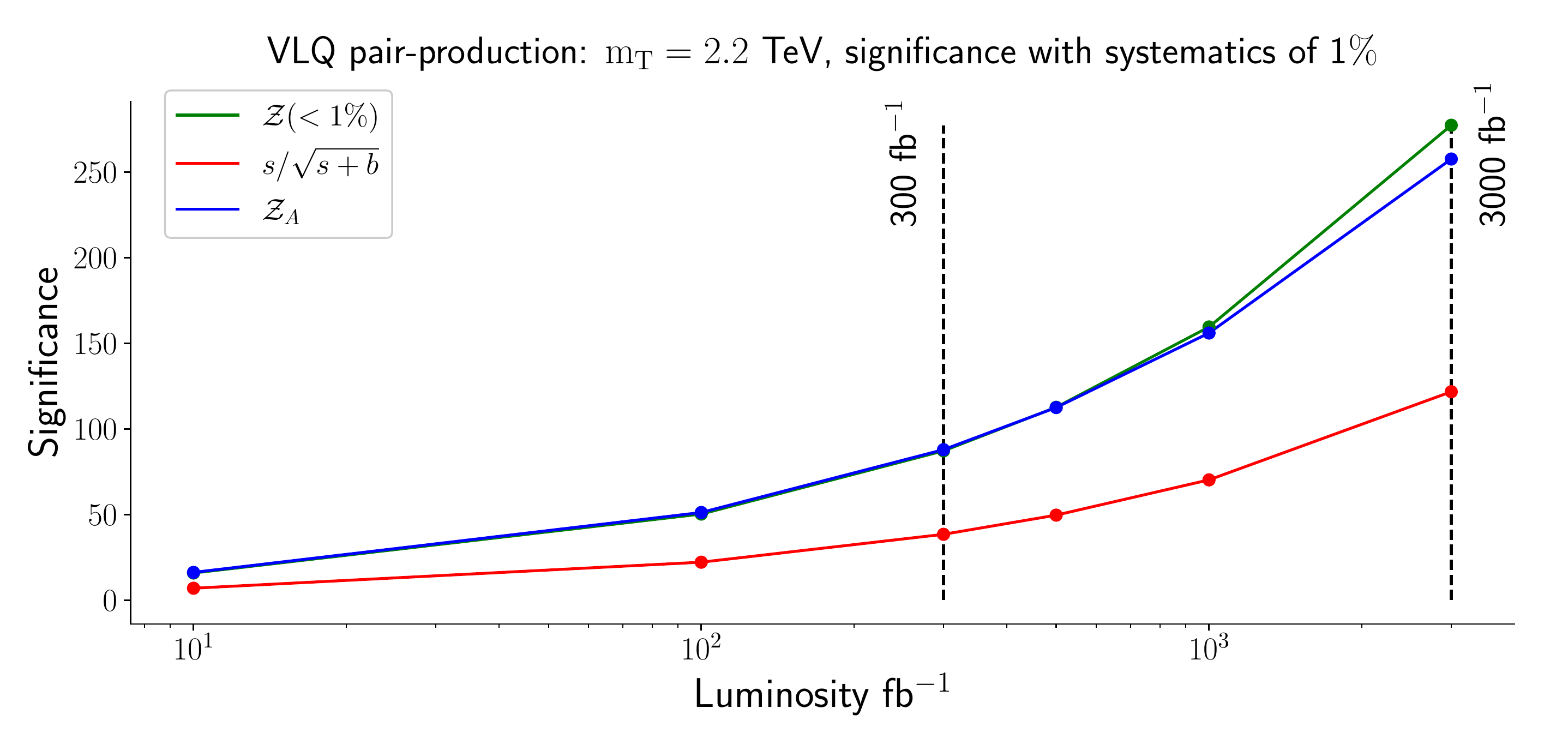}
			\caption{Statistical significance as a function of the integrated luminosity for the three statistical metrics under consideration. We fix the the lightest VLQ mass as $m_{\mathrm{T}} = 2.2~\mathrm{TeV}$. The green curve represents the $\mathcal{Z}(<1\%)$ metric, the red curve indicates the $s/\sqrt{s+b}$ one while the blue curve shows our results for the Asimov significance $\mathcal{Z}_A$.}
			\label{fig:VLQ_lum_sig}
		\end{figure}

A similar analysis was performed for VLQ searches pair produced via gluon-gluon fusion. In Fig.~\ref{fig:ACC-Sig-plots_1} of appendix \ref{app:sig_plots_DL} we show the significance in terms of the NN score where it a massive significance increase is immediately noticeable in comparison to the VLL case. This is mostly due to a far superior cross-section which, for $\mathcal{L}=3000~\mathrm{fb^{-1}}$, promptly results in a large Asimov significance, i.e.~$\mathcal{Z}_A = 257.53\sigma$, as well as equally large values for the other two metrics, $s/\sqrt{s+b} = 121.71\sigma$ and $\mathcal{Z}(<1\%) = 275.76\sigma$. As we show in Fig.~\ref{fig:VLQ_lum_sig}, the statistical significance for the VLQ searches is such that we can probe a large range of masses from $2.2~\mathrm{TeV}$ up to about $4.0~\mathrm{TeV}$.
The corresponding numerical results can be found in Tab.~\ref{tab:VLQs}. From this, we note that for a high-luminosity run, $\mathcal{L}=3000~\mathrm{fb^{-1}}$, one can exclude, or claim a discovery, by over $5$ standard deviations, of VLQ masses up to $3.8~\mathrm{TeV}$. More interestingly, the VLQ sector of the model under consideration, can already be probed at the forthcoming LHC Run-III, $\mathcal{L}=300~\mathrm{fb^{-1}}$, for VLQ masses of around $3.4~\mathrm{TeV}$.
		
			\begin{table}[h!]
		\centering
		\captionsetup{justification=raggedright,singlelinecheck=false}
		\resizebox{0.85\textwidth}{!}{\begin{tabular}{c||ccc||ccc||ccc||}
				\multirow{2}{*}{Mass of VLQ} & \multicolumn{3}{c||}{$300$ $\mathrm{fb^{-1}}$}                                               & \multicolumn{3}{c||}{$1000$ $\mathrm{fb^{-1}}$}                              & \multicolumn{3}{c||}{$3000$ $\mathrm{fb^{-1}}$}                                            \\ \cline{2-10} 
				& $s/\sqrt{s+b}$                          & $\mathcal{Z}(<1\%)$                         & $\mathcal{Z}_A$                & $s/\sqrt{s+b}$                           & $\mathcal{Z}(<1\%)$                         & $\mathcal{Z}_A$   & $s/\sqrt{s+b}$                          & $\mathcal{Z}(<1\%)$                          & $\mathcal{Z}_A$               \\ \hline
				$2.2$ TeV                      & \multicolumn{1}{c|}{$38.49$} & \multicolumn{1}{c|}{$87.19$} & \multicolumn{1}{c||}{$87.91$}       &\multicolumn{1}{c|}{$70.27$} & \multicolumn{1}{c|}{$159.51$} & \multicolumn{1}{c||}{$155.98$}& \multicolumn{1}{c|}{$121.71$} & \multicolumn{1}{c|}{$275.76$} & \multicolumn{1}{c||}{$257.53$}               \\
				$2.4$ TeV                      & \multicolumn{1}{c|}{$31.22$} & \multicolumn{1}{c|}{$67.13$} & \multicolumn{1}{c||}{$67.96$}       &\multicolumn{1}{c|}{$57.00$} & \multicolumn{1}{c|}{$122.70$} & \multicolumn{1}{c||}{$121.85$}& \multicolumn{1}{c|}{$98.73$} & \multicolumn{1}{c|}{$212.29$} & \multicolumn{1}{c||}{$202.30$}               \\
				$2.6$ TeV                      & \multicolumn{1}{c|}{$24.94$} & \multicolumn{1}{c|}{$50.01$} & \multicolumn{1}{c||}{$49.84$}       &\multicolumn{1}{c|}{$45.53$} & \multicolumn{1}{c|}{$91.34$} & \multicolumn{1}{c||}{$89.28$}& \multicolumn{1}{c|}{$78.87$} & \multicolumn{1}{c|}{$158.17$} & \multicolumn{1}{c||}{$147.40$}               \\
				$2.8$ TeV                      & \multicolumn{1}{c|}{$19.51$} & \multicolumn{1}{c|}{$36.10$} & \multicolumn{1}{c||}{$35.94$}       &\multicolumn{1}{c|}{$35.62$} & \multicolumn{1}{c|}{$65.92$} & \multicolumn{1}{c||}{$64.62$}& \multicolumn{1}{c|}{$61.70$} & \multicolumn{1}{c|}{$114.16$} & \multicolumn{1}{c||}{$107.57$}               \\
				$3.0$ TeV                      & \multicolumn{1}{c|}{$14.97$} & \multicolumn{1}{c|}{$25.33$} & \multicolumn{1}{c||}{$25.22$}       &\multicolumn{1}{c|}{$27.33$} & \multicolumn{1}{c|}{$27.33$} & \multicolumn{1}{c||}{$45.48$}& \multicolumn{1}{c|}{$47.34$} & \multicolumn{1}{c|}{$80.10$} & \multicolumn{1}{c||}{$76.23$}               \\
				$3.2$ TeV                      & \multicolumn{1}{c|}{$11.22$} & \multicolumn{1}{c|}{$17.28$} & \multicolumn{1}{c||}{$17.21$}       &\multicolumn{1}{c|}{$20.49$} & \multicolumn{1}{c|}{$31.55$} & \multicolumn{1}{c||}{$31.11$}& \multicolumn{1}{c|}{$35.49$} & \multicolumn{1}{c|}{$54.65$} & \multicolumn{1}{c||}{$52.42$}               \\
				$3.4$ TeV                      & \multicolumn{1}{c|}{$8.16$} & \multicolumn{1}{c|}{$11.43$} & \multicolumn{1}{c||}{$11.39$}       &\multicolumn{1}{c|}{$14.89$} & \multicolumn{1}{c|}{$20.87$} & \multicolumn{1}{c||}{$20.62$}& \multicolumn{1}{c|}{$25.79$} & \multicolumn{1}{c|}{$36.15$} & \multicolumn{1}{c||}{$34.88$}               \\
				$3.6$ TeV                      & \multicolumn{1}{c|}{$4.51$} & \multicolumn{1}{c|}{$4.91$} & \multicolumn{1}{c||}{$4.98$}       &\multicolumn{1}{c|}{$8.52$} & \multicolumn{1}{c|}{$8.10$} & \multicolumn{1}{c||}{$9.11$}& \multicolumn{1}{c|}{$16.32$} & \multicolumn{1}{c|}{$17.10$} & \multicolumn{1}{c||}{$15.43$}               \\
				$3.8$ TeV                      & \multicolumn{1}{c|}{$1.20$} & \multicolumn{1}{c|}{$3.25$} & \multicolumn{1}{c||}{$2.93$}       &\multicolumn{1}{c|}{$3.00$} & \multicolumn{1}{c|}{$7.11$} & \multicolumn{1}{c||}{$5.02$}& \multicolumn{1}{c|}{$6.01$} & \multicolumn{1}{c|}{$8.13$} & \multicolumn{1}{c||}{$10.05$}               \\
				$4.0$ TeV                      & \multicolumn{1}{c|}{$0.44$} & \multicolumn{1}{c|}{$0.66$} & \multicolumn{1}{c||}{$0.47$}       &\multicolumn{1}{c|}{$1.01$} & \multicolumn{1}{c|}{$1.33$} & \multicolumn{1}{c||}{$1.28$}& \multicolumn{1}{c|}{$2.20$} & \multicolumn{1}{c|}{$2.51$} & \multicolumn{1}{c||}{$1.91$}               \\
		\end{tabular}}
		\caption{Signal significance for the lightest VLQ-pair production. The computation follows a genetic algorithm that maximizes the Asimov significance metric. All significances are computed for proton-proton collisions at the centre-of-mass energy of $\sqrt{s} = 14$ TeV.}
		\label{tab:VLQs}
	\end{table}	

\begin{table}[htb!]    
	\centering
	\captionsetup{justification=raggedright,singlelinecheck=false}
	\resizebox{0.60\textwidth}{!}{\begin{tabular}{c|c c c c c c}
			$\mathrm{M_{VLQ}}=2.2~\mathrm{TeV}$ & $5\%$ & $10\%$ & $20\%$ & $40\%$ & $80\%$ & $96\%$\\[0.1CM] \hline
			$\mathrm{sys}: 1\%$ & $58.39\sigma$ & $56.32\sigma$ & $52.02\sigma$ & $42.74\sigma$ & $19.33\sigma$ & $5.21\sigma$ \\[0.05CM]
			$\mathrm{sys}: 10\%$ & $43.14\sigma$ & $41.80\sigma$ &  $39.00\sigma$ & $32.83\sigma$ & $16.13\sigma$ & $4.70\sigma$
	\end{tabular}}
	\caption{Asimov significance, $\mathcal{Z}_A$, for VLQ pair-production. The last six columns show the Asimov significance assuming a suppression of $5\%$ to $96\%$ in the signal cross section coming from unaccounted effects. In the first row a systematic uncertainty in $\mathcal{Z_A}$ of 1\% is considered, whereas in the second row we assume a systematic uncertainty of 10\%. We consider an integrated luminosity of $\mathcal{L} = 139~\mathrm{fb^{-1}}$ corresponding the acquired data after the LHC Run-II.}
	\label{tab:Xsecs}
\end{table}

It is relevant to mention that the obtained significance at low luminosity, in particular, for Run-II data, may naively suggest that all such scenarios are excluded. However, one must note that direct searches at the LHC are mostly focused in VLQ decays to third generation quarks, so far not considering channels with light jets and di-leptons as we propose in this article, making a comparison of our results with available data not suitable. With this in mind, the key point of our analysis relies on the fact that the process in Fig.~\ref{fig:VLQ-events} results in a cross-section significantly larger than that of the corresponding irreducible backgrounds, thus yielding a large significance for this particular channel. In fact, the evolutive algorithm employed in this work was engineered to find Neural Network models that further enhance such a discovery (or exclusion) significance. Since the architectures obtained with this methodology can easily find regions on the feature phase space, in our case the kinematic and angular observables described in tables \ref{tab:vars_Zp} and \ref{tab:vars_VLQ}, the separation between signal and background can be maximized as it is shown in Fig.~\ref{fig:ACC-Sig-plots_1}(a). For completeness of information we show in Tab.~\ref{tab:Xsecs} how the significance drops with a decreasing cross-section and with increasing systematic uncertainties. While our calculation has already been subject to detector effects with \texttt{Delphes} as well as to systematic uncertainties in the definition of the Asimov metric, one can take a conservative approach and consider that further unaccounted effects can impose a larger suppression in the cross-section than the one obtained from the \texttt{Delphes} output. It is remarkable to note that even for a suppression factor of $96\%$, a $2.2~\mathrm{TeV}$ VLQ can still be probed with a statistical significance of o $5.21\sigma$ with current LHC Run-II data. If we further increase the systematic uncertainties up to $10\%$, we obtain a worst case scenario lower bound on the discovery (or exclusion) significance of $4.7$ standard deviations.

\section{Conclusions}
\label{sec:conclusions}
	
In this paper, we have proposed a novel model where the SM gauge symmetry is enlarged by the $\mathcal{Q}_6 \times \mathcal{Z}_2$ discrete group. Within this framework, new exotic VLFs are emergent, both of quark and lepton types, as well as RH Majorana neutrinos. Furthermore, the scalar sector is enlarged by the inclusion of a new doublet and singlet scalar fields. We show that tree-level masses for third-generation fermions (top and tau) due to interactions with doublet scalars ($H_1$ and $H_2$) are generated after the spontaneous breaking of the electroweak gauge and the flavour $\mathcal{Q}_6 \times \mathcal{Z}_2$ symmetries. The remaining SM charged fermions gain their masses via a Universal seesaw mechanism mediated by VLFs. The tiny masses of the light active neutrinos arise from a tree-level type I seesaw mechanism mediated by heavy right handed Majorana neutrinos.
	
Due to sizeable couplings between the exotic VLL and the new scalar fields, contributions to the anomalous magnetic moment of the muon are generated. For certain benchmark scenarios for the couplings, we demonstrate that the model can successfully accommodate the measured muon $(g-2)$ anomaly. More specifically, considering the two benchmark scenarios $y_{E}=y_{E2}=0.2$ and $y_{E}=y_{E2}=0.3$ we can explain the observed muon $(g-2)$ anomaly within 2$\sigma$ error bars for masses of the $E_2$ between 200 GeV and 2 TeV. 

Phenomenological studies, in the context of collider physics at the LHC, are conducted for both VLLs and their quark counterparts. For this purpose, we employ the genetic algorithms to optimize the construction of neural networks, whose objective is to maximise the statistical significance of an hypothetical discovery of these particles at future experiments. For VLLs, we consider double production channels, either via production of a $Z^0$ boson or virtual photon, or via vector-boson fusion. We also consider the channel for single production. Using kinematic information of the final states, we determine that we can not exclude VLLs with more than five standard deviations for masses above $200~\mathrm{GeV}$, at the high-luminosity phase of the LHC, $\mathcal{L} = 3000$ $\mathrm{fb}^{-1}$. Assuming a hypothetical extension towards $\mathcal{L} = 9000$ $\mathrm{fb}^{-1}$, one can exclude the lightest VLL with masses up to approximately $200~\mathrm{GeV}$. We also determine the impact from the increasing center-of-mass energy at future colliders. For a mass of $m_{E_2} = 500$ GeV, we show that the combined significance improves when moving from $\sqrt{s} = 14$ TeV to $\sqrt{s} = 28$ TeV. Specifically, the significance increases from $0.86\sigma$ to $1.20\sigma$ for $s\sqrt{s+b}$, $1.23\sigma$ to $1.92\sigma$ for $\mathcal{Z}(<1\%)$ and $0.0562\sigma$ to $0.019\sigma$ for $\mathcal{Z}_A$. A similar analysis is made for VLQs, focusing on double production via strong-interaction channels, which is characterized by four leptons and two light jets in the final states. We found that VLQ masses up to $3.8~\mathrm{TeV}$ can be excluded at a luminosity of $\mathcal{L} = 3000$ $\mathrm{fb^{-1}}$ and up to $3.4~\mathrm{TeV}$ for $\mathcal{L} = 300$ $\mathrm{fb^{-1}}$.  To the best of our knowledge, the VLQ production channel proposed in this article has so far not been adopted in direct searches by experimental collaborations and must be considered both with currently available as well as with future data to be collected in forthcoming LHC runs.

	\section*{Acknowledgments}
	\noindent
	The authors acknowledge Ulises Saldaña Salazar for fruitful discussions at early stages of this work. The authors are also very grateful with Martin Hirsch for providing very useful comments and suggestions. A.E.C.H. acknowledges support by FONDECYT (Chile) under grant 
	No.~1210378, Milenio-ANID-ICN2019\_044 and ANID PIA/APOYO AFB180002. The work of C.B. was supported by FONDECYT grant No. 11201240. R.P.~is supported in part by the Swedish Research Council grant, contract number 2016-05996, as well as by the European Research Council (ERC) under the European Union's Horizon 2020 research and innovation programme (grant agreement No 668679). J.G., F.F.F., and A.P.M. are supported by the Center for Research and Development in Mathematics and Applications (CIDMA) through the Portuguese Foundation for Science and 
	Technology (FCT - Funda\c{c}\~{a}o para a Ci\^{e}ncia e a Tecnologia), references UIDB/04106/2020 and UIDP/04106/2020. A.P.M., F.F.F. and J.G. are supported by the project PTDC/FIS-PAR/31000/2017. A.P.M.~is also supported by national funds (OE), through FCT, I.P., in the scope of the framework contract foreseen in the numbers 4, 5 and 6 of the article 23, of the Decree-Law 57/2016, of August 29, changed by Law 57/2017, of July 19. The authors also would like to acknowledge the FCT Advanced Computing Project to provide computational resources via the project CPCA/A00/7395/2020. This work was partially produced with the support of INCD funded by FCT and FEDER under the project 01/SAICT/2016 nº 022153. J.G. is also directly funded by FCT through the doctoral program grant with the reference 2021.04527.BD.
	C.B. and R.P. would like to thank the members of the UTFSM particle physics group in Valpara\'iso for their hospitality during their visit, where part of this work was made.

	\appendix
	
	\section{$\mathcal{Q}_6$ multiplication rules}\label{app}

	The $\mathcal{Q}_6$ group contains 12 elements, ${a^m b^n}$, with $m=0,1,2,3,4,5$
	and $n=0,1$, where the group generators $a$ and $b$ satisfy $a^6=e$, $b^2=a^3$ and $b^{-1}ab=a^{-1}$. This discrete group has four 1-dimensional irreducible representations, $\mathbf{1}_{++}$, $\mathbf{1}_{+-}$, $\mathbf{1%
	}_{-+}$, and $\mathbf{1}_{--}$, and two 2-dimensional, $\mathbf{2}_1$ and $%
	\mathbf{2}_2$. The representation matrices of $a$ and $b$ for each irreducible representation are given by,
	
		\begin{eqnarray}
\begin{array}{lllll}
\label{generators}
{\bf 1}_{++} & : & a= 1 & & b= 1 \\
{\bf 1}_{--}   & : & a= 1 & & b=-1 \\
{\bf 1}_{+-}  & : & a=-1 & & b=-i \\
{\bf 1}_{-+} & : & a=-1 & & b= i \\
{\bf 2}_1  & : & a= \bmx{cc}
\alpha & 0 \\
0 & \alpha^{-1} \emx  & & b= \bmx{cc}
0 & i \\
i & 0
\emx \\
{\bf 2}_2 & : & a= \bmx{cc}
\alpha^2 & 0 \\
0 & \alpha^{-2} \emx  & & b^{}= \bmx{cc}
0 & 1 \\
1 & 0 \emx
\end{array}
\end{eqnarray}
with $\alpha=\exp{\left\{i\frac{2\pi}{6}\right\}}$.	The tensor products for the $\mathcal{Q}_6$ representations are given
	by 
	\begin{eqnarray}
	\label{22products}
	\left( 
	\begin{array}{c}
	a \\ 
	b%
	\end{array}
	\right)_{\mathbf{2}_2} \otimes \left( 
	\begin{array}{c}
	c \\ 
	d%
	\end{array}
	\right)_{\mathbf{2}_{1}} = \left(a c- b d \right)_{\mathbf{1}_{+-}} \oplus
	\left(a c+ b d \right)_{\mathbf{1}_{-+}} \oplus \left( 
	\begin{array}{c}
	a d \\ 
	b c%
	\end{array}
	\right)_{\mathbf{2}_{1}},
	\end{eqnarray}
	\vspace{-1.5em}
	\begin{eqnarray}
	\left( 
	\begin{array}{c}
	a \\ 
	b%
	\end{array}
	\right)_{\mathbf{2}_k} \otimes \left( 
	\begin{array}{c}
	c \\ 
	d%
	\end{array}
	\right)_{\mathbf{2}_{k}} = \left(a d- b c \right)_{\mathbf{1}_{++}} \oplus
	\left(a d+ b c \right)_{\mathbf{1}_{--}} \oplus \left( 
	\begin{array}{c}
	a c \\ 
	- b d%
	\end{array}
	\right)_{\mathbf{2}_{k^{\prime }}},
	\end{eqnarray}
	for $k,k^{\prime }=1,2$ and $k^{\prime }\neq k$, 
	\begin{eqnarray}
	\label{12products}
	\hspace{-3mm} & & \left( w\right)_{\mathbf{1}_{++}} \otimes \left( 
	\begin{array}{c}
	a \\ 
	b%
	\end{array}
	\right)_{\mathbf{2}_k}= \left( 
	\begin{array}{c}
	wa \\ 
	wb
	\end{array}
	\right)_{\mathbf{2}_k}, \quad \left( w\right)_{\mathbf{1}_{--}} \otimes
	\left( 
	\begin{array}{c}
	a \\ 
	b%
	\end{array}
	\right)_{\mathbf{2}_k}= \left( 
	\begin{array}{c}
	wa \\ 
	-w b%
	\end{array}
	\right)_{\mathbf{2}_k},  \notag \\
	\hspace{-3mm} & & \left( w\right)_{\mathbf{1}_{+-}} \otimes \left(
	\begin{array}{c}
	a \\ 
	b%
	\end{array}
	\right)_{\mathbf{2}_k}= \left( 
	\begin{array}{c}
	wb \\ 
	wa%
	\end{array}
	\right)_{\mathbf{2}_k}, \quad \left( w\right)_{\mathbf{1}_{-+}} \otimes
	\left( 
	\begin{array}{c}
	a \\ 
	b%
	\end{array}
	\right)_{\mathbf{2}_k}= \left( 
	\begin{array}{c}
	wb \\ 
	- wa%
	\end{array}
	\right)_{\mathbf{2}_k},
	\end{eqnarray}
	\begin{eqnarray}
	\label{11products}
	\mathbf{1}_{s_1s_2} \otimes \mathbf{1}_{s^{\prime }_1s^{\prime }_2} = 
	\mathbf{1}_{s^{\prime \prime }_1s^{\prime \prime }_2},
	\end{eqnarray}
	where $s_{1,2}\in \{+,-\}$, $s^{\prime \prime }_1=s_1s^{\prime }_1$ and $s^{\prime \prime
	}_2=s_2s^{\prime }_2$.

   The invariant Lagrangian of the theory is constructed with the use of eqs. (\ref{generators})-(\ref{11products}). To give an example of the field transformation under $\mathcal{Q}_6$, we take the $y_T$ term in eq. (\ref{Lyu}) where
   \begin{equation}
     \overline{Q}_{L_D}\sim {\bf 2}_2 \sim 
     \bmx{c}
\overline{Q}_{L_1}  \\
\overline{Q}_{L_2}
\emx \ \ 
T_R\sim {\bf 2}_1 \sim
\bmx{c}
T_{R_1}  \\
T_{R_2}
\emx \ \
H_1\sim 1_{+-}
   \end{equation}
   
   The generators $a$ and $b$ act in these fields as follows,
   
\begin{equation}
a: \overline{Q}_{L_D}\rightarrow a \overline{Q}_{L_D}= 
\bmx{c}
\alpha^{2} \overline{Q}_{L_1}  \\
\alpha^{-2} \overline{Q}_{L_2}
\emx,\ \
T_{R}\rightarrow a T_{R}=
\bmx{c}
\alpha^{} T_{R_1}  \\
\alpha^{-1} T_{R_2}
\emx,
\ \ \text{and} \ \
\tilde{H}_1\rightarrow a H_1= - \tilde{H}_1
\end{equation}

where $\alpha^{3}=\alpha^{-3}-1$ and

\begin{equation}
b: Q_{L_D}\rightarrow b Q_{L_D}= 
\bmx{c}
 \overline{Q}_{L_2}  \\
 \overline{Q}_{L_1}
\emx,\ \
T_{R}\rightarrow b T_{R}=
i\bmx{c}
 T_{R_2}  \\
 T_{R_1}
\emx
\ \ \text{and} \ \
\tilde{H}_1\rightarrow b \tilde{H}_1= i \tilde{H}_1
\end{equation}
   
Therefore, the term $y_T$ invariant under the SM gauge group and $\mathcal{Q}_6\times \mathcal{Z}_2$ is formed with the tensor product between the representation ${\bf1}_{+-}= (\overline{Q}_{L_1} T_{R_1} - \overline{Q}_{L_2} T_{R_2}) \supset {\bf2}_2\otimes {\bf2}_1$ and ${\bf1}_{+-}\sim \tilde{H}_1$, i.e. $\overline{Q}_{L_1} T_{R_1}\tilde{H}_1 - \overline{Q}_{L_2} T_{R_2}\tilde{H}_1$ as in eq. (\ref{Lyu2}). 
This term is invariant under $a$ and $b$ transformations,
\begin{equation}
a: \left(\overline{Q}_{L_1} T_{R_1}\tilde{H}_1 - \overline{Q}_{L_2} T_{R_2}\tilde{H}_1\right)\rightarrow (\alpha^2 \overline{Q}_{L_1})(\alpha \overline{T}_{R_1})(-\tilde{H}_1)-(\alpha^{-2} \overline{Q}_{L_2})
(\alpha^{-1} \overline{T}_{R_2})(-\tilde{H}_1)   
\end{equation}

\begin{equation}
b: \left(\overline{Q}_{L_1} T_{R_1}\tilde{H}_1 - \overline{Q}_{L_2} T_{R_2}\tilde{H}_1\right)\rightarrow (\overline{Q}_{L_2})(i \overline{T}_{R_2})(i\tilde{H}_1)-(\overline{Q}_{L_1})(i\overline{T}_{R_1})(i\tilde{H}_1)   
\end{equation}

\section{$\chi^2$-fit in the quark sector}
\label{app:chiQuarks}
In order to find the best fit point that successfully reproduces the SM quark masses and CKM parameters, we proceed to minimize the following $\chi ^{2}$ function:
\begin{equation}
\chi ^{2} = \sum_{f}\frac{(m_{f}^{\text{th}}-m_{f}^{\text{exp}})^{2}}{\sigma_{f}^{2}}
+\frac{(|\mathbf{V}_{12}^{\text{th}}|-|\mathbf{V}_{12}^{\text{exp}}|)^{2}}{\sigma _{12}^{2}}+\frac{(|\mathbf{V}_{23}^{\text{th}}|-|\mathbf{V}_{23}^{\text{exp}}|)^{2}}{\sigma _{23}^{2}}+\frac{(|\mathbf{V}_{13}^{\text{th}}|-|\mathbf{V}_{13}^{\text{exp}}|)^{2}}{\sigma _{13}^{2}} + \frac{(J_{q}^{\text{th}}-J_{q}^{\text{exp}})^{2}}{\sigma _{J}^{2}}\, \;,
\end{equation}
where $f=u,c,t,d,s,b$ and $J_q$ is the Jarlskog parameter. The experimental values for the quark masses are given by~\cite{Xing:2020ijf},
\begin{align}
\begin{split}
    m_u^{\text{exp}} (M_Z) & = 1.24\pm 0.22 \text{ MeV} \;, \\
	m_c^{\text{exp}} (M_Z) & = 0.626 \pm 0.020 \text{ GeV} \;, \\
    m_t^{\text{exp}} (M_Z) & = 172.9 \pm 0.04 \text{ GeV} \;, \\
    m_d^{\text{exp}} (M_Z) & = 2.69\pm 0.19 \text{ MeV} \;, \\
    m_s^{\text{exp}} (M_Z) & = 53.5\pm 4.6 \text{ MeV} \;, \\
    m_b^{\text{exp}} (M_Z) & = 2.86 \pm 0.03 \text{ GeV}  \;,
\end{split} 
\end{align}
and CKM parameters are~\cite{Zyla:2020zbs} 
\begin{align}
\begin{split}
    |\mathbf{V}_{12}^{\text{exp}}|  = 0.22452 \pm 0.00044 \;, \quad
    &|\mathbf{V}_{23}^{\text{exp}}|  = 0.04214 \pm 0.00076  \;, \quad
    |\mathbf{V}_{13}^{\text{exp}}|  = 0.00365 \pm 0.00012 \;,\\
    &J_{q}^{\text{exp}}  = (3.18\pm 0.15) \times 10^{-5}\;.
\end{split}
\end{align}

The parameters in the quark sector are varied considering real up-type quark Yukawa couplings while the down-type ones are taken as complex.
Under this assumption, we find that the minimization of the $\chi ^{2}$ function yields the following quark mass matrices:
\begin{eqnarray}
M_U&=&\left(
\begin{array}{ccccc}
 0 & 0 & 0 & 13.2667 & 0 \\
 0 & 0 & 0 & 0 & -13.2667 \\
 0 & 0 & 210.385 & 0 & 0 \\
 88.4086 & 0 & -217.233 & -847345. & -314.094 \\
 0 & -88.4086 & -1358.2 & 314.094 & -1932.57 \\
\end{array}
\right)GeV\\
M_D&=&\left(
\begin{array}{ccccc}
 0 & 0 & 0 & 1.97436\, +1.22717 i & 0 \\
 0 & 0 & 0 & 0 & -1.97436-1.22717 i \\
 0 & 0 & 2.8349 & 0 & 0 \\
 -7.81061+35.4833 i & 0 & -98.684+14.9212 i & 18807.\, -19287.6 i & 4650.77\, -4769.62 i \\
 0 & 7.81061\, -35.4833 i & 29.0364\, -6.86904 i & -4650.77+4769.62 i & 183.845\, -188.543 i \\
\end{array}
\right)\notag
\label{benchmarkquark}
\end{eqnarray}
It is important to mention that this solution is not unique. Therefore,
given the large amount of free parameters,
there are many more viable solutions solutions that provide consistent values for the SM quark masses and CKM parameters and that give rise to same values of vector like quark masses used in our benchmark scenario. Furthermore, let us note that the lepton sector observables are easily fit using a similar procedure to the one described in this section.
\section{Kinematic and angular distributions for the VLLs and VLQs}
\label{app2}

\begin{figure}[h!]
	\vspace*{-2.5cm}
	\centering
	\captionsetup{justification=raggedright}
	\includegraphics[width=0.85\textwidth]{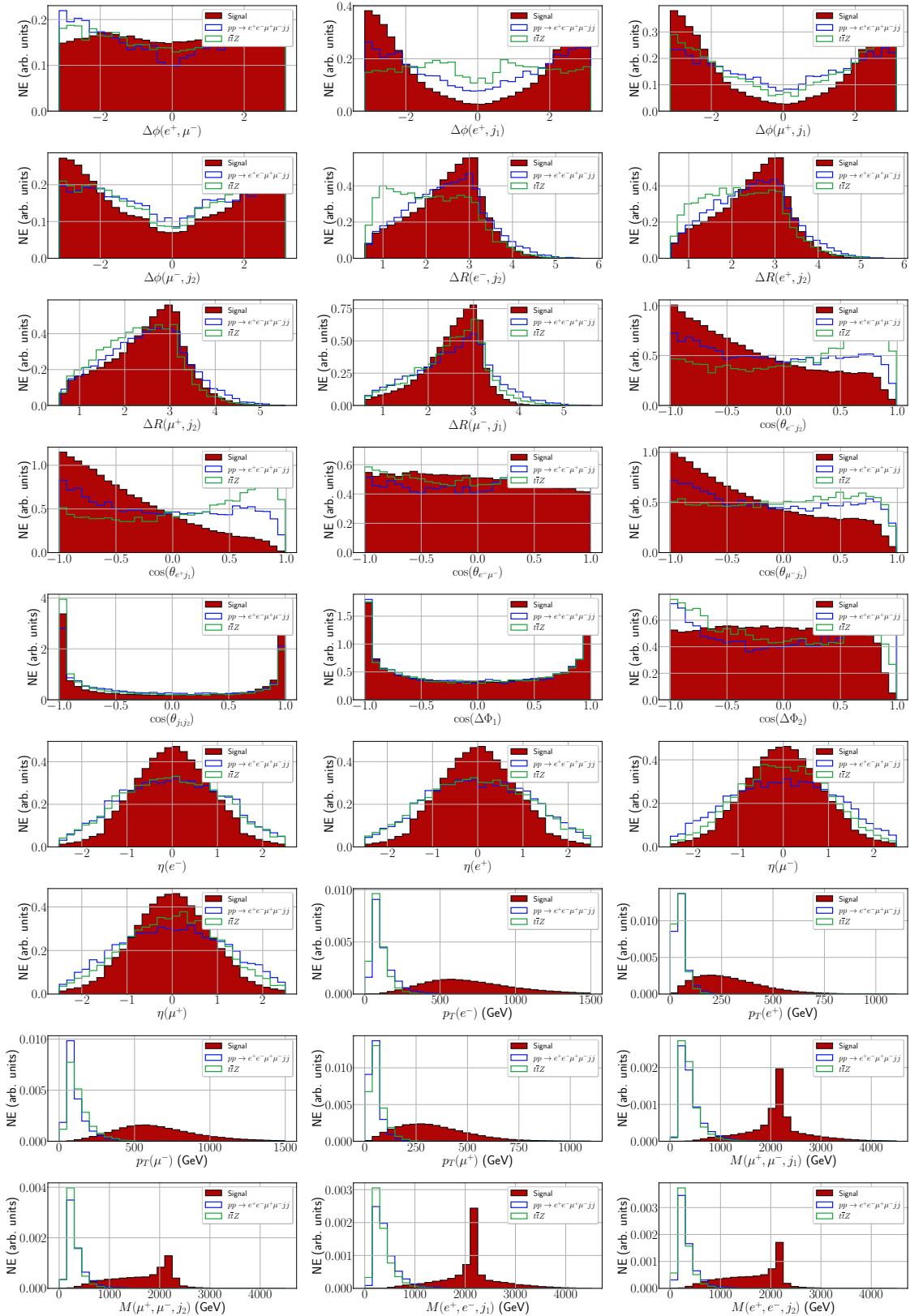}
	\caption{Set of angular and kinematic observables for VLQ pair-production with a mass $m_{T_1} = 2.2$ TeV. All histograms are normalized, with each having 30 bins. Signal distributions are filled and shown in red, while background distributions are unfilled and shown in blue ($pp \rightarrow e^+ e^- \mu^+ \mu^- j_1 j_ 2$) and green ($t\bar{t}Z$). From top-left to bottom-right, we have $\Delta \phi$ angles for final states combinations $(e^+,\mu^-)$, $(e^+,j_1)$, $(\mu^+,j_1)$ and $(\mu^-,j_2)$; $\Delta R$ distributions for the final states combinations $(e^-,j_2)$, $(e^+,j_2)$, $(\mu^+,j_2)$ and $(\mu^-,j_1)$; distributions of the cosine of the angle within the pair of states $(e^-,j_2)$, $(e^+,j_1)$, $(e^-,\mu^-)$, $(\mu^-,j_2)$ and $(j_1,j_2)$; distributions for the cosine of the angle $\Delta\Phi_1$ and $\Delta\Phi_2$; pseudo-rapidity for $e^-$, $e^+$, $\mu^-$ and $\mu^+$; transverse momentum for $e^-$, $e^+$, $\mu^-$, $\mu^+$ and invariant mass distributions for the combinations $(\mu^+,\mu^-,j_1)$, $(\mu^+,\mu^-,j_2)$, $(e^+,e^-,j_1)$ and $(e^+,e^-,j_2)$.}
	\label{fig:VLQ_vars}
\end{figure}

\begin{figure}[h!]
	\vspace*{-2.5cm}
	\centering
	\captionsetup{justification=raggedright}
	\includegraphics[width=0.85\textwidth]{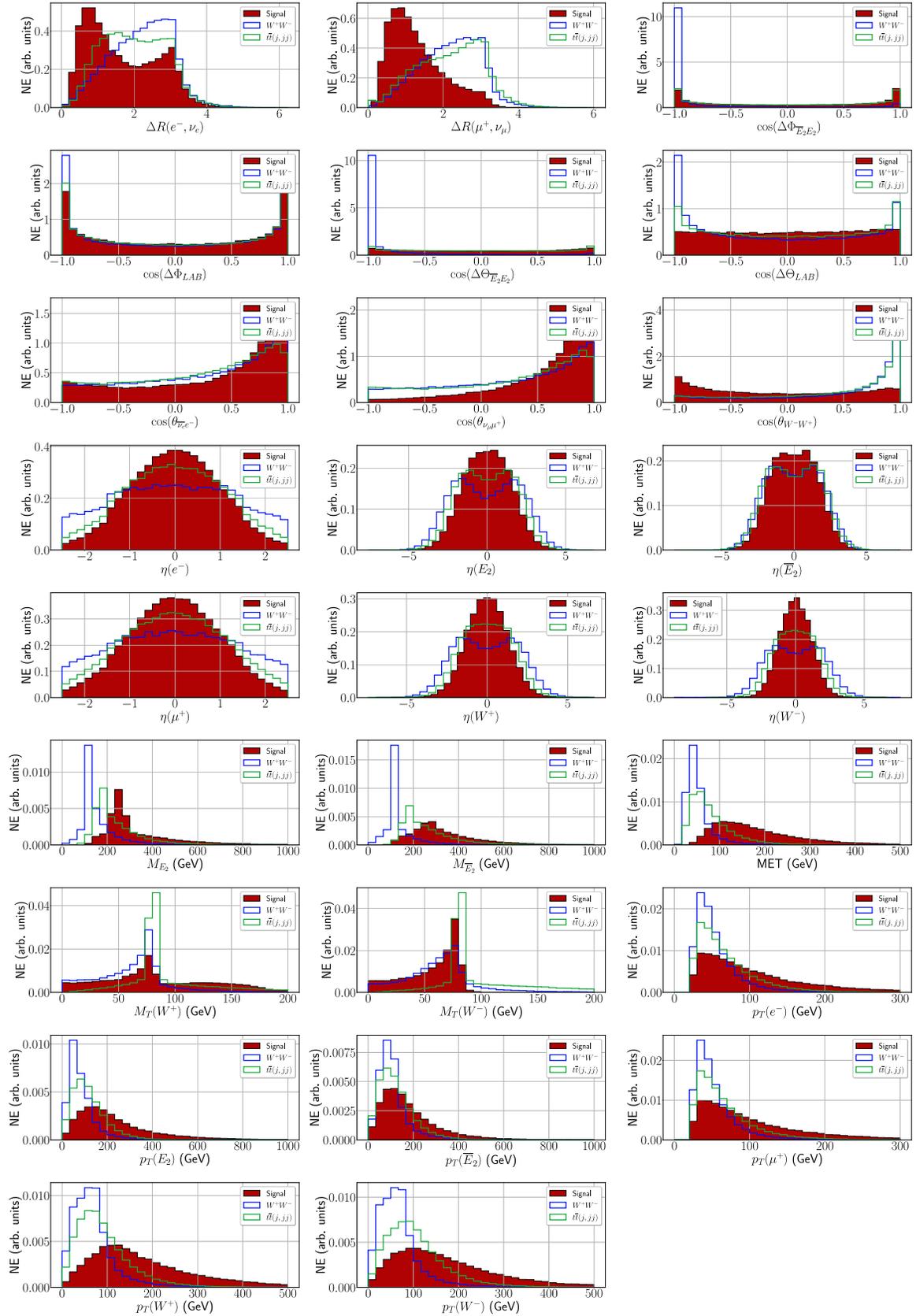}
	\caption{Set of angular and kinematic observables for VLL pair-production via the VBF topology with a mass $m_{E_2} = 200$ GeV. All histograms are normalized, with each having 30 bins. Signal distributions are filled and shown in red, while background distributions are unfilled and shown in blue ($W^+W^-$) and green ($t\bar{t} + (j,jj)$). From top-left to bottom-right, we have $\Delta R$ distributions for final states combinations $(e^-,\nu)$ and $(\mu^+,\nu)$; distributions for cosine of angle $\Delta \Phi$ and $\Delta\Theta$ in the laboratory and $E_2\bar{E}_2$ centre-of-mass frames; distributions of the cosine of the angle between the pairs of particles $(\nu,e^-)$, $\nu,\mu^+$ and $W^+W^-$; pseudo-rapidity for the particles $e^-$, $E_2$, $\bar{E}_2$, $\mu^+$, $W^+$ and $W^-$; mass distributions for $E_2$ and $\bar{E}_2$; MET; transverse mass distributions for $W^+$ and $W^-$ and transverse momentum distributions for $e^-$, $E_2$, $\bar{E}_2$, $\mu^+$, $W^+$ and $W^-$.}
	\label{fig:VBF_vars_LabFrame}
\end{figure}

\begin{figure}[h!]
	\vspace*{-2.5cm}
	\centering
	\captionsetup{justification=raggedright}
	\includegraphics[width=0.85\textwidth]{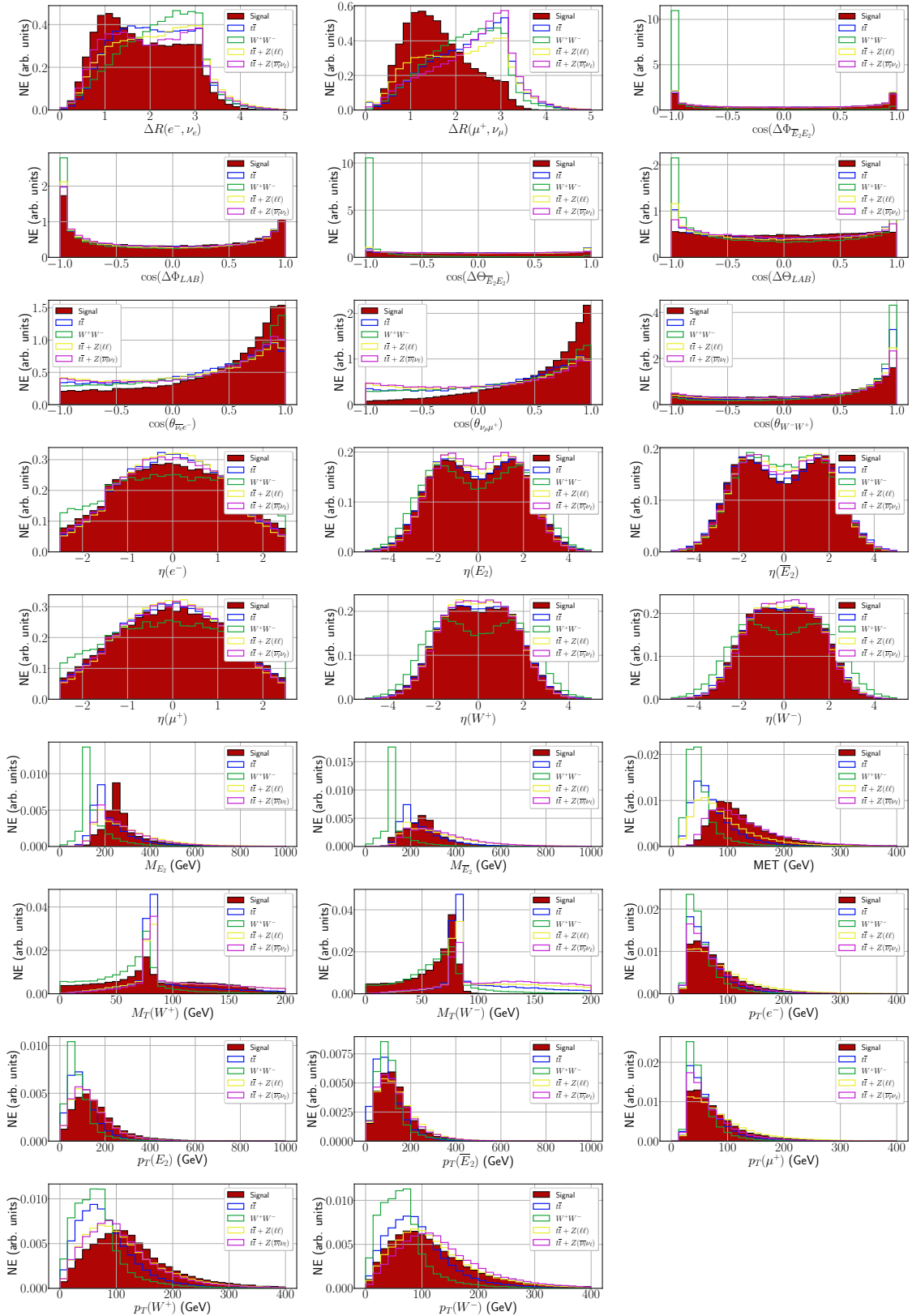}
	\caption{Set of angular and kinematic observables for VLL pair-production via the ZA topology with a mass $m_{E_2} = 200$ GeV. All histograms are normalized, with each having 30 bins. Signal distributions are filled and shown in red, while background distributions are unfilled and shown in blue ($t\bar{t}^-$), green ($W^+W^-$), yellow ($t\bar{t} + Z(\ell\ell)$) and pink ($t\bar{t} + Z(\bar{\nu}_\ell\nu_\ell)$). From top-left to bottom-right, we have $\Delta R$ distributions for final states combinations $(e^-,\nu)$ and $(\mu^+,\nu)$; distributions for cosine of angle $\Delta \Phi$ and $\Delta\Theta$ in the laboratory and $E_2\bar{E}_2$ centre-of-mass frames; distributions of the cosine of the angle between the pairs of particles $(\nu,e^-)$, $\nu,\mu^+$ and $W^+W^-$; pseudo-rapidity for the particles $e^-$, $E_2$, $\bar{E}_2$, $\mu^+$, $W^+$ and $W^-$; mass distributions for $E_2$ and $\bar{E}_2$; MET; transverse mass distributions for $W^+$ and $W^-$ and transverse momentum distributions for $e^-$, $E_2$, $\bar{E}_2$, $\mu^+$, $W^+$ and $W^-$.}
	\label{fig:ZA_vars_LabFrame}
\end{figure}

\begin{figure}[h!]
	\vspace*{-2.5cm}
	\centering
	\captionsetup{justification=raggedright}
	\includegraphics[width=0.85\textwidth]{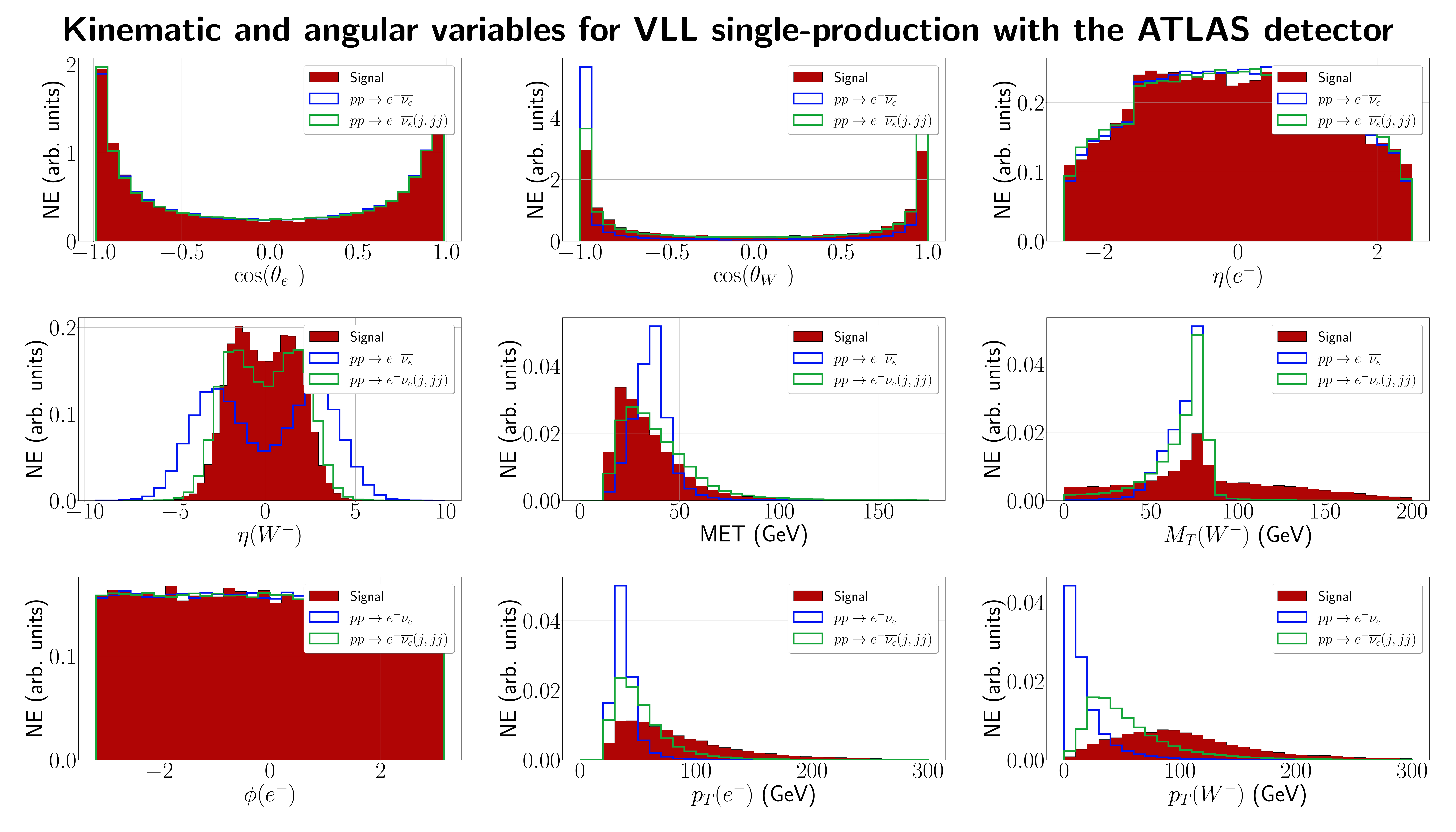}
	\caption{Set of angular and kinematic observables for VLL single-production via the VLBSM topology with a mass $m_{E_2} = 200$ GeV. All histograms are normalized, with each having 30 bins. Signal distributions are filled and shown in red, while background distributions are unfilled and shown in blue ($pp\rightarrow e^-\bar{\nu}_e$) and green ($pp\rightarrow e^-\bar{\nu}_e + (j,jj)$). From top-left to bottom-right, we have distributions for cosine of the angle for particles $e^-$ and $W^-$; pseudo-rapidity for $e^-$ and $W^-$; MET; transverse mass of the $W^-$ boson, azimuthal angle for $e^-$ and transverse momentum for $e^-$ and $W^-$.}
	\label{fig:VLBSM_vars_LabFrame}
\end{figure}

\begin{figure*}[]
	\centering
	\captionsetup{justification=raggedright}
	\includegraphics[width=0.85\textwidth]{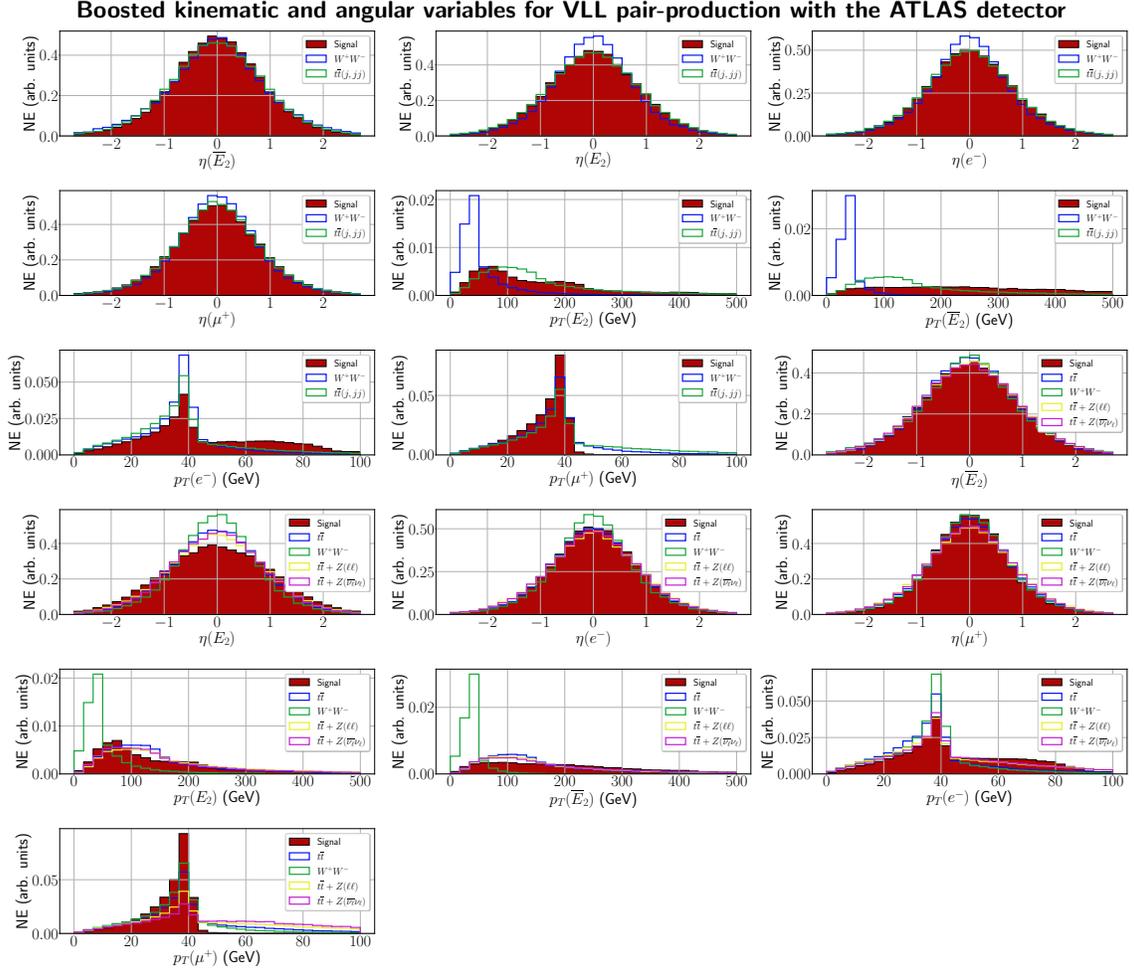}
	\caption{Set of angular and kinematic observables for VLL double-production for both VBF and ZA topologies with a mass $m_{E_2} = 200$ GeV. All histograms are normalized, with each having 30 bins. VBF channel corresponds to the plots where only two backgrounds are present, whereas ZA plots show four backgrounds. From top-left to bottom-right, for both signals, we have pseudo-rapidity distributions for $\bar{E}_2$, $E_2$, $e^-$ and $\mu^+$ and transverse momentum distributions for $E_2$, $\bar{E}_2$, $e^-$ and $\mu^+$.}
	\label{fig:Vars_BoostedFrame}
\end{figure*}

\cleardoublepage
\section{Deep Learning significance plots}\label{app:sig_plots_DL}
In this appendix, plots of the statistical significance as a function of the cut on the classifier score are presented, for both a VLL with mass 200 GeV (for every topology studied in this paper) and for a 2.2 TeV VLQ. The values for the cut on the classifier score are assigned by the NN architecture. Such values are real numbers (often referred to as decision boundary or decision surface) and can be interpreted as a continuous label that identifies whether a set of features (kinematic and angular observables) is classified as signal or as background. As an example, in Fig.~\ref{fig:ACC-Sig-plots_1} (a), the values for the classifier score around $\sim 0.56$ to $\sim 0.61$ correspond to the region where the significance is maximized, \textit{i.e.}, where the NN has found a maximal number of signal events for a non-zero background. Notice that from the definition of the Asimov significance in Eq.~\eqref{eq:Asimov_sig}, for a region with zero background $\mathcal{Z_A}$ would diverge and therefore is not considered in our analysis. On the other hand, for scores above $0.61$ the number of background events increases and the statistical significance quickly drops to zero.

		\begin{figure*}[htb!]
			\captionsetup{justification=raggedright}
			\subfloat[ZA topologies]{{\includegraphics[width=0.32\textwidth]{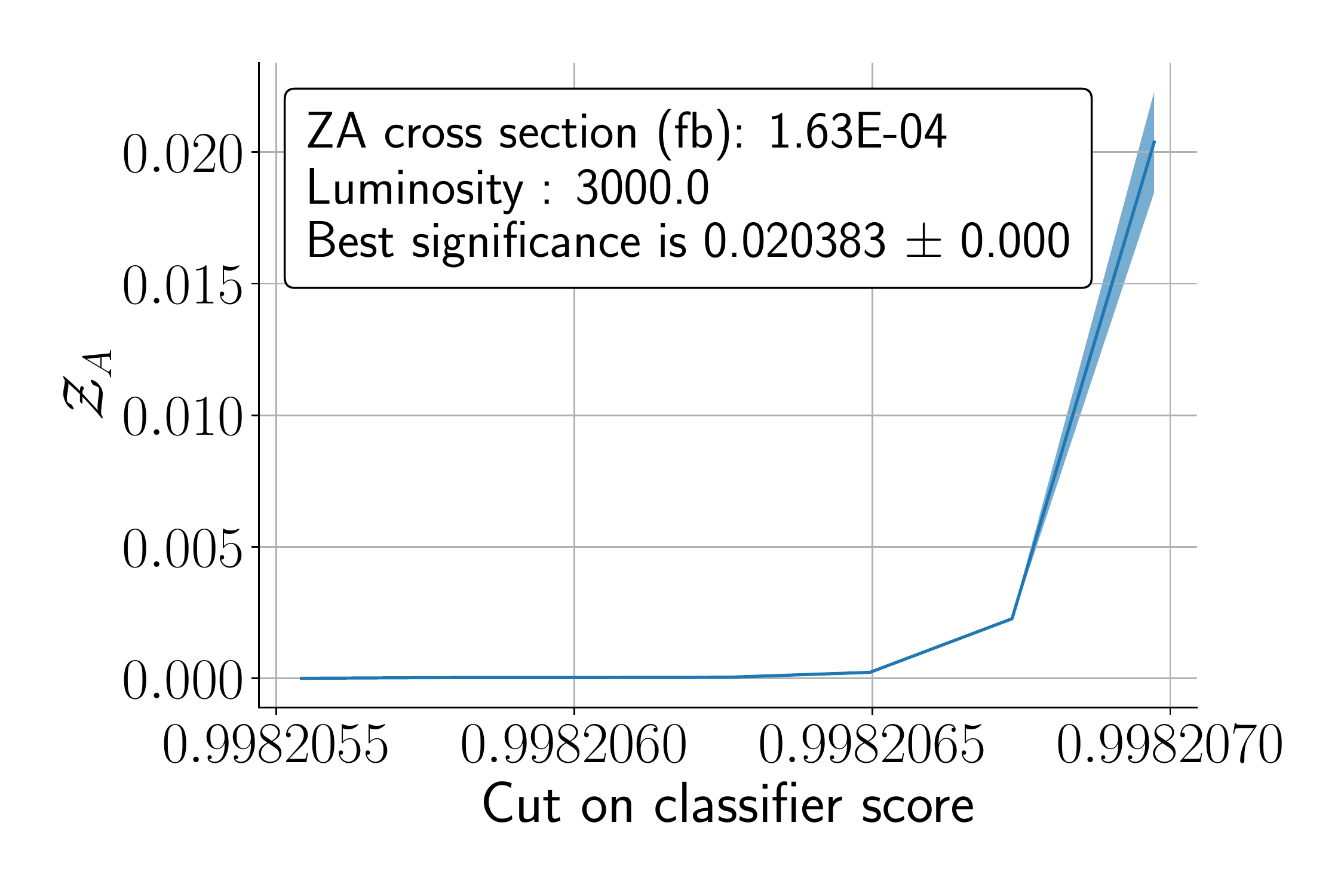} }} 
			\subfloat[ZA topologies]{{\includegraphics[width=0.32\textwidth]{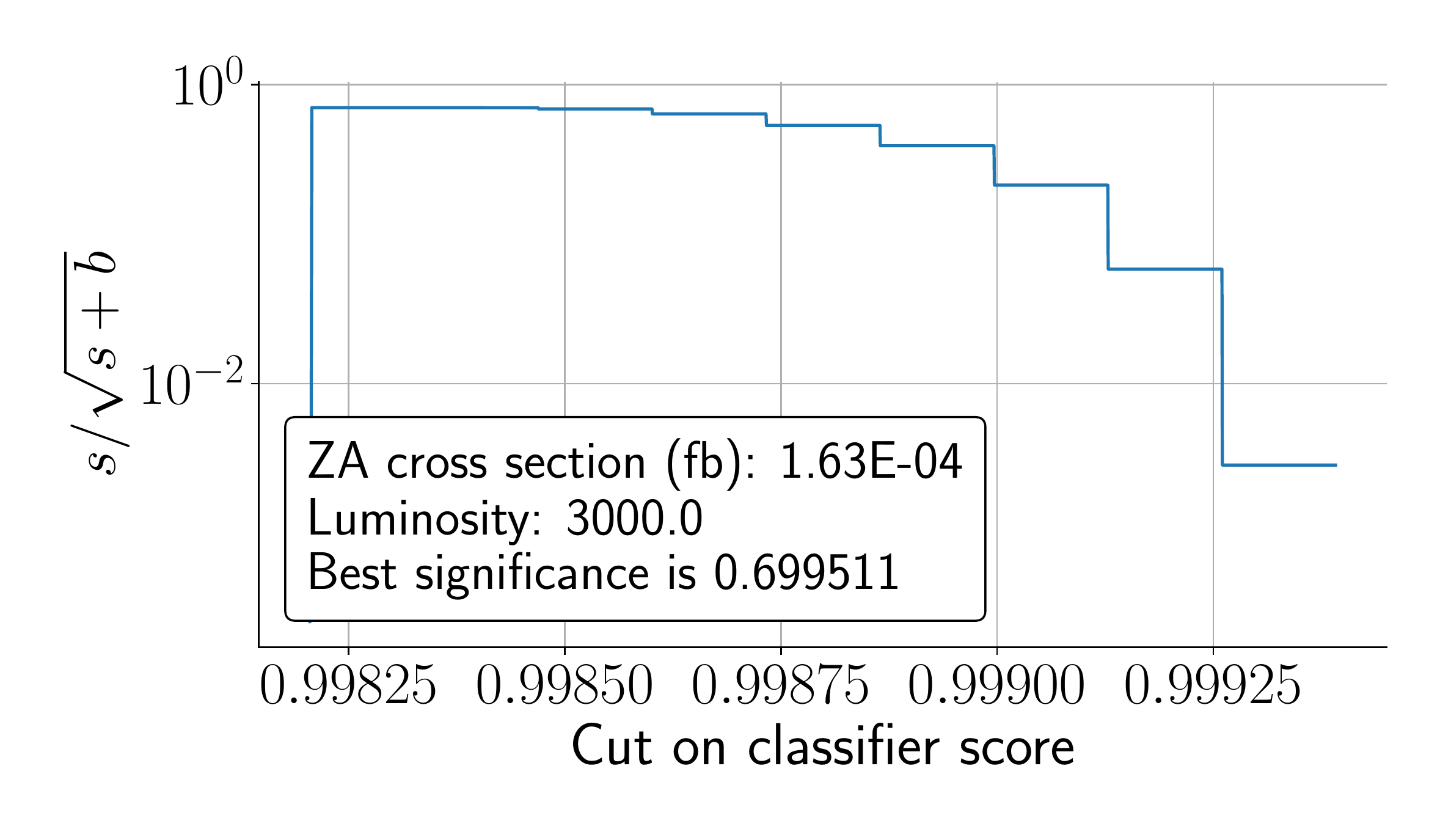} }} 
			\subfloat[ZA topologies]{{\includegraphics[width=0.32\textwidth]{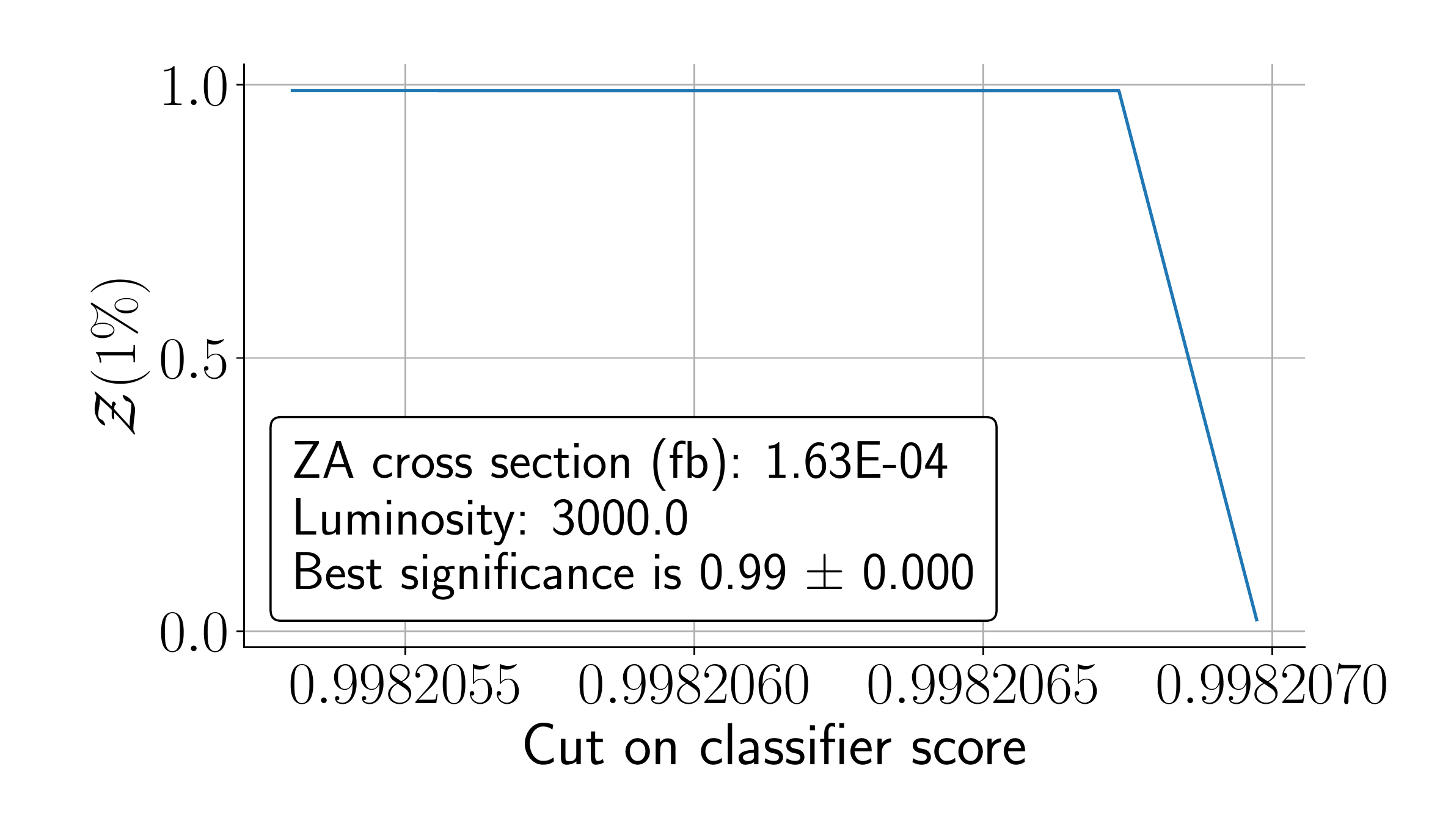} }}\\
			\hspace*{-1.0cm}
			\subfloat[VBF topologies]{{\includegraphics[width=0.32\textwidth]{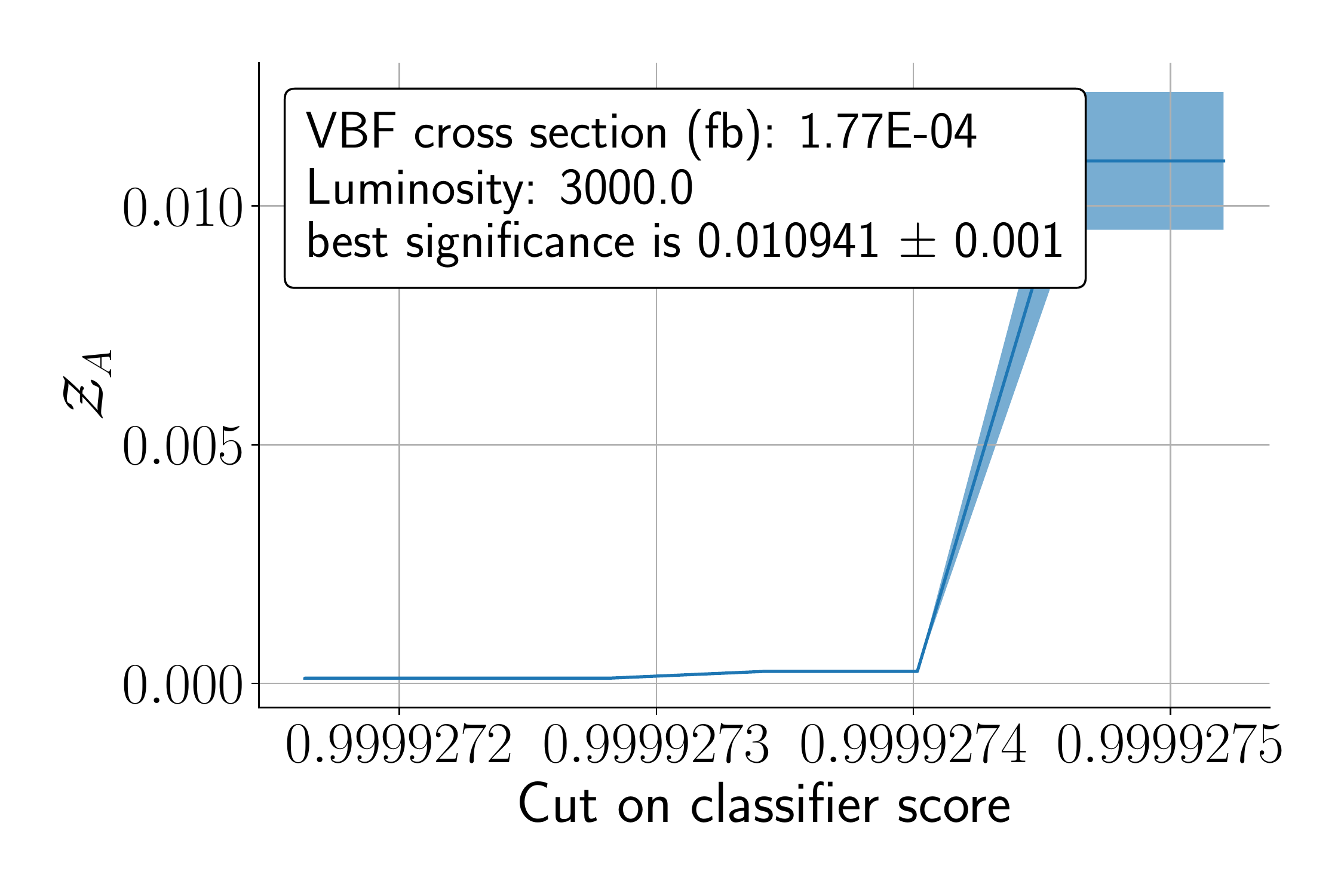} }} 
			\subfloat[VBF topologies]{{\includegraphics[width=0.32\textwidth]{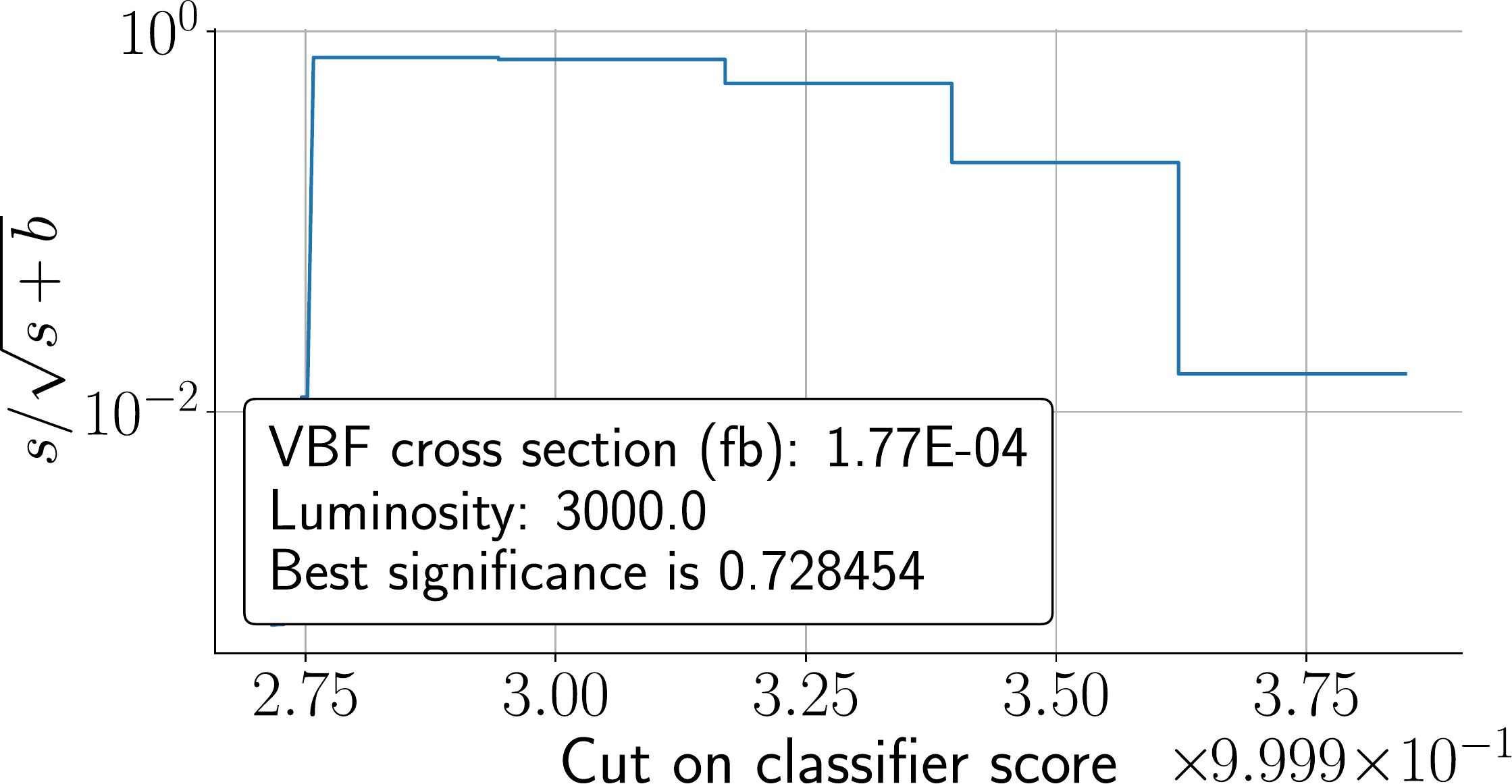} }} 
			\subfloat[VBF topologies]{{\includegraphics[width=0.32\textwidth]{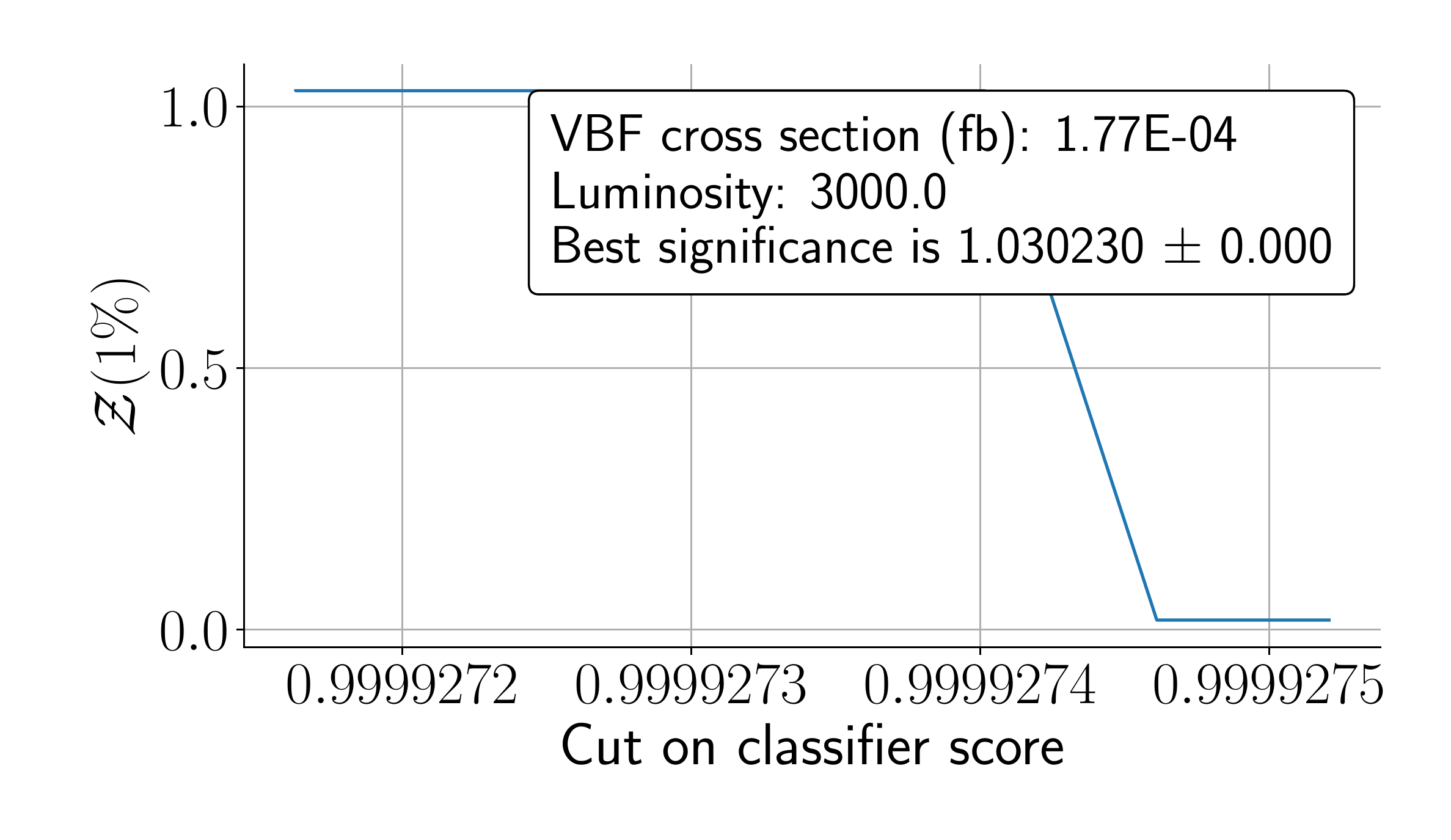} }}\\
			\hspace*{-1.0cm}
			\subfloat[VLBSM topologies]{{\includegraphics[width=0.32\textwidth]{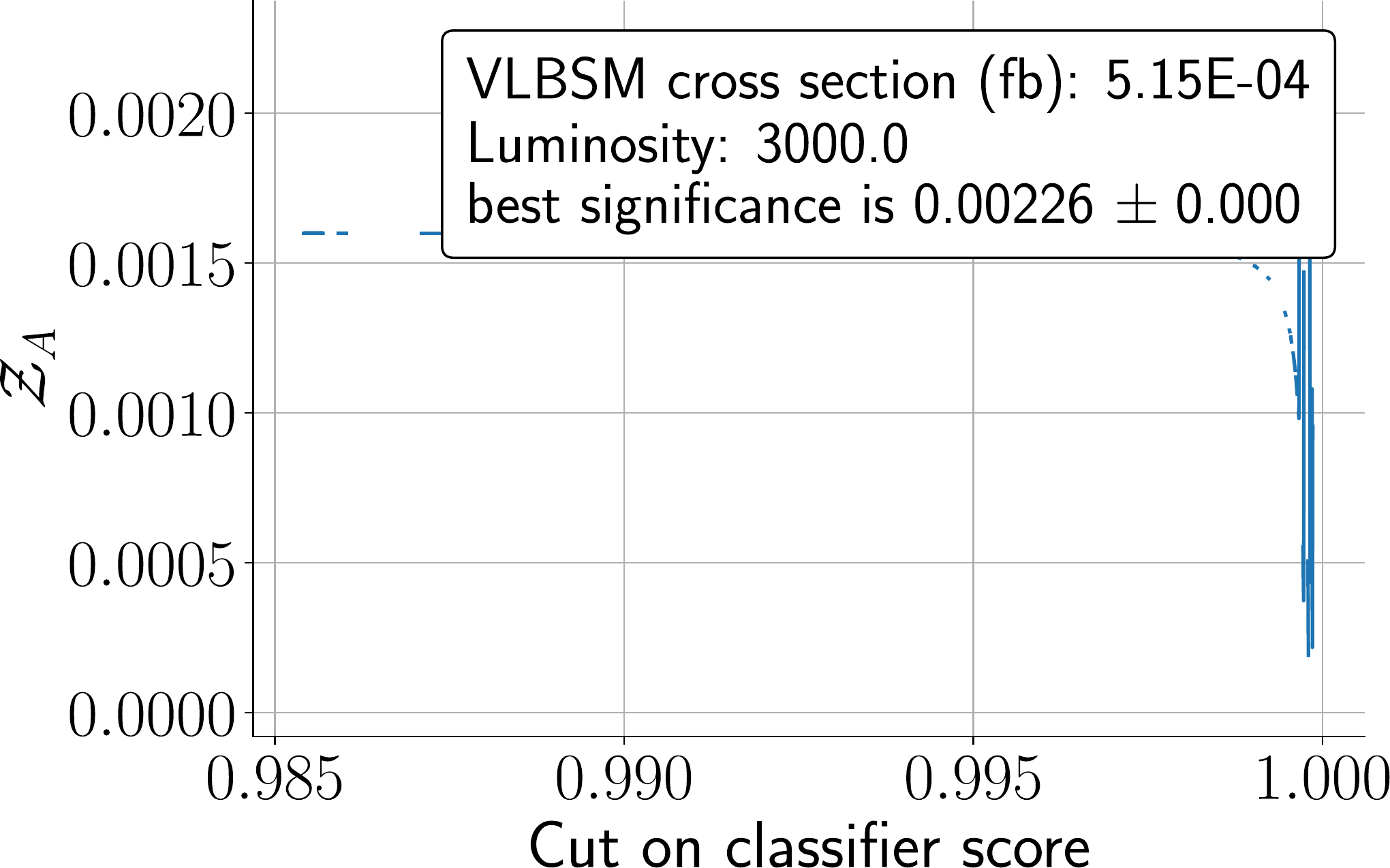} }} 
			\subfloat[VLBSM topologies]{{\includegraphics[width=0.32\textwidth]{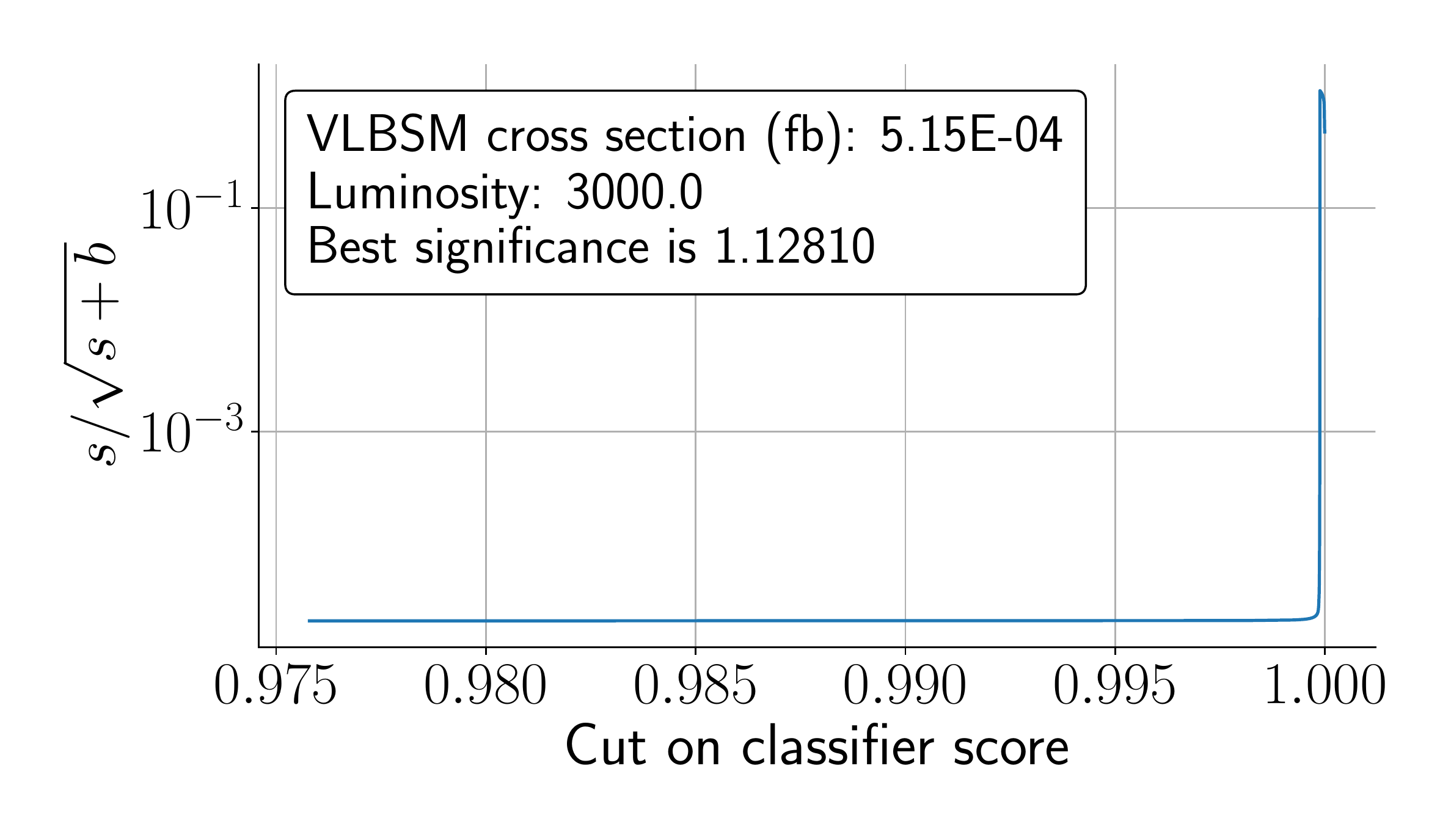} }} 
			\subfloat[VLBSM topologies]{{\includegraphics[width=0.36\textwidth]{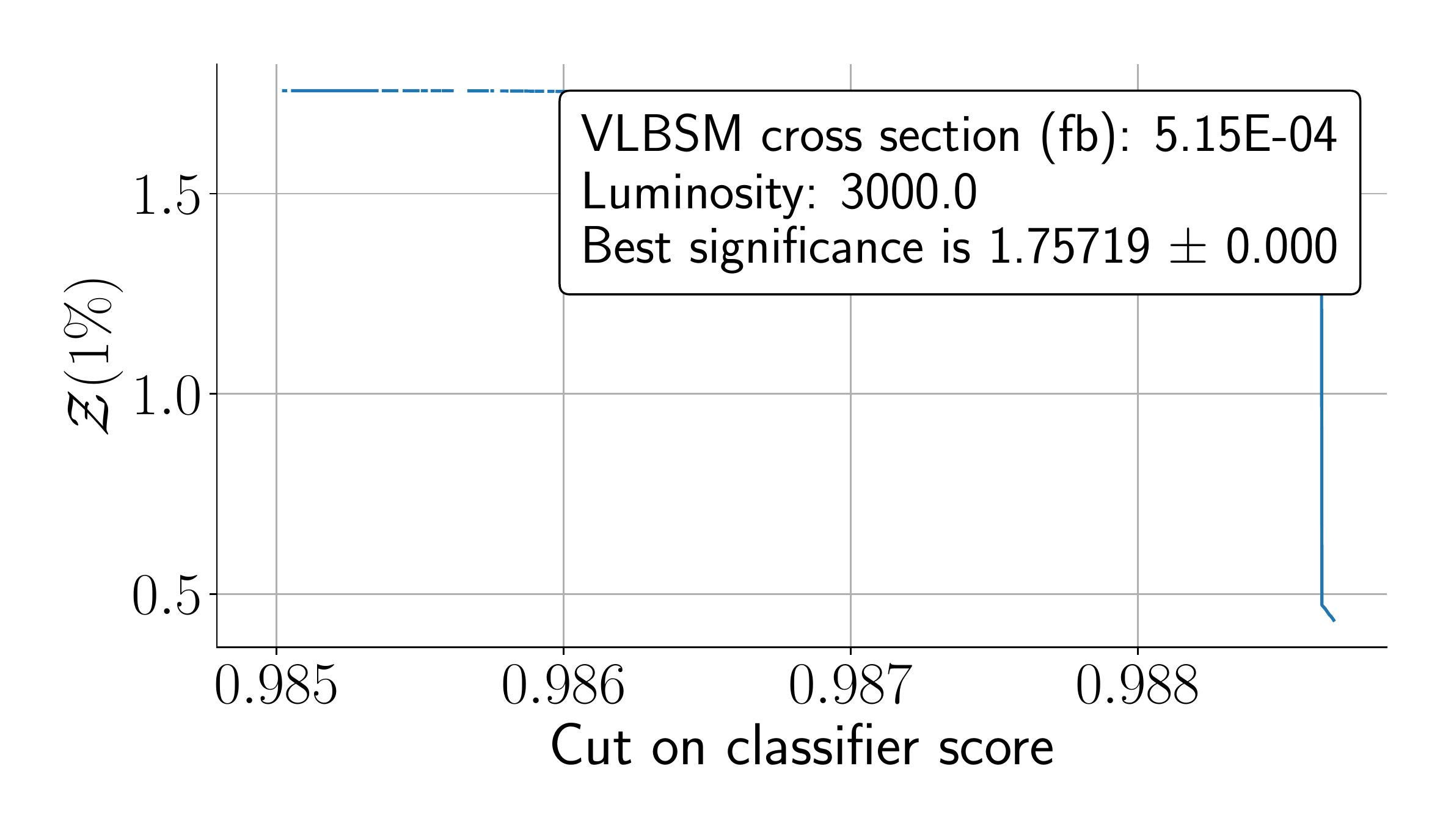} }}\\
			\caption{Statistical significance as a function of the classifier score given by NN for each signal graph, assuming the lightest VLL with mass $m_{E_2} = 200~\mathrm{GeV}$ and a luminosity $\mathcal{L} = 3000$ $\mathrm{fb}^{-1}$. All showcased plots are representative of the best NN found, following a genetic algorithm that maximizes the Asimov significance. For plots (a), (d) and (g) we present the Asimov significance, with 1\% systematics, for (b), (c) and (h) -- the standard significance $s/\sqrt{s+b}$ and for (c), (f) and (i) -- the adapted Asimov significance, where it is assumed that backgrounds are known with error $10^{-3}$.
				\label{fig:ACC-Sig-plots}}
		\end{figure*}
			\begin{figure*}[]
			\captionsetup{justification=raggedright}
			\subfloat[]{{\includegraphics[width=0.30\textwidth]{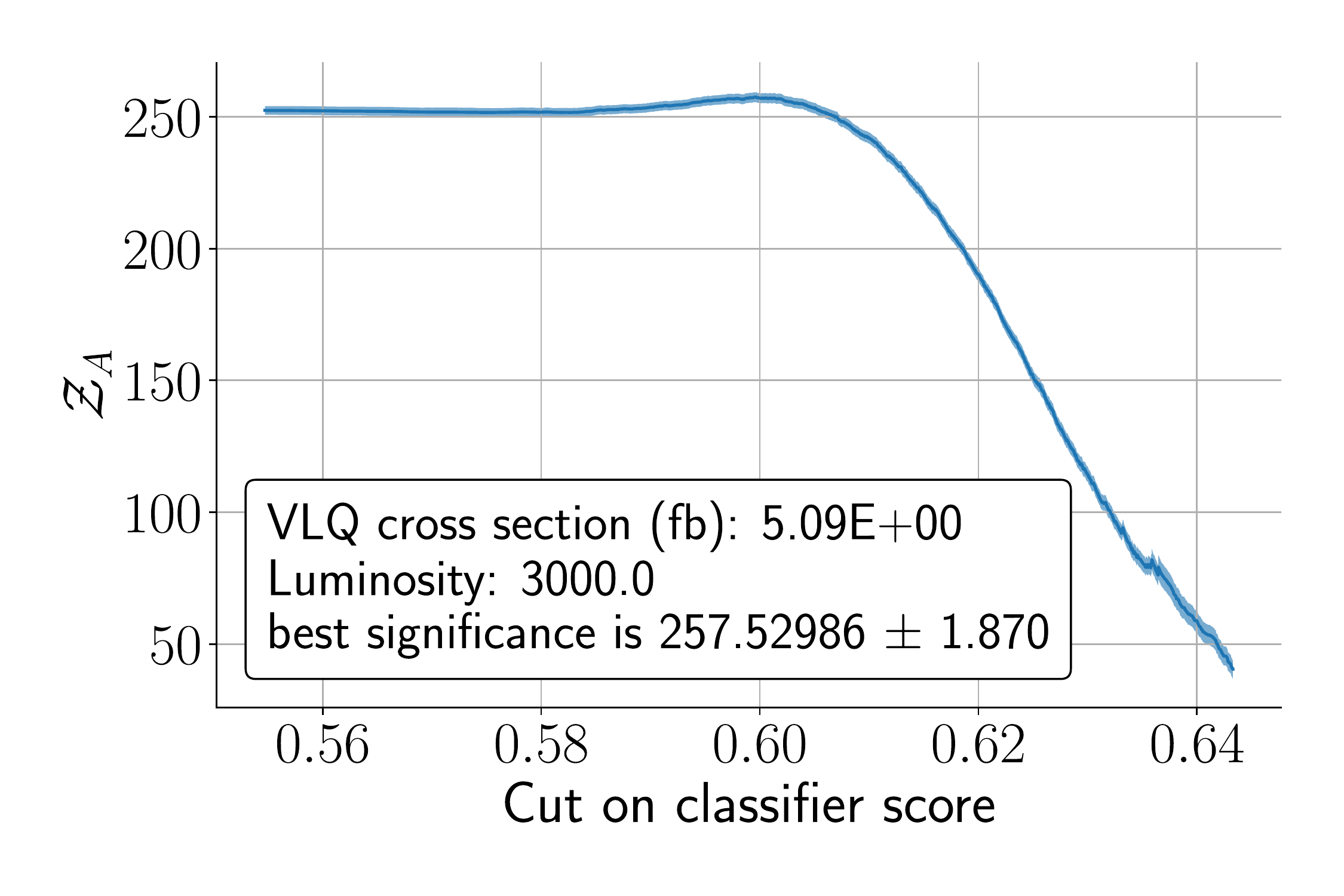} }} 
			\subfloat[]{{\includegraphics[width=0.32\textwidth]{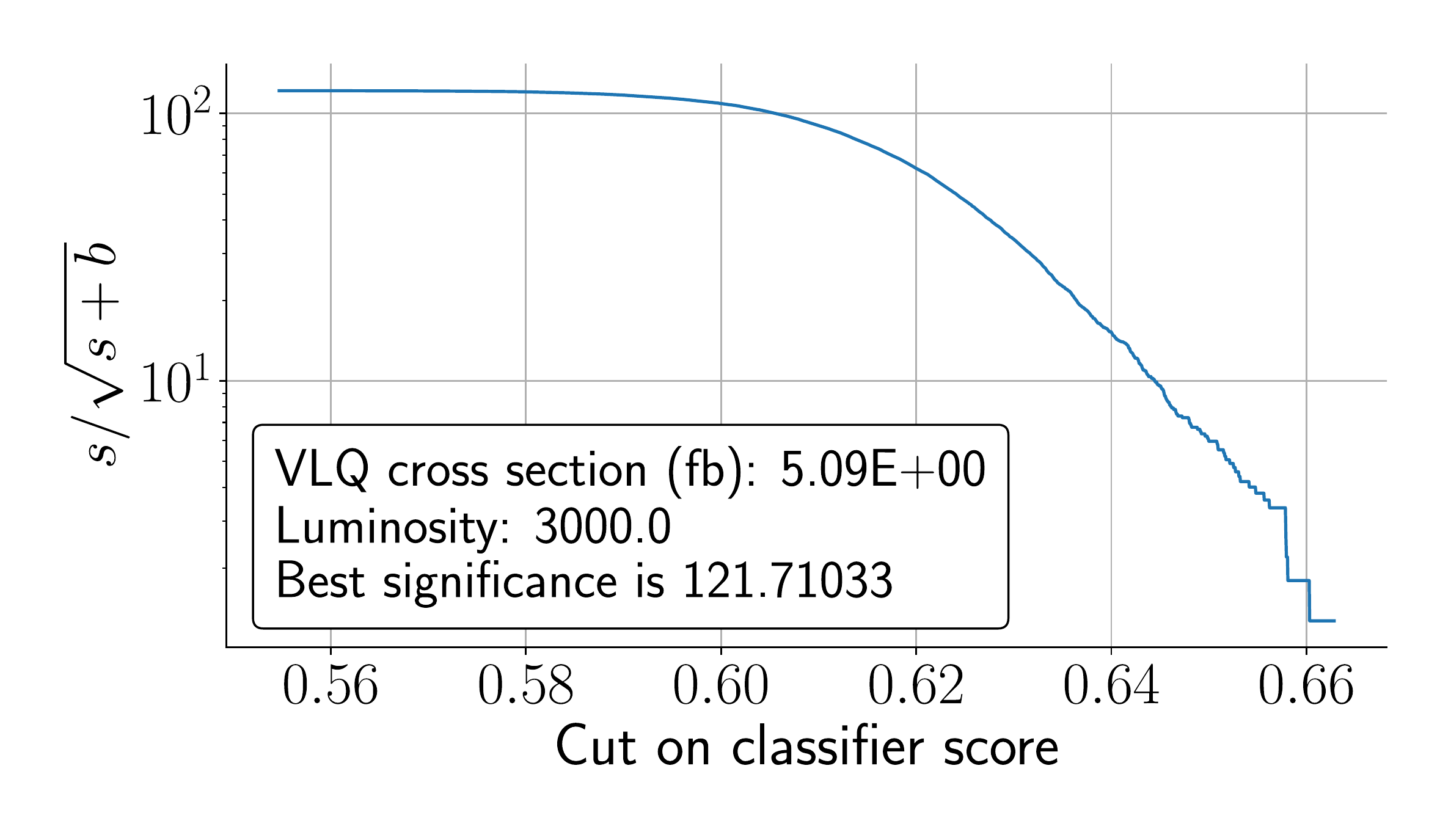} }} 
			\subfloat[]{{\includegraphics[width=0.32\textwidth]{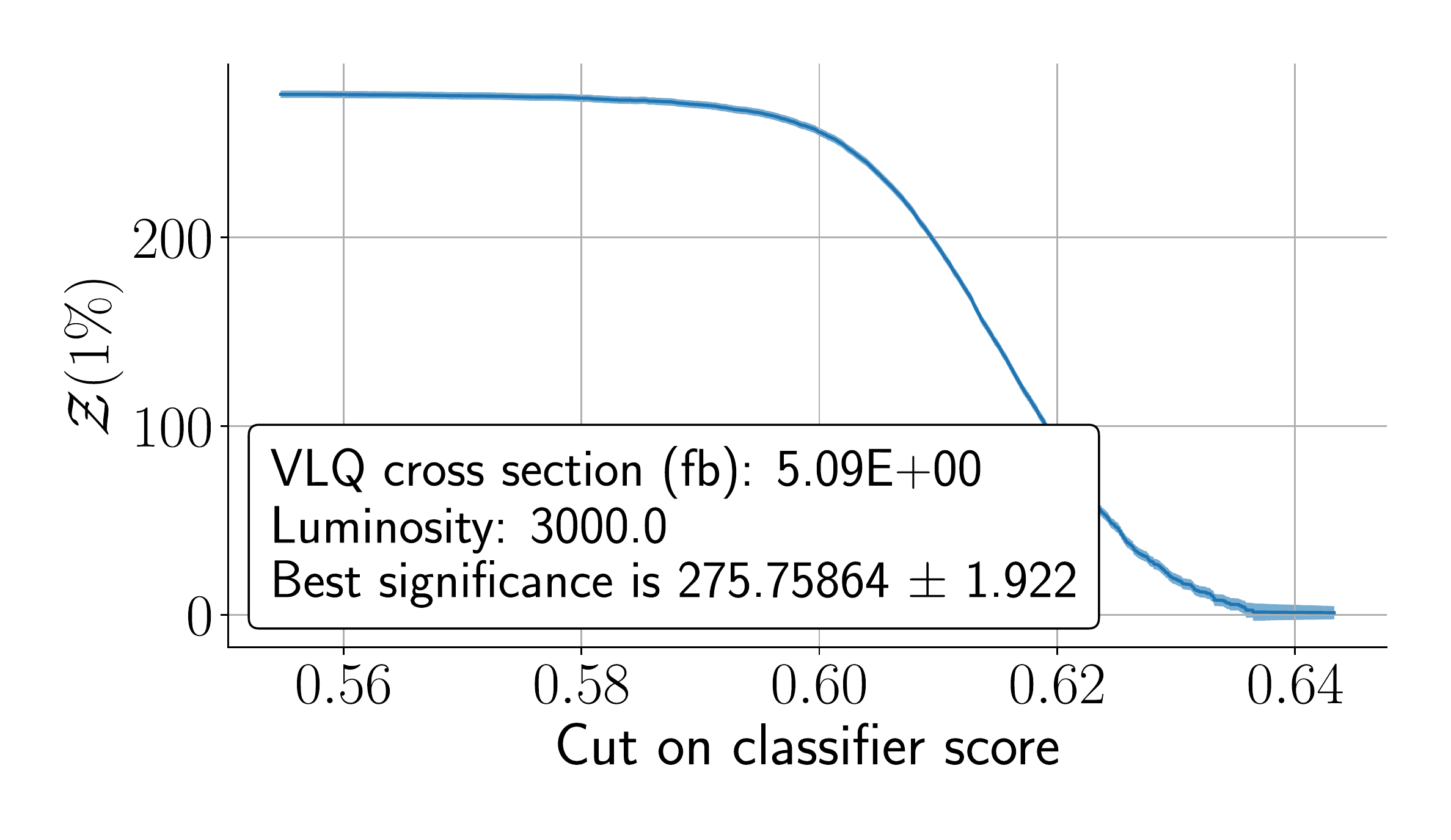} }}\\
			\caption{Statistical significance as a function of the classifier score given by NN for each signal graph, assuming the lightest VLQ with mass $m_{T_1} = 2.2~\mathrm{TeV}$ and a luminosity $\mathcal{L} = 3000$ $\mathrm{fb}^{-1}$. All showcased plots are representative of the best NN found, following an genetic algorithm that maximizes the Asimov significance. For (a) we present the Asimov significance, with 1\% systematics, for (b) -- the standard significance $s/\sqrt{s+b}$ and for (c) -- the adapted Asimov significance, where it is assumed that backgrounds are known with error $10^{-3}$.
				\label{fig:ACC-Sig-plots_1}}
		\end{figure*}

	\vspace*{1cm}
    \section{Relevant Feynman Rules for the collider analysis}\label{app:feynman}

    For the diagrams that we have included in this appendix, we have defined $\theta_W$ as the Weinberg angle, $U^u$, $U^e$ and $U_\nu$ are up-quark, charged lepton and neutrino mixing matrices, respectively, $g_2$ the SU(2) coupling constant, $g_1$ the U(1) gauge coupling constant and $\gamma_\mu$/$\gamma_5$ are Dirac gamma matrices. $\delta_{\alpha\beta}$ is a Kronecker delta in color space and $\delta_{ij}$ is a Kronecker delta in generation space.
    \begin{center} 
    \includegraphics[]{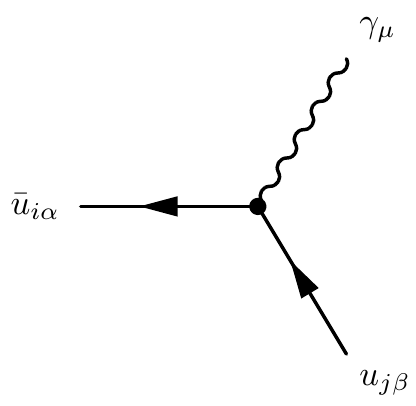} \hspace{3cm}
    \includegraphics[]{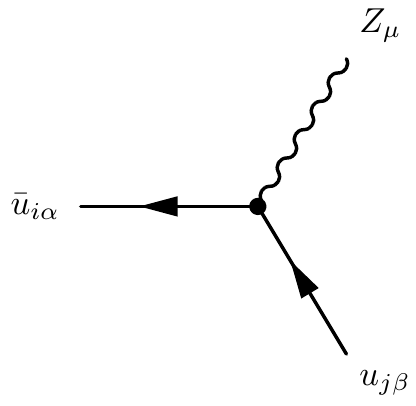}
	\end{center}
	
	The Feynman rule for the quark-antiquark-photon interactions has the form:	
	\begin{align} 
	&-\frac{i}{6} \delta_{\alpha \beta} \Big(U^{u,*}_{L,{j 1}} \Big(3 g_2 \sin\theta_W   + g_1 \cos\theta_W  \Big)U_{L,{i 1}}^{u} +U^{u,*}_{L,{j 2}} \Big(3 g_2 \sin\theta_W   + g_1 \cos\theta_W  \Big)U_{L,{i 2}}^{u} \nonumber \\ 
	&+g_1 U^{u,*}_{L,{j 3}} \cos\theta_W  U_{L,{i 3}}^{u} +3 g_2 U^{u,*}_{L,{j 3}} \sin\theta_W  U_{{L},{i 3}}^{u} +4 g_1 U^{u,*}_{L,{j 4}} \cos\theta_W  U_{L,{i 4}}^{u} \nonumber \\ 
	&+4 g_1 U^{u,*}_{L,{j 5}} \cos\theta_W  U_{L,{i 5}}^{u} \Big)\Big(\gamma_{\mu}\cdot\frac{1-\gamma_5}{2}\Big) - \frac{2 i}{3} g_1 \cos\theta_W  \delta_{\alpha \beta} \delta_{ij}  \Big(\gamma_{\mu}\cdot\frac{1+\gamma_5}{2}\Big),  
	\end{align}
	whereas the $Z$-boson quark-antiquark interaction is written as
	\begin{align} 
     &-\frac{i}{6} \delta_{\alpha \beta} \Big(U^{u,*}_{L,{j 1}} \Big(3 g_2 \cos\theta_W   - g_1 \sin\theta_W  \Big)U_{L,{i 1}}^{u} +U^{u,*}_{L,{j 2}} \Big(3 g_2 \cos\theta_W   - g_1 \sin\theta_W  \Big)U_{L,{i 2}}^{u} \nonumber \\ 
     &+3 g_2 U^{u,*}_{L,{j 3}} \cos\theta_W  U_{L,{i 3}}^{u} - g_1 U^{u,*}_{L,{j 3}} \sin\theta_W  U_{L,{i 3}}^{u} -4 g_1 U^{u,*}_{L,{j 4}} \sin\theta_W  U_{L,{i 4}}^{u} \nonumber \\ 
     &-4 g_1 U^{u,*}_{L,{j 5}} \sin\theta_W  U_{L,{i 5}}^{u} \Big)\Big(\gamma_{\mu}\cdot\frac{1-\gamma_5}{2}\Big)
     + \,\frac{2 i}{3} g_1 \delta_{\alpha \beta} \sin\theta_W  \delta_{ij} \Big(\gamma_{\mu}\cdot\frac{1+\gamma_5}{2}\Big).\end{align} 

	From the expressions given above, using the unitarity of the $U_{L}^{u}$ matrix and taking into account that $U^{u}_{L,{j n}}\lesssim 10^{-3}$ $(n=4,3)$,  (see eq.~\eqref{eq:quark_mixing}), it follows that the above given flavor violating photon interaction is strongly suppressed. Furthermore, it is worth mentioning that there is also a quark flavor violating $Z$ boson interaction, however its corresponding amplitude is suppressed by the same unitarity arguments and that amplitude is lower than the one corresponding to the photon by a factor of $\sim\tan\theta_W$. Because of this reason we do not show in Figure 5 the  subleading contribution arising from $Z$ violating quark interaction to the VLQ pair-production via gluon fusion mechanism.
    \begin{center} 
    \includegraphics[]{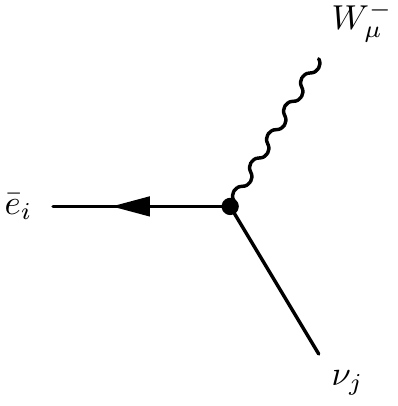}
	\end{center}  
	\begin{align} 
	&-i \frac{1}{\sqrt{2}} g_2 \sum_{a=1}^3 U^*_{{\nu},{j a}} U_{L,{i a}}^{e}  \Big(\gamma_{\mu}\cdot\frac{1-\gamma_5}{2}\Big)\end{align} 
    \cleardoublepage

	\bibliographystyle{apsrev4-1}
	\bibliography{Bib/Q6model}
	
\end{document}